\documentclass[a4paper,11pt]{article}
\usepackage{jheppub}
\usepackage{duckuments}
\usepackage[T1]{fontenc}
\usepackage[utf8]{inputenc}
\usepackage{lipsum}
\usepackage{amsmath,amssymb,amsfonts,bbold}
\usepackage{braket}
\usepackage{natbib}
\usepackage[dvipsnames,svgnames,x11names]{xcolor}
\usepackage{graphicx,subfigure}
\usepackage{tikz}
\usetikzlibrary{patterns, decorations.pathmorphing, arrows.meta, positioning}
\usepackage{hyperref}
\usepackage{mathtools}
\usepackage[normalem]{ulem}
\newcommand{\be}{\begin{equation}}
\newcommand{\ee}{\end{equation}}
\newcommand{\bse}{\begin{subequations}}
\newcommand{\ese}{\end{subequations}}
\newcommand{\ba}{\begin{eqnarray}}
\newcommand{\ea}{\end{eqnarray}}
\newcommand{\bea}{\begin{eqnarray}}
\newcommand{\eea}{\end{eqnarray}}

\begin{document}


\title{Creation of the Universe in Lorentzian Classical Gravity}

\author[1,2]{Avipsa Chakrabarti,}
\emailAdd{avipsachak.vimp@gmail.com}
\affiliation[1]{Instituto de F\'{\i}sica, Pontificia Universidad Cat\'{o}lica de Valpara\'{\i}so,
Avenida Universidad 330, Valpara\'{\i}so, Chile}
\affiliation[2]{Departamento de F\'{\i}sica, Universidad T\'{e}cnica Federico Santa Mar\'{\i}a,
Casilla 110-V, Valpara\'{\i}so, Chile}
\author[3]{Avik Chakraborty,}
\emailAdd{avik.phys88@gmail.com}
\affiliation[3]{Departamento de Ciencias F\'isicas, Facultad de Ciencias Exactas, Universidad Andr\'{e}s Bello,
Sazi\'e 2212, Piso 7, Santiago, Chile}
\author[1,2]{Mart\'in Molina,}
\emailAdd{martinmolinaramos95@gmail.com}
\author[1]{Ayan Mukhopadhyay,}
\emailAdd{ayan.mukhopadhyay@pucv.cl}
\author[1,2]{and Salom\'{o}n Romero}
\emailAdd{salo.romero1996@gmail.com}

\date{\today}

\abstract{
The singularity theorems and the past incompleteness of inflationary spacetimes suggest that the beginning of the Universe lies beyond the domain of semiclassical gravity. We show that the creation of the Universe can nevertheless be understood within Lorentzian gravity by allowing gravitational junctions, so that the full spacetime need not be a smooth manifold. We construct rotationally invariant cosmic creation gates joining any finite number $n\geq2$ of locally de Sitter regions in $D\geq3$ spacetime dimensions at a timelike perfect-fluid junction with $P=w\rho$. For $w>-(D-3)/(D-2)$, they are geodesically complete in both temporal directions. The total spatial volume approaches finite, positive limits in the infinite past and future on the global-time foliation adapted to the junction fluid. We also construct exact deformed cosmic creation gates through nonlinear homogeneous backreaction that retain the initial volume while the future volume grows without bound. In these solutions, the junction fluid always expands, cooling for $w>0$ and heating for allowed $w<0$. Keeping the bulk regions locally de Sitter, we determine the linearized perturbations: in three dimensions, the generic inhomogeneous sector contains one hybrid sound mode and $n-1$ relative transverse modes; in four and higher dimensions, only the homogeneous hybrid sound mode survives. A finite-volume beginning requires fine-tuned homogeneous and inhomogeneous initial data, while causal response generically leads to unbounded future spatial volume. We also discuss a broader creation criterion allowing bounded past-volume oscillations. The gravitational equations determine conditioned admissible histories. We discuss implications for cosmology.
}
\maketitle

\section{Introduction}
\label{Sec:intro}

The singularity theorems of Hawking, and subsequently Hawking and Penrose, have profoundly shaped our understanding of the origin of the Universe \cite{Hawking1966,HawkingPenrose1970}. They establish that, under suitable assumptions about the causal structure of spacetime and the energy conditions satisfied by matter, an expanding universe cannot be geodesically complete in the past. Inflation does not, by itself, resolve this difficulty. In particular, Borde, Guth and Vilenkin showed that spacetimes with sufficiently positive average expansion are necessarily past incomplete \cite{BordeGuthVilenkin2003}. These results have led to the widespread expectation that the beginning of the Universe lies beyond the domain of semiclassical gravity and requires a quantum description of spacetime, as envisaged, for instance, in the no-boundary and quantum tunneling proposals \cite{HartleHawking1983,Vilenkin1982}.

In this paper, we show that this limitation can be overcome by allowing gravitational junctions gluing multiple dynamical spacetime fragments, thereby relaxing the requirement that the global spacetime be a smooth manifold. We construct such spacetimes by joining a finite number of locally de Sitter regions in three, four and higher spacetime dimensions at a common junction (interface). The resulting geometries are complete to both the past and the future: inextendible timelike and null geodesics have infinite proper and affine length, respectively, in both temporal directions. Nevertheless, these spacetimes admit a precise sense of \textit{creation}: on the natural global-time foliation synchronized by the junction fluid clock, the total spatial volume approaches a finite, strictly positive value in the infinite past. They satisfy the classical Einstein equations with positive cosmological constant in each bulk region, together with the gravitational junction conditions at the interfaces.

Specifically, we assume that the equation of state of the junction fluid is $P=w\rho$, with $P$ and $\rho$ denoting the pressure and energy density, respectively. We find that for any $w>-(D-3)/(D-2)$, there exists a rotationally invariant \textit{cosmic creation gate} solution in which $n\geq2$ locally dS$_D$ regions can be joined such that the total spatial volume is finite \textit{both} in the infinite past and future. This is the simplest solution in the class described above. The temperature of the junction fluid remains constant for $w\neq0$, and its clock synchronizes the global-time foliations of the $n$ locally dS$_D$ regions defined by comoving geodesic observers at fixed spatial coordinates on the spheres. Incidentally, the two-way solution belongs to the class of \textit{cosmic gate} geometries studied in \cite{ForghaniMazharimousavi2022}, and related perfect-fluid shell constructions in arbitrary spacetime dimensions were also studied in \cite{deBoerLiskaSalehiVaziri2026}. However, as far as we are aware, the property of finite spatial volume in the infinite past and future on these natural cosmological foliations has not been previously established. Our multiway construction further allows three or more locally dS$_D$ spacetimes to be seeded together at a common perfect-fluid junction. These cosmic creation gates attain a maximal volume at an intermediate time in the natural foliation.

We also find a new class of exact rotationally invariant \textit{deformed cosmic creation gate} solutions. They approach the cosmic creation gates in the infinite past, retaining their finite, strictly positive past volume, and therefore also describe creation. Their total spatial volume, measured on the same natural global-time foliation, grows without bound in the future. The junction fluid temperature also evolves with its proper time in these solutions. For $w>0$, it decreases from a finite, strictly positive value in the infinite past to zero in the infinite future. For the allowed negative values of $w$ in four and higher dimensions, it instead increases from its finite past value and grows without bound in the infinite future.  These exact solutions describe the nonlinear backreaction of the homogeneous hybrid sound mode of the undeformed cosmic creation gate that decays towards the infinite past and grows towards the future.

Our general constructions follow the methodology for defining and solving gravitational junctions developed in \cite{Banerjee:2024sqq,Chakraborty:2025jtj} based on an earlier formulation in \cite{Bachas:2020yxv}. For purely tensile junctions in three spacetime dimensions, the number of independent junction conditions equals the number of variables that specify the embeddings of the common junction in the $n$ bulk regions \cite{Banerjee:2024sqq,Chakraborty:2025jtj}. In four and higher dimensions, the junction conditions generically form an overdetermined system \cite{Chakraborty:2025jtj}, and sufficient symmetry allows well-behaved solutions.

For a purely tensile $n$-way junction in three dimensions, there are $3n-2$ embedding variables, of which $n-1$ are emergent dynamical degrees of freedom. They correspond to independent pairwise differences or averages of the transverse junction displacements, depending on the orientations of the normals. Remarkably, they are described in terms of solutions of the non-linear Nambu--Goto equations for $n-1$ strings coupled through specific Monge--Amp\`ere interactions. Once these $n-1$ degrees of freedom and the rigid parameters associated with isometries are specified, the remaining $2n-1$ embedding variables are determined by the junction conditions. These constructions were examined specifically for purely tensile junctions joining $n$ dS$_3$ spacetimes in \cite{Chakraborty:2025baw}. Transient Nambu--Goto vibrations of the $n-1$ strings about the equator, with amplitudes decaying in both the infinite past and future, were shown to yield well-behaved, single-valued solutions of the junction conditions.

We replace the purely tensile junction with a perfect fluid junction and analyse perturbations of the simplest rotationally invariant $n$-way solution with finite total volume in the infinite past. Remarkably, the equality between the number of independent junction conditions and the number of variables persists for the $n$-way junction in three spacetime dimensions. In this case, the velocity and temperature fields of the junction fluid add two more variables, so that the $n$-way junction has $3n$ variables in total along with an equal number of independent junction conditions. 

For generic inhomogeneous perturbations, these $3n$ variables contain $n$ independent dynamical degrees of freedom. Of these, $n-1$ are associated with independent pairwise differences or averages of the transverse junction displacements, depending on the orientations of the normals. The additional degree of freedom is a hybrid sound mode with dispersion relation
\[
\omega_m^2=(1+w)(m^2-2)
=w k_m^2-2(1+w),
\qquad
k_m=\frac{m}{R_0},
\qquad
R_0^2=\frac{w}{1+w}.
\]
Here the de Sitter radius is set to unity, $\omega_m$ is measured in junction proper time, and $m\in\mathbb Z$ labels the Fourier harmonics on the spatial circle of the junction. The coupling to the junction geometry thus shifts the usual sound dispersion relation $\omega_m^2=w k_m^2$, while preserving the short-wavelength sound speed $c_s^2=w$. The homogeneous mode $m=0$ has two exponential branches, growing towards the infinite past and the infinite future, respectively. The analogous homogeneous radial stability problem for four-dimensional two-way cosmic gate geometries was studied in \cite{ForghaniMazharimousavi2022} without the hybrid sound interpretation. The $|m|=1$ modes are associated with isometries, while the $|m|\geq2$ hybrid sound modes are physical oscillatory modes. For these higher harmonics, the remaining $n-1$ relative transverse modes grow towards at least one of the two temporal ends. At linear order, there are therefore no nonzero free modes whose physical amplitudes decay in both the infinite past and future, unlike the transient modes of the purely tensile construction. The relative transverse modes satisfy fluid-dependent equations with elliptic principal part for $w>0$, and the coupled Nambu--Goto--Monge--Amp\`ere description of purely tensile junctions does not directly extend to them.

The modes that grow towards the infinite past show that the cosmological histories constructed here require fine-tuned initial conditions. Requiring a controlled semiclassical description arbitrarily far back in time, with finite spatial volume on the natural foliation synchronized by the junction fluid clock, therefore selects a special family of solutions. The initial-value problem incorporates this condition on the past while allowing future growth, with no corresponding bound on the late-time spatial volume. In three dimensions, we compute the second-order backreaction of the inhomogeneous sound modes and find an additional restriction when a definite past-volume limit is required: each harmonic present in the infinite past must be chiral, with the direction of propagation chosen independently for different harmonics. The permitted waves give an amplitude-dependent correction to the limiting volume. We also discuss a broader interpretation of creation in which the past volume remains bounded and bounded away from zero while continuing to oscillate.

We further study perturbations of the rotationally invariant perfect-fluid cosmic creation gate joining $n$ locally dS$_4$ regions, whose total spatial volume approaches finite, positive values in both the infinite past and future on the natural cosmological foliation. As in the purely tensile case, the junction conditions form a formally overdetermined system. At linear order, only the homogeneous ($\ell=0$) hybrid sound modes survive, apart from deformations associated with isometries. Their two independent exponential branches grow towards the infinite past and future, respectively, and maintaining the prescribed past requires excluding the former. In both three and four dimensions, the branch that decays towards the past admits an exact nonlinear continuation to the $n$-way deformed cosmic creation gate discussed above. This homogeneous backreaction leaves the past volume unchanged while the future volume grows without bound.

The existence of a regular past and the presence of instabilities are thus compatible features of our construction. Together, they illustrate how \textit{the gravitational equations and junction conditions determine the allowed cosmological histories, conditioned on the past spatial volume remaining between finite, strictly positive bounds on the natural foliation synchronized by the junction fluid clock.} We show that a causal response can be consistently defined while allowing the future spatial volume to grow without bound.

The plan of the paper is as follows. In Sec~\ref{Sec:setup}, we formulate the junction conditions and identify the independent variables and dynamical degrees of freedom. Sec~\ref{Sec:creation-gates} presents the rotationally invariant creation gates and determines their limiting and maximal spatial volumes. Their linear perturbations in three dimensions are analysed for two-way and multiway junctions in Sec~\ref{Sec:two-way-perturbations-three} and Sec~\ref{Sec:n-way-perturbations-three}, respectively. Sec~\ref{Sec:higher-dimensional-perturbations} extends the analysis to four and higher dimensions, where only the homogeneous hybrid sound mode survives, apart from isometries. In Sec~\ref{Sec:admissible-histories}, we construct exact deformed creation gates with unbounded future volume, study the restrictions imposed by second-order inhomogeneous backreaction, and discuss conditioned admissible histories, the Cauchy problem and retarded response. We conclude in Sec~\ref{Sec:conclusions} where we discuss the implications for cosmology.

\section{General formulation and properties of the gravitational junctions in three and more dimensions}
\label{Sec:setup}
We introduce the setup by reviewing the construction of multiway junctions in \cite{Banerjee:2024sqq,Chakraborty:2025jtj,Chakraborty:2025baw}. We then include the perfect fluid in the gravitational action and determine the numbers of variables, independent junction conditions and dynamical degrees of freedom. The construction applies in all spacetime dimensions greater than or equal to three.

\subsection{Gravitational $2$-way junction} 
\label{Sec:two_jn}

\label{Sec:set_up}

We begin the discussion with a gravitational 2-way junction in dS$_3$. The extension to four and higher dimensions will follow directly. Consider the global dS$_3$ spacetime $\mathcal{M}$, a solution of the three-dimensional Einstein equations with positive cosmological constant $\Lambda=L^{-2}$. Its metric is
\begin{equation} \label{Eq:dsmetric}
ds^2=-dt^2+L^2 \cosh^2 \left(\frac{t}{L}\right) \left( d\theta^2+\sin^2\theta d\phi^2 \right)\,.
\end{equation}
Here $t$ is the global time, $\theta$ and $\phi$ are the polar and azimuthal angles on the spatial $S^2$, and $L$ is the de Sitter radius. The sphere contracts from infinite area in the infinite past, $t\to-\infty$, to a minimum area $4\pi L^2$ at $t=0$, before expanding again without bound.

Let us consider two \textit{identical} copies $\mathcal{M}_{1,2}$ of this spacetime. Each of the copies $\mathcal{M}_{1,2}$ is divided into two fragments, northern ($N$) and southern ($S$), by distinct codimension-$1$ hypersurfaces $\Sigma_{1,2}$. The coordinate charts inherited from $\mathcal{M}$ distinguish the northern and southern fragments $\mathcal{M}_{iN}$ and $\mathcal{M}_{iS}$ in each $\mathcal{M}_{i}$. To construct a gravitational junction $\Sigma$, a codimension-one hypersurface, we glue one fragment of $\mathcal{M}_1$, say $N$, to one fragment of $\mathcal{M}_2$, say $S$. We denote the retained fragments by $\mathcal{M}_{i \alpha_i}$, with $i=1,2$ and the corresponding $\alpha_i = N,S$. The full spacetime $\widetilde{\mathcal{M}}$ together with the junction $\Sigma$ is formed by identifying a point $P_1$ on $\Sigma_1$ with a point $P_2$ on $\Sigma_2$. Therefore, $\Sigma_{1,2}$ should be regarded as the images of the junction $\Sigma$ in $\mathcal{M}_{1,2}$, respectively. The \textit{embeddings} of $\Sigma_i$ in $\mathcal{M}_i$ and the \textit{identifications} of the points $P_i$ of $\Sigma_i$ should satisfy the gravitational junction conditions at $\Sigma$, and the full spacetime $\widetilde{\mathcal{M}}$ must satisfy Einstein's equations.

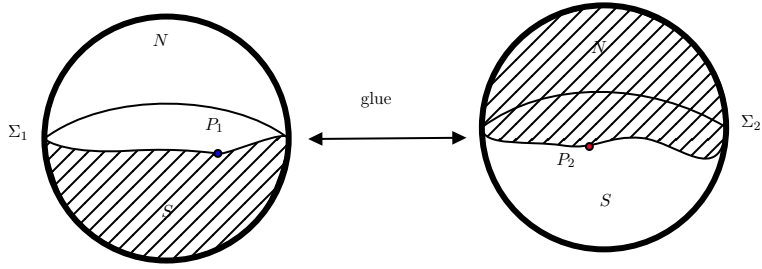
\begin{figure}[ht]
    \centering
    \tikzset{
    pattern size/.store in=\mcSize, 
    pattern size = 5pt,
    pattern thickness/.store in=\mcThickness, 
    pattern thickness = 0.3pt,
    pattern radius/.store in=\mcRadius, 
    pattern radius = 1pt}
    \makeatletter
    \pgfutil@ifundefined{pgf@pattern@name@_fl5un0t35}{
    \pgfdeclarepatternformonly[\mcThickness,\mcSize]{_fl5un0t35}
    {\pgfqpoint{0pt}{0pt}}
    {\pgfpoint{\mcSize+\mcThickness}{\mcSize+\mcThickness}}
    {\pgfpoint{\mcSize}{\mcSize}}
    {
    \pgfsetcolor{\tikz@pattern@color}
    \pgfsetlinewidth{\mcThickness}
    \pgfpathmoveto{\pgfqpoint{0pt}{0pt}}
    \pgfpathlineto{\pgfpoint{\mcSize+\mcThickness}{\mcSize+\mcThickness}}
    \pgfusepath{stroke}
    }}
    \makeatother

    \tikzset{
    pattern size/.store in=\mcSize, 
    pattern size = 5pt,
    pattern thickness/.store in=\mcThickness, 
    pattern thickness = 0.3pt,
    pattern radius/.store in=\mcRadius, 
    pattern radius = 1pt}
    \makeatletter
    \pgfutil@ifundefined{pgf@pattern@name@_3pcaj0uq4}{
    \pgfdeclarepatternformonly[\mcThickness,\mcSize]{_3pcaj0uq4}
    {\pgfqpoint{0pt}{0pt}}
    {\pgfpoint{\mcSize+\mcThickness}{\mcSize+\mcThickness}}
    {\pgfpoint{\mcSize}{\mcSize}}
    {
    \pgfsetcolor{\tikz@pattern@color}
    \pgfsetlinewidth{\mcThickness}
    \pgfpathmoveto{\pgfqpoint{0pt}{0pt}}
    \pgfpathlineto{\pgfpoint{\mcSize+\mcThickness}{\mcSize+\mcThickness}}
    \pgfusepath{stroke}
    }}
    \makeatother

    \tikzset{every picture/.style={line width=0.75pt}} 

    \begin{tikzpicture}[scale=0.8, every node/.style={scale=0.6}, x=0.75pt,y=0.75pt,yscale=-1,xscale=1]

    \draw  [line width=2.25]  (81,127.5) .. controls (81,85.25) and (115.25,51) .. (157.5,51) .. controls (199.75,51) and (234,85.25) .. (234,127.5) .. controls (234,169.75) and (199.75,204) .. (157.5,204) .. controls (115.25,204) and (81,169.75) .. (81,127.5) -- cycle ;
    \draw  [line width=2.25]  (355,120.5) .. controls (355,78.25) and (389.25,44) .. (431.5,44) .. controls (473.75,44) and (508,78.25) .. (508,120.5) .. controls (508,162.75) and (473.75,197) .. (431.5,197) .. controls (389.25,197) and (355,162.75) .. (355,120.5) -- cycle ;
    \draw    (81,127.5) .. controls (121,97.5) and (197.5,99.17) .. (234,127.5) ;
    \draw  [pattern=_fl5un0t35,pattern size=6pt,pattern thickness=0.75pt,pattern radius=0pt, pattern color={rgb, 255:red, 0; green, 0; blue, 0}] (81,127.5) .. controls (83.17,130.17) and (97.84,137.25) .. (133.5,135.17) .. controls (169.16,133.08) and (185.74,137.06) .. (191.83,136.83) .. controls (197.92,136.6) and (233.63,120.9) .. (234,127.5) .. controls (234.37,134.1) and (224.5,172.17) .. (207.5,186.17) .. controls (190.5,200.17) and (174.82,204.69) .. (154.5,203.83) .. controls (134.18,202.98) and (117.5,191.83) .. (103.83,183.17) .. controls (90.17,174.5) and (78.83,124.83) .. (81,127.5) -- cycle ;
    \draw  [fill={rgb, 255:red, 15; green, 12; blue, 222 }  ,fill opacity=0.99 ] (187.5,136.83) .. controls (187.5,135.64) and (188.47,134.67) .. (189.67,134.67) .. controls (190.86,134.67) and (191.83,135.64) .. (191.83,136.83) .. controls (191.83,138.03) and (190.86,139) .. (189.67,139) .. controls (188.47,139) and (187.5,138.03) .. (187.5,136.83) -- cycle ;
    \draw    (355,120.5) .. controls (395,90.5) and (460,91.17) .. (508,120.5) ;
    \draw  [pattern=_3pcaj0uq4,pattern size=6pt,pattern thickness=0.75pt,pattern radius=0pt, pattern color={rgb, 255:red, 0; green, 0; blue, 0}] (392.67,54.83) .. controls (402.96,49.69) and (421.59,43.01) .. (438.67,44.83) .. controls (455.74,46.66) and (472.9,57.8) .. (476,58.83) .. controls (479.1,59.87) and (511,81.83) .. (508,120.5) .. controls (505,159.17) and (476.53,129.09) .. (455,127.17) .. controls (433.47,125.25) and (426.51,135.08) .. (405.67,131.83) .. controls (384.83,128.59) and (355.12,133.18) .. (355,120.5) .. controls (354.88,107.82) and (356.41,99.7) .. (363,86.5) .. controls (369.59,73.3) and (382.38,59.98) .. (392.67,54.83) -- cycle ;
    \draw  [fill={rgb, 255:red, 222; green, 12; blue, 41 }  ,fill opacity=0.99 ] (420.17,132.5) .. controls (420.17,131.3) and (421.14,130.33) .. (422.33,130.33) .. controls (423.53,130.33) and (424.5,131.3) .. (424.5,132.5) .. controls (424.5,133.7) and (423.53,134.67) .. (422.33,134.67) .. controls (421.14,134.67) and (420.17,133.7) .. (420.17,132.5) -- cycle ;
    \draw    (249.2,127.78) -- (342.4,127.02) ;
    \draw [shift={(345.4,127)}, rotate = 179.54] [fill={rgb, 255:red, 0; green, 0; blue, 0 }  ][line width=0.08]  [draw opacity=0] (8.93,-4.29) -- (0,0) -- (8.93,4.29) -- cycle    ;
    \draw [shift={(246.2,127.8)}, rotate = 359.54] [fill={rgb, 255:red, 0; green, 0; blue, 0 }  ][line width=0.08]  [draw opacity=0] (8.93,-4.29) -- (0,0) -- (8.93,4.29) -- cycle    ;

    \draw (147.83,60.57) node [anchor=north west][inner sep=0.75pt]    {$N$};
    \draw (153.17,168.23) node [anchor=north west][inner sep=0.75pt]    {$S$};
    \draw (57.17,117.9) node [anchor=north west][inner sep=0.75pt]    {$\Sigma _{1}$};
    \draw (180.5,113.23) node [anchor=north west][inner sep=0.75pt]    {$P_{1}$};
    \draw (422.33,65.23) node [anchor=north west][inner sep=0.75pt]    {$N$};
    \draw (427.67,160.9) node [anchor=north west][inner sep=0.75pt]    {$S$};
    \draw (516.33,111.9) node [anchor=north west][inner sep=0.75pt]    {$\Sigma _{2}$};
    \draw (400.67,135.23) node [anchor=north west][inner sep=0.75pt]    {$P_{2}$};
    \draw (278,97.2) node [anchor=north west][inner sep=0.75pt]    {glue};

    \end{tikzpicture}

    \caption{A gravitational junction formed by gluing two dS$_3$ manifolds $\mathcal{M}_1$ and $\mathcal{M}_2$ ($2$-way junction). The points $P_1$ and $P_2$ on $\Sigma_1$ and $\Sigma_2$, respectively, are identified. The shaded portions of $\mathcal{M}_1$ and $\mathcal{M}_2$ are discarded  before gluing.}
    \label{fig:setup}
\end{figure}

Let the coordinates of $\mathcal{M}$ be $(t,\theta,\phi)$ (with $t$ being the time coordinate and $\theta,\phi$ being the angular coordinates), and so the fragments $\mathcal{M}_{i \alpha_i}$ inherit the coordinates $(t_i$, $\theta_i$, $\phi_i)$.  Now, as shown in Fig. \ref{fig:setup}, we identify unique points $P_i$ in $\Sigma_i$  to each point $P$ in $\Sigma$ with coordinates $(\tau,\sigma)$. We fix the worldsheet gauge freedom by choosing the junction coordinates $(\tau,\sigma)$ as follows
\begin{equation} \label{WSgauge}
     \tau(P)=\frac{t_1(P_1)+t_2(P_2) }{2}\,, \ \sigma(P)=\frac{\phi_1(P_1)+\phi_2 (P_2)}{2}\,.  
\end{equation}
Therefore, the embeddings of $\Sigma_i$ in $\mathcal{M}_i$ assume the form
\begin{align} \label{Eq:embed}
    &\Sigma_{1}: t_1 = \tau + \tau_d(\tau,\sigma), \,\, \phi_1 = \sigma + \sigma_d(\tau,\sigma),\,\, \theta_{1} = \mathfrak{f}_{1}(\tau,\sigma),\nonumber\\
    &\Sigma_{2}: t_2 = \tau - \tau_d(\tau,\sigma), \,\, \phi_2 = \sigma - \sigma_d(\tau,\sigma),\,\, \theta_{2} = \mathfrak{f}_{2}(\tau,\sigma).
\end{align}
Thus, in total, we have the following four functions, namely 
\begin{equation} \label{Eq:varbs}
    \tau_d = \frac{t_1 - t_2}{2}, \,\ \sigma_d = \frac{\phi_1 - \phi_2}{2},\, \ \theta_{s} = \frac{\mathfrak{f}_1 + \mathfrak{f}_2}{2},\,\,\theta_{d} = \frac{\mathfrak{f}_1 - \mathfrak{f}_2}{2},
\end{equation}
that need to be determined as functions of the worldsheet coordinates $\tau$ and $\sigma$, by solving the gravitational junction conditions. The relative shifts $\tau_d$ and $\sigma_d$ are invariant under bulk diffeomorphisms whose induced time and angular shifts agree across the junction. If these diffeomorphisms vanish at the junction, fixing it pointwise, the average coordinates $\tau$ and $\sigma$ are also invariant.

The full gravitational action that yields the junction conditions is
\begin{align}\label{Eq:bulk-action}
   & S=\frac{1}{16\pi G_N}\int_{\widetilde{\mathcal{M}}}d^3x \sqrt{-g}(R - 2\Lambda)+\int_{\Sigma}{\rm d}\tau{\rm d}\sigma\,\sqrt{-\gamma} \, P(T)+{\rm GHY \ terms}\,,
\end{align}
where the bulk metric $g$ and the fluid variables on $\Sigma$ are the \textit{only} degrees of freedom, GHY denotes the Gibbons--Hawking--York terms corresponding to the boundaries of $\mathcal{M}_{i \alpha_i}$, and $P(T)$ is the pressure of the junction fluid, fixed by its equation of state. The variation of the action with respect to $g$ away from the junction $\Sigma$ implies that each fragment $\mathcal{M}_{i \alpha_i}$ is an Einstein manifold. The action \eqref{Eq:bulk-action} assumes that the induced metrics on $\Sigma_i$ are identical at each point $P(\tau,\sigma)$ on the junction $\Sigma$ (first junction condition), thus defining the worldsheet metric via
\begin{equation}\label{Eq:hcont}
    \gamma_{\mu\nu}(\tau,\sigma):=\gamma_{1,\mu\nu}(\tau,\sigma)= \gamma_{2,\mu\nu}(\tau,\sigma)\,.
\end{equation}
Variation of the action with respect to $g$ at the junction $\Sigma$ gives
\begin{equation}\label{Eq:Kdisc}
    \sum_{i=1}^2 (-1)^{s(\alpha_i)}\left(K_{i,\mu\nu} - K_i\,\gamma_{i,\mu\nu}\right) = 8\pi G_N S_{\mu\nu},
\end{equation}
with $s(\alpha_i) = 0$ if $\alpha_i = N$ and $s(\alpha_i) = 1$ if $\alpha_i = S$. Here, $K_{i,\mu\nu}$ are the extrinsic curvatures of $\Sigma_i$ in $\mathcal{M}_{i\alpha_i}$, computed as the pullbacks of $\nabla_a n_b$ with the reference normals pointing towards increasing $\theta_i$, and $K_i = \gamma^{\mu\nu}K_{i,\mu\nu}$. $S_{\mu\nu}$ is the perfect fluid stress tensor, which will be determined below. The first condition equates the intrinsic metrics induced from the two sides. Since $\gamma_{\mu\nu}$ is a symmetric two-dimensional metric, \eqref{Eq:hcont} gives three equations. The left hand side of \eqref{Eq:Kdisc} is the oriented sum of the Brown--York tensors of the two fragments.

In order to obtain $S_{\mu\nu}$, we need to clarify the degrees of freedom of the fluid first. Since the fluid is in local equilibrium, it chooses a frame $u^\mu$ locally which satisfies $u^\mu u^\nu\gamma_{\mu\nu}=-1$ and in which the energy momentum tensor $S^\mu_{\,\,\nu}=\gamma^{\mu\rho}S_{\rho\nu}$ is diagonal. Assuming the possibility of Euclidean continuation of the spacetime, we set the time circle to have period $\beta = T^{-1}$, so that we can define the local inverse-temperature vector $\beta^\mu = T^{-1}u^\mu$. Note that $T = (-\gamma_{\mu\nu}\beta^\mu\beta^\nu)^{-1/2}$. We can combine the degrees of freedom $T$ and $u^\mu$ into the vector $\beta^\mu$. The variation of temperature with respect to the metric is then obtained by taking the vector $\beta^\mu$ fixed, so that 
\[
\delta T=\frac{T^3}{2}\beta^\mu\beta^\nu \delta\gamma_{\mu\nu}= \frac{T}{2} u^\mu u^\nu \delta\gamma_{\mu\nu}.
\]
The thermodynamic identities
\be
{\rm d}\rho = T {\rm d} s, \quad \rho + P = Ts,
\ee
with $s$ denoting the entropy density, yield 
\[
{\rm d}P = s {\rm d}T = \frac{\rho + P}{T} {\rm d}T.
\]
Therefore, writing $S_\Sigma=\int_\Sigma {\rm d}\tau\,{\rm d}\sigma\,\sqrt{-\gamma}P(T)$, we find the stress tensor as
\[
S^{\mu\nu}=\frac{2}{\sqrt{-\gamma}}\frac{\delta S_\Sigma}{\delta\gamma_{\mu\nu}} =2 \left(\frac{{\rm d}P}{{\rm d}T}\frac{\delta T}{\delta{\gamma_{\mu\nu}}}+ \frac{1}{2} P \gamma^{\mu\nu}\right)
\]
with the metric variation taken at fixed \(\beta^\mu\). Substitution gives
\begin{equation}
S^{\mu\nu} = (\rho + P)u^\mu u^\nu + P\gamma^{\mu\nu}. 
\end{equation}
The conservation law $\nabla_\mu S^{\mu\nu} =0 $ gives the relativistic Euler equations, with $\nabla$ the covariant derivative on the junction:
\begin{align}\label{Eq:Euler}
    & D\, \ln T = - c_s^2 \nabla\cdot u \Rightarrow\nabla_\mu (su^\mu) =0,\nonumber\\
    &\nabla_{\perp\mu} \ln T =  - D u_\mu = -a_\mu,   
\end{align}
where $c_s^2 = {\rm d}P/{\rm d}\rho$ is the square of the speed of sound, $D = (u\cdot \nabla)$ is the convective derivative, $\nabla_{\perp\mu}= (u_\mu u^\nu + \delta_\mu^{\,\,\nu})\nabla_\nu$ is the spatial derivative in the direction orthogonal to the flow and $a^\mu$ is the fluid acceleration. We choose the equation of state
\begin{equation}
    P(T) = P_0\left(\frac{T}{T_0}\right)^{\frac{1+w}{w}} \,\,\Rightarrow\,\,\rho = T\frac{{\rm d}P}{{\rm d}T}-P =\frac{P}{w}\,\, \Rightarrow\,\, P = w\rho,
\end{equation}
so that
\begin{equation}
    c_s^2 = w.
\end{equation}
Stable causal sound propagation demands that $0 \leq c_s^2 = w \leq 1$. We take $w<1$ below and discuss the regular limit $w\to1^-$.\footnote{The temperature parametrization above assumes $w\neq0$. We include $w=0$ separately as pressureless dust, described by its conserved density and geodesic velocity field.} The junction modifies sound propagation, so the analysis of causality and stability should be repeated for the coupled system. For $D\geq 4$, the junction conditions also admit background solutions with $-\frac{D-3}{D-2}<w<0$.

The contracted Codazzi equations, together with the bulk Einstein equations, imply that the signed sum of the trace-adjusted extrinsic curvatures on the left hand side of \eqref{Eq:Kdisc} is conserved along the junction. This conservation law gives the Euler equations \eqref{Eq:Euler}. Therefore, in total we have $6$ independent junction conditions, 3 each from \eqref{Eq:hcont} and \eqref{Eq:Kdisc}. The six independent variables are $u^\phi$, $T$, $\theta_s$, $\theta_d$, $\tau_d$ and $\sigma_d$.

The construction extends directly to two-way junctions joining locally
dS$_D$ regions, with $D\geq4$. The bulk metric and cosmological constant are
\begin{equation}
ds^2=-dt^2+L^2\cosh^2\left(\frac{t}{L}\right)
\left(d\theta^2+\sin^2\theta\,d\Omega_{D-2}^2\right),
\qquad
\Lambda=\frac{(D-1)(D-2)}{2L^2},
\end{equation}
where $d\Omega_{D-2}^2$ is the metric on the unit $S^{D-2}$.
Writing its angular coordinates as $\varphi^A$, with
$A=1,\ldots,D-2$, we choose junction coordinates $(\tau,\sigma^A)$.
The embeddings take the local form
\begin{equation}
t_{1,2}=\tau\pm\tau_d,\qquad
\varphi_{1,2}^A=\sigma^A\pm\sigma_d^A,\qquad
\theta_{1,2}=\theta_s\pm\theta_d,
\end{equation}
with all embedding functions depending on $(\tau,\sigma^A)$.
The action \eqref{Eq:bulk-action} has bulk measure ${\rm d}^Dx$
and junction measure ${\rm d}\tau\,{\rm d}^{D-2}\sigma$.
The junction conditions \eqref{Eq:hcont} and \eqref{Eq:Kdisc}
retain their tensorial form.

Within this locally de Sitter class, the junction coordinates chosen above leave $D+1$ independent embedding variables. The temperature and
$D-2$ independent spatial velocity components supply another $D-1$
variables, giving $2D$ in total. Each junction condition has
$D(D-1)/2$ components, giving $D(D-1)$ equations altogether.
The Euler equations again follow from the contracted Codazzi equations
and impose no additional independent conditions. In four dimensions,
there are therefore eight variables and twelve junction equations.
For $D>3$, the junction conditions form a formally overdetermined system.

\subsection{Gravitational $n$-way junction}
\label{Sec:n_jn}
We now consider a gravitational $n$-way junction in dS$_3$, with $n\geq3$. Let us consider $n$ \textit{identical} copies $\mathcal{M}_i$, with $i=1,\cdots,n$, of the spacetime $\mathcal{M}$ with metric \eqref{Eq:dsmetric}. Each copy is divided into northern and southern fragments $\mathcal{M}_{iN}$ and $\mathcal{M}_{iS}$ by a codimension-one hypersurface $\Sigma_i$. We retain one fragment $\mathcal{M}_{i\alpha_i}$ from each copy, with $\alpha_i=N,S$. The full spacetime $\widetilde{\mathcal{M}}$ together with the junction $\Sigma$ is formed by identifying a point $P_i$ on each $\Sigma_i$ with a common point $P$ on $\Sigma$. Therefore, the $\Sigma_i$ should be regarded as the images of the junction $\Sigma$ in the $\mathcal{M}_i$. The embeddings of these images and the identifications of their points should satisfy the gravitational junction conditions.

The fragments $\mathcal{M}_{i\alpha_i}$ inherit the coordinates $(t_i,\theta_i,\phi_i)$ of $\mathcal{M}$. The embeddings of $\Sigma_i$ in $\mathcal{M}_i$ are specified by the $n$ functions
\begin{equation}\label{Eq:embed_n}
  \Sigma_i:\,\,  \theta_i = \mathfrak{f}_{i}(t_i, \phi_i)\,, \ \ i = 1, 2,\cdots,n\,.
\end{equation}
We choose coordinates $(\tau,\sigma)$ on $\Sigma$ and fix the worldsheet gauge freedom as follows
\begin{align}\label{Eq:WSgauge}
    \tau(P) = \frac{1}{n}\sum_{i=1}^n t_i(P_i), \quad \sigma(P) = \frac{1}{n}\sum_{i=1}^n \phi_i(P_i)\,.
\end{align}
After this choice, there are $n-1$ independent relative time shifts and $n-1$ independent relative azimuthal shifts between the identified points. Together with the $n$ transverse embedding functions $\mathfrak f_i$, these give $n+2(n-1)=3n-2$ embedding variables, all functions of $(\tau,\sigma)$.

The full gravitational action has the form \eqref{Eq:bulk-action}. Its bulk integral runs over all $n$ retained fragments, each fragment contributes a GHY term, and the perfect fluid contributes its junction action.  The action assumes that the induced metrics on all the $\Sigma_i$ are identical at the identified points, thus defining the worldsheet metric via the first junction condition
\begin{align}\label{Eq:hcont-n}
&\gamma_{\mu\nu}(\tau, \sigma) := \gamma_{1,\mu\nu}(\tau,\sigma)=\cdots = \gamma_{n,\mu\nu}(\tau,\sigma)\,.
\end{align}
Since $\gamma_{\mu\nu}$ is a symmetric two-dimensional metric, equating the metric induced by each of the remaining $n-1$ fragments to that of the first gives $3(n-1)$ equations. Variation of the action with respect to the bulk metric at $\Sigma$ gives the second junction condition\footnote{The $n$-way gravitational junctions were first constructed in AdS$_2$ in \cite{Shen:2024dun}, assuming generalized continuity of the normals across the junction. However, the general constructions in \cite{Chakraborty:2025jtj,Chakraborty:2025baw} do not impose this restriction. Even for a two-way purely tensile junction in three dimensions, allowing independent normals is necessary to obtain the general correspondence between the junction conditions and the solutions of the non-linear Nambu--Goto equation of a string.}
\begin{equation}\label{Eq:Kdisc-n}
    \sum_{i=1}^n (-1)^{s(\alpha_i)}\left(K_{i,\mu\nu} - K_i\,\gamma_{i,\mu\nu}\right) = 8\pi G_N S_{\mu\nu}\,,
\end{equation}
where $s(\alpha_i)=0$ for a northern fragment and $s(\alpha_i)=1$ for a southern fragment, as in the two-way case. The extrinsic curvatures $K_{i,\mu\nu}$ are computed with the same reference orientation before including the factors $(-1)^{s(\alpha_i)}$, and $K_i=\gamma^{\mu\nu}K_{i,\mu\nu}$. The stress tensor of the fluid on $\Sigma$ is
\[
S_{\mu\nu}=(\rho+P)u_\mu u_\nu+P\gamma_{\mu\nu},
\qquad P=w\rho,
\]
with the thermodynamic description given in Sec.~\ref{Sec:two_jn}. The temperature $T$ and the spatial velocity $u^\phi$ provide two more variables, while the normalization of $u^\mu$ determines its temporal component. There are therefore $3n$ independent variables.

The contracted Codazzi equations imply that the signed sum of the trace-adjusted extrinsic curvatures on the left hand side of \eqref{Eq:Kdisc-n} is conserved along the junction. Using \eqref{Eq:Kdisc-n}, this conservation law gives the Euler equations \eqref{Eq:Euler}, which impose no additional independent conditions. Therefore, \eqref{Eq:Kdisc-n} supplies three junction equations, and the total number is
\begin{equation}\label{Eq:3d-junction-count}
 3(n-1)+3=3n\,.
\end{equation}
As in the two-way case, the independent junction conditions and variables are equal in number. For example, for $n=3$ we have nine functions and nine junction equations.

The construction extends directly to an $n$-way junction joining locally dS$_D$ regions. A codimension-one junction now has $D-1$ intrinsic coordinates. With the junction coordinates chosen above, the $n$ transverse embedding functions and the $(D-1)(n-1)$ relative time and angular shifts give $Dn-D+1$ embedding variables. The temperature and the $D-2$ independent spatial velocity components provide another $D-1$ variables. Hence the total number of variables is
\begin{equation}\label{Eq:D-junction-variables}
 N_{\rm variables}=n+(D-1)(n-1)+(D-1)=Dn\,.
\end{equation}
A symmetric metric on the $(D-1)$-dimensional junction has $D(D-1)/2$ components. Its continuity gives $D(D-1)(n-1)/2$ equations, while the extrinsic-curvature junction condition gives another $D(D-1)/2$ equations. The Euler equations again follow from the contracted Codazzi equations and impose no additional independent conditions. Thus
\begin{equation}\label{Eq:D-junction-conditions}
 N_{\rm conditions}=\frac{D(D-1)(n-1)}{2}+\frac{D(D-1)}{2}
 =\frac{Dn(D-1)}{2}\,.
\end{equation}
The difference is
\begin{equation}\label{Eq:D-junction-difference}
 N_{\rm conditions}-N_{\rm variables}=\frac{Dn(D-3)}{2}\,.
\end{equation}
The numbers agree when $D=3$. In four dimensions, there are $4n$ variables and $6n$ junction equations. For $D>3$, the junction conditions therefore form a formally overdetermined system. Solutions with sufficient symmetry can nevertheless exist, as the rotationally invariant constructions discussed in this paper demonstrate. This count alone does not determine the dynamical degrees of freedom, which we identify in the next subsection.

Returning to dS$_3$, for the $n$-way junctions with $n\geq3$ considered below, we work with the \textit{folded picture} in which $\alpha_i=N$ for all $i$.

Following \cite{Chakraborty:2025jtj}, we define $3n-3$ independent relative shifts of the time ($\tau_{d_i}$), azimuthal angle ($\sigma_{d_i}$) and transverse polar coordinate ($\theta_{d_i}$) across the junction as
\begin{eqnarray}\label{Eq:xsxd}
\tau_{d_i} &=&\begin{cases}
        \frac{1}{n}(t_n-t_{i+1}) \,\, {\rm for}\,\, i = 1, \cdots, n-2\\
        \frac{1}{n}(t_n-t_1) \,\, {\rm for}\,\, i = n-1
    \end{cases}\,,\nonumber\\
\sigma_{d_i} &=&\begin{cases}
        \frac{1}{n}(\phi_n-\phi_{i+1}) \,\, {\rm for}\,\, i = 1, \cdots, n-2\\
        \frac{1}{n}(\phi_n-\phi_1) \,\, {\rm for}\,\, i = n-1
    \end{cases}\,,\nonumber\\
\theta_{d_i} &=&\begin{cases}
        \frac{1}{n}(\theta_n-\theta_{i+1}) \,\, {\rm for}\,\, i = 1, \cdots, n-2\\
        \frac{1}{n}(\theta_n-\theta_1) \,\, {\rm for}\,\, i = n-1
    \end{cases}\,.
\end{eqnarray}
These, together with the average transverse coordinate
\begin{equation}
    \theta_s = \frac{1}{n}\sum_i \theta_{i}
\end{equation}
give the $3n-2$ embedding functions to be determined. Together with $T$ and $u^\phi$, they constitute the $3n$ independent variables of the perfect-fluid junction. For $n=3$, the definitions reduce to $\tau_{d_1}=(t_3-t_2)/3$, $\tau_{d_2}=(t_3-t_1)/3$, and similarly for $\sigma_{d_i}$ and $\theta_{d_i}$. For gluings involving southern fragments, the transverse displacements are assigned signs according to the orientations of the normals. The independent combinations then involve pairwise differences or averages of the transverse displacements.

\subsection{Degrees of freedom of the $n$-way gravitational perfect fluid junction}
\label{Sec:prev}
It has been shown in \cite{Banerjee:2024sqq, Chakraborty:2025jtj} that the general solutions of the $n$-way purely tensile gravitational junction gluing $n$ identical three-dimensional spacetimes that are locally flat or locally AdS correspond to $n-1$ coupled strings, up to $3n$ rigid parameters that are related to the ambient spacetime and worldsheet isometries. 

The generalization to dS$_3$ was studied in \cite{Chakraborty:2025baw} by expanding about the equatorial junction with tension $\lambda=O(\epsilon)$ and transverse displacement amplitudes of $O(\epsilon)$. After setting the rigid isometry parameters to zero, the relative clock and azimuthal shifts begin at $O(\epsilon^2)$. The results, to the orders established in that expansion, have the following properties:
\begin{enumerate}
    \item The $n-1$ hypersurfaces $$\Sigma_{NG_i}: t=\tau, \,\, \phi=\sigma,\,\, \theta = \frac{\pi}{2}+\theta_{d_i}(\tau,\sigma)$$ correspond to solutions of the \textit{non-linear} Nambu-Goto equations coupled by \textit{non-linear} Monge-Amp\`{e}re like terms for their embeddings in the background global  dS$_3$ metric when the tension $\lambda$ and the rigid parameters are taken to zero for each solution of the junction conditions.
    
    \item The $n-1$ relative transverse displacements $\theta_{d_i}$ are the only independent dynamical variables. Once they are specified, $\tau_{d_i}$, $\sigma_{d_i}$ and $\theta_s$ are determined from the junction conditions, up to rigid parameters, order by order in the perturbative expansion. Thus, $n-1$ correlated clocks $\tau_{d_i}$ emerge dynamically \textit{without the need for external observers}. These clocks are invariant under bulk diffeomorphisms that are continuous at the junction for any pair of fragments $\mathcal{M}_{i\alpha_i}$ and $\mathcal{M}_{j\alpha_j}$ glued at the junction. 
    
    \item For $n\geq3$, the $n-1$ degrees of freedom described by the Nambu-Goto-Monge-Amp\`{e}re equations survive the tensionless limit. Thus degrees of freedom emerge even in \textit{pure three-dimensional gravity} when three or more spacetimes are glued at the junction.
\end{enumerate}

Above we have assumed that we are gluing the northern fragments of each $\mathcal{M}_i$. For a two-way junction we have chosen to glue the northern and southern fragments of $\mathcal{M}_1$ and $\mathcal{M}_2$, respectively. However, the sign convention in our definition of $\theta_{d_i}$ and $\theta_s$ in the previous subsection automatically exchanges the roles of $\theta_s$ and $-\theta_d$. In particular, for the 2-way junction, $\theta_s$ satisfies the non-linear Nambu--Goto equation in dS$_3$ when the tension and rigid parameters are taken to zero.

The correspondence reviewed above concerns purely tensile junctions. Our present analysis replaces the purely tensile junction by a perfect-fluid junction and examines the perturbations of the rotationally invariant solutions with finite asymptotic spatial volume. We keep all bulk regions locally de Sitter. In three dimensions, solving the linearized junction conditions shows that the $3n$ variables contain $n$ independent field degrees of freedom in the generic inhomogeneous sector. Of these, $n-1$ correspond to the independent relative transverse displacements, with pairwise differences or averages chosen according to the orientations of the normals. The additional degree of freedom is a \textit{hybrid sound mode}, in which the average transverse displacement is coupled to the temperature and velocity of the fluid. Once these $n$ fields are specified, the remaining $2n$ variables are determined by the junction conditions, up to rigid parameters.

To state this more precisely, let $\eta_i$ denote the linear proper normal displacement of the image $\Sigma_i$, with all orientations chosen in the folded picture. We define the average normal displacement $\eta$ via the average displacement $n^{-1}\sum_i\eta_i$ and the $n-1$ independent relative displacements $\eta_I^{\rm rel}$ via $\eta_I-\eta_n$, with $I=1,\cdots,n-1$ as in Sec.~\ref{Sec:two-way-perturbations-three} and Sec.~\ref{Sec:n-way-perturbations-three}. The background junction is a circle of constant areal radius $R_0$, with $R_0^2=L^2w/(1+w)$ and $w>0$ as will be shown in Sec~\ref{Sec:creation-three-dimensional}. As demonstrated in Sec.~\ref{Sec:two-way-perturbations-three} and Sec.~\ref{Sec:n-way-perturbations-three}, expanding the displacements in Fourier harmonics $e^{im\sigma}$, the linearized junction conditions give
\begin{align}\label{Eq:fluid-mode-equations}
&\ddot\eta_m+\frac{1+w}{L^2}(m^2-2)\eta_m=0,\nonumber\\
&\ddot\eta_{I,m}^{\rm rel}
-\frac{1+w}{L^2}m^2\eta_{I,m}^{\rm rel}=0,
\qquad I=1,\cdots,n-1,
\end{align}
where dots denote derivatives with respect to the junction proper time $\tau_{\rm p}$, and $m\in\mathbb Z$ labels the harmonics on the spatial circle. The average normal displacement therefore gives the hybrid sound mode described in the Introduction. For $|m|\geq2$, it is oscillatory, while each relative transverse mode has two exponential branches, growing towards the infinite past and the infinite future, respectively. The $|m|=1$ solutions correspond to isometries. For homogeneous perturbations ($m=0$), metric continuity and periodicity require the relative normal displacements $\eta_{I,0}^{\rm rel}$ to vanish, leaving only the average transverse displacement $\eta_0$.

Thus the perfect-fluid junction has one more independent field degree of freedom than the purely tensile junction in three dimensions. The relative transverse modes now satisfy fluid-dependent equations, whose principal part is elliptic for $w>0$. Section~\ref{Sec:admissible-histories} discusses the resulting Cauchy problem and how the gravitational equations determine admissible histories. The Nambu--Goto--Monge--Amp\`ere correspondence reviewed above does not directly extend to these modes. The relative clock and angular shifts are still determined by the junction conditions, together with the fluid variables, once the independent normal displacements and rigid parameters are specified.

In four and higher dimensions, we consider smooth perturbations of the rotationally invariant junction, whose spatial sections are round $S^{D-2}$, while keeping the bulk regions locally de Sitter. Let $\Omega_{AB}$ be the metric on the unit $S^{D-2}$ and $\mathcal D_A$ its covariant derivative, with $A,B=1,\cdots,D-2$. The traceless spatial part of \eqref{Eq:Kdisc-n} requires the average normal displacement to satisfy
\begin{equation}\label{Eq:fluid-angular-collective}
\left(\mathcal D_A\mathcal D_B
-\frac{1}{D-2}\Omega_{AB}\mathcal D^2\right)\eta=0,
\qquad D\geq4,
\end{equation}
where $\mathcal D^2=\Omega^{AB}\mathcal D_A\mathcal D_B$. The smooth scalar solutions contain only the $\ell=0$ and $\ell=1$ spherical harmonics. The traceless spatial part of \eqref{Eq:hcont-n} requires each independent relative angular shift $\xi_{I,A}$ to satisfy
\begin{equation}\label{Eq:fluid-angular-relative}
\mathcal D_A\xi_{I,B}+\mathcal D_B\xi_{I,A}
-\frac{2}{D-2}\Omega_{AB}\mathcal D_C\xi_I^C=0.
\end{equation}
These shifts are therefore conformal Killing fields on the sphere. The trace of the same junction condition fixes the relative normal displacements in terms of their divergences, which contain only $\ell=1$ harmonics. The remaining junction conditions identify these $\ell=1$ deformations and the Killing parts of the angular shifts with isometries. The mixed components of \eqref{Eq:Kdisc-n} determine the fluid velocity and leave no independent vector mode. Thus all physical inhomogeneous perturbations are excluded at linear order when all the ambient fragments are locally de Sitter.

The only physical mode that remains is the homogeneous hybrid sound mode, which changes the radius of every junction image by the same amount. Writing the areal radius shared by all images as $R(\tau_{\rm p})=R_0+\delta R(\tau_{\rm p})$, the background radius and the linearized radial equation are
\begin{equation}\label{Eq:fluid-radial-zero-mode}
R_0^2=L^2\left(1-\frac{1}{(D-2)(1+w)}\right),
\qquad
\delta\ddot R-\frac{2(D-2)(1+w)}{L^2}\delta R=0.
\end{equation}
For $w>-(D-3)/(D-2)$, its general linearized solution is
\begin{equation}\label{Eq:fluid-radial-zero-solution}
\delta R=C_+e^{\Gamma_D\tau_{\rm p}}
+C_-e^{-\Gamma_D\tau_{\rm p}},
\qquad
\Gamma_D=\frac{\sqrt{2(D-2)(1+w)}}{L},
\end{equation}
where $C_+$ and $C_-$ are integration constants. One branch grows towards the infinite future and the other towards the infinite past. This homogeneous average transverse displacement is fixed by finitely many integration constants and supplies no local field degree of freedom on $S^{D-2}$. Therefore, for the perfect-fluid backgrounds considered here, the linearized junction conditions yield $n$ independent field degrees of freedom in three dimensions, while in four and higher dimensions only the homogeneous radial zero mode of the average transverse coordinate survives in the folded picture, apart from isometries.

\section{Exact $n$-way cosmic creation gates with rotational symmetry}
\label{Sec:creation-gates}

We construct the simplest rotationally symmetric solutions of the junction conditions whose total spatial volume approaches finite, strictly positive values in both the infinite past and future on the natural global time foliation synchronized by the junction fluid. We begin in three dimensions, then consider four dimensions and finally give the construction in arbitrary $D\geq3$. We use the global coordinates and the embedding variables introduced in Sec.~\ref{Sec:setup} throughout. The global times of all the identified points will agree, so that $\tau=t_i$ defines a common time foliation of the retained bulk regions.

\subsection{The creation gate in three dimensions}
\label{Sec:creation-three-dimensional}

Let us consider $n\geq2$ identical copies of the global dS$_3$ spacetime with metric \eqref{Eq:dsmetric}. We first work in the folded picture, retaining the northern fragment of each copy. We choose the embeddings and the identifications of their points as follows:
\begin{equation}\label{Eq:gate-embedding-three}
t_i=\tau,\qquad \phi_i=\sigma,\qquad
\mathfrak f_i(\tau,\sigma)=\mathfrak f(\tau)
=\arcsin\left(\frac{R_0}{L\cosh(\tau/L)}\right),
\qquad i=1,\cdots,n,
\end{equation}
where $0<R_0<L$ is a constant to be determined from the equation of state. Thus the relative shifts $\tau_{d_i}$, $\sigma_{d_i}$ and $\theta_{d_i}$ vanish, while $\theta_s=\mathfrak f(\tau)$. The retained regions are $0\leq\theta_i\leq\mathfrak f(\tau)$. Their boundary circles have the same constant areal radius,
\begin{equation}\label{Eq:gate-radius-global}
L\cosh(\tau/L)\sin\mathfrak f(\tau)=R_0.
\end{equation}

For the northern--southern two-way gluing of Sec.~\ref{Sec:two_jn}, the same solution is written as $\mathfrak f_1=\mathfrak f$ and $\mathfrak f_2=\pi-\mathfrak f$. In the variables of \eqref{Eq:varbs}, it is
\begin{equation}\label{Eq:gate-unfolded-two}
\tau_d=\sigma_d=0,\qquad
\theta_s=\frac{\pi}{2},\qquad
\theta_d=\mathfrak f(\tau)-\frac{\pi}{2}.
\end{equation}
The second retained region is then $\pi-\mathfrak f(\tau)\leq\theta_2\leq\pi$. Reflection of the second parent sphere relates this description to the folded one. The orientation factors in \eqref{Eq:Kdisc} ensure that both give the same junction stress tensor.

Since
\begin{equation}\label{Eq:gate-angle-derivative}
\frac{{\rm d}\mathfrak f}{{\rm d}\tau}
=-\frac{1}{L}\tanh(\tau/L)\tan\mathfrak f,
\end{equation}
the metric induced on every image of the junction is
\begin{equation}\label{Eq:gate-induced-three}
{\rm d}s_\Sigma^2
=-\frac{1-R_0^2/L^2}
{1-(R_0^2/L^2)\operatorname{sech}^2(\tau/L)}\,{\rm d}\tau^2
+R_0^2\,{\rm d}\sigma^2.
\end{equation}
This solves the first junction condition \eqref{Eq:hcont-n} identically. Its temporal component is strictly negative for every finite $\tau$, so that the junction is timelike throughout.

We next determine the extrinsic curvatures directly from the global embeddings. In the folded picture, the unit normal pointing towards increasing $\theta_i$ is
\begin{equation}\label{Eq:gate-global-normal}
n^{(i)}_a\,{\rm d}x_i^a
=L\cosh(\tau/L)\frac{1}{\sqrt{-\gamma_{\tau\tau}}}
\left({\rm d}\theta_i-\frac{{\rm d}\mathfrak f}{{\rm d}\tau}
{\rm d}t_i\right),
\end{equation}
where $a$ labels the bulk coordinates and the expression is evaluated on $\Sigma_i$. With the extrinsic-curvature convention of Sec.~\ref{Sec:setup}, the nonzero mixed components are
\begin{equation}\label{Eq:gate-curvatures-three}
K_i{}^\tau{}_\tau
=-\frac{R_0}{L\sqrt{L^2-R_0^2}},
\qquad
K_i{}^\sigma{}_\sigma
=\frac{\sqrt{L^2-R_0^2}}{LR_0}.
\end{equation}
They are independent of $\tau$ and identical for all the retained northern fragments. In the northern--southern description, reflection reverses the reference extrinsic curvature of the second image, which is compensated by its orientation factor.

The temporal and spatial components of \eqref{Eq:Kdisc-n} now give  
\begin{equation}\label{Eq:gate-stress-three}
\rho=\frac{n}{8\pi G_N L}
\frac{\sqrt{L^2-R_0^2}}{R_0},
\qquad
P=\frac{n}{8\pi G_N L}
\frac{R_0}{\sqrt{L^2-R_0^2}}\,
\end{equation}
respectively. All off-diagonal components vanish. Requiring $P=w\rho$ therefore fixes
\begin{equation}\label{Eq:gate-eos-three}
w=\frac{R_0^2}{L^2-R_0^2},
\qquad
R_0^2=L^2\frac{w}{1+w},
\qquad
\rho=\frac{n}{8\pi G_N L\sqrt w}.
\end{equation}
In three dimensions, a nondegenerate timelike solution consequently requires $w>0$. The radius depends on $w$ and $L$, while the energy density is proportional to the number of bulk regions joined at the junction.

The spatial fluid velocity vanishes, and normalization fixes its temporal component:
\begin{equation}\label{Eq:gate-fluid-three}
u^\phi=0,\qquad
u^\tau=\sqrt{1+w\tanh^2(\tau/L)},\qquad
T=T_0\left(\frac{w\rho}{P_0}\right)^{\frac{w}{1+w}}.
\end{equation}
Here $T$, $\rho$ and $P$ are constant on the junction. In terms of $w$, the temporal component of \eqref{Eq:gate-induced-three} is simply $\gamma_{\tau\tau}=-[1+w\tanh^2(\tau/L)]^{-1}$. Since the spatial metric is independent of $\tau$ and $\sqrt{-\gamma}\,u^\tau=R_0$, we have $\nabla_\mu u^\mu=0$. The fluid acceleration also vanishes. Both Euler equations \eqref{Eq:Euler} are therefore satisfied. This configuration satisfies the bulk equations, both junction conditions and the fluid equations.

We now compute the spatial volume on the common global-time slices. In three dimensions this is the total area of the $n$ retained northern spherical caps. The junction itself has zero measure in this volume. Hence
\begin{align}\label{Eq:gate-volume-three}
V_{n,3}(\tau)
&=2\pi n L^2\cosh^2(\tau/L)
\int_0^{\mathfrak f(\tau)}\sin\theta\,{\rm d}\theta
\nonumber\\
&=\frac{2\pi n R_0^2}
{1+\sqrt{1-(R_0^2/L^2)\operatorname{sech}^2(\tau/L)}}.
\end{align}
The angular openings of the caps vanish in both temporal limits, while the radius of each parent sphere diverges. The shrinking angular openings compensate for the diverging parent radii, giving finite, strictly positive initial and final volumes,
\begin{equation}\label{Eq:gate-volume-three-limits}
V_{n,3}^{\rm in}=V_{n,3}^{\rm out}
=n\pi L^2\frac{w}{1+w}.
\end{equation}
The volume increases from its past limit until $\tau=0$, and then decreases towards the same limit in the future. Its maximum is
\begin{equation}\label{Eq:gate-volume-three-maximum}
V_{n,3}^{\rm max}
=2\pi n L^2\left(1-\frac{1}{\sqrt{1+w}}\right),
\qquad {\rm when}\,\,\, \tau=0.
\end{equation}
Thus the contraction of each parent de Sitter sphere in the past accompanies an increase of the volume retained by the junction.

For the range $0<w<1$ considered in Sec.~\ref{Sec:two_jn}, both the asymptotic volume and the maximum volume increase with $w$. Their limiting values as $w\to1^-$ are
\begin{equation}\label{Eq:gate-stiff-three}
R_0\longrightarrow\frac{L}{\sqrt2},\qquad
V_{n,3}^{\rm in,out}\longrightarrow\frac{n\pi L^2}{2},\qquad
V_{n,3}^{\rm max}\longrightarrow
2\pi n L^2\left(1-\frac{1}{\sqrt2}\right).
\end{equation}
The limiting geometry remains regular and timelike.

\begin{figure}[tbp]
\centering
\resizebox{\textwidth}{!}{%
\begin{tikzpicture}[
  x=0.01cm,y=-0.01cm,line cap=round,line join=round,
  every node/.style={inner sep=0pt,outer sep=0pt},
  gate two label/.style={font=\fontsize{10.5}{12}\selectfont},
  gate two tick/.style={font=\fontsize{8.5}{10}\selectfont}
]
\definecolor{gateTwoBlue}{RGB}{25,91,179}
\definecolor{gateTwoFill}{RGB}{220,234,251}
\definecolor{gateTwoHorizon}{RGB}{106,118,130}
\definecolor{gateTwoSlice}{RGB}{154,169,190}
\definecolor{gateTwoArrow}{RGB}{84,100,122}
\definecolor{gateTwoOutline}{RGB}{32,40,51}
\definecolor{gateTwoFrame}{RGB}{36,36,36}
\path[use as bounding box] (0,0) rectangle (2048,1634);

\node at (1078,56) {%
  \resizebox{15.24cm}{!}{%
    \bfseries Two-way timelike perfect-fluid junction in global de Sitter time}};

\foreach \gateleft/\gateside in {153.5/0,1238.5/1}{
  \begin{scope}
    \clip (\gateleft,194) rectangle ({\gateleft+765},959);
    \ifnum\gateside=0
      \path[fill=gateTwoFill]
        plot[domain=-90:90,samples=241,variable=\gatev]
        ({\gateleft+4.25*asin(0.72*cos(\gatev))},
         {576.5-4.25*\gatev})
        -- (\gateleft,194) -- cycle;
    \else
      \path[fill=gateTwoFill]
        plot[domain=-90:90,samples=241,variable=\gatev]
        ({\gateleft+765-4.25*asin(0.72*cos(\gatev))},
         {576.5-4.25*\gatev})
        -- ({\gateleft+765},194) -- cycle;
    \fi
    \foreach \gateheight in {385.25,576.5,767.75}
      \draw[gateTwoSlice,line width=0.43pt,
            dash pattern=on 1.8pt off 2.6pt]
        (\gateleft,\gateheight) -- ({\gateleft+765},\gateheight);
    \draw[gateTwoHorizon,line width=0.85pt,
          dash pattern=on 4.5pt off 3.4pt]
      (\gateleft,194) -- ({\gateleft+765},959);
    \draw[gateTwoHorizon,line width=0.85pt,
          dash pattern=on 4.5pt off 3.4pt]
      (\gateleft,959) -- ({\gateleft+765},194);
    \ifnum\gateside=0
      \draw[gateTwoBlue,line width=1.85pt]
        plot[domain=-90:90,samples=241,variable=\gatev]
        ({\gateleft+4.25*asin(0.72*cos(\gatev))},
         {576.5-4.25*\gatev});
    \else
      \draw[gateTwoBlue,line width=1.85pt]
        plot[domain=-90:90,samples=241,variable=\gatev]
        ({\gateleft+765-4.25*asin(0.72*cos(\gatev))},
         {576.5-4.25*\gatev});
    \fi
  \end{scope}
  \draw[gateTwoFrame,line width=0.85pt]
    (\gateleft,194) rectangle ({\gateleft+765},959);
  \node[gate two tick,anchor=east]
    at ({\gateleft-16},194) {$\pi/2$};
  \node[gate two tick,anchor=east]
    at ({\gateleft-16},576.5) {$0$};
  \node[gate two tick,anchor=east]
    at ({\gateleft-16},959) {$-\pi/2$};
  \node[gate two tick] at (\gateleft,981) {$0$};
  \node[gate two tick] at ({\gateleft+382.5},981) {$\pi/2$};
  \node[gate two tick] at ({\gateleft+765},981) {$\pi$};
  \node[font=\fontsize{9.5}{11}\selectfont]
    at ({\gateleft+382.5},1024) {$\theta$};
  \node[font=\fontsize{8.2}{10}\selectfont,rotate=90]
    at ({\gateleft-106},577.5) {conformal time $\eta$};
  \node[gate two tick] at ({\gateleft+27},152) {\resizebox*{0.26cm}{0.21cm}{$\mathcal I^+$}};
  \node[gate two tick] at ({\gateleft+27},1021) {\resizebox*{0.26cm}{0.21cm}{$\mathcal I^-$}};
}
\node at (536.5,154.5)
  {\resizebox*{3.08cm}{0.36cm}{(a) Northern cap}};
\node at (1621,154.5)
  {\resizebox*{3.14cm}{0.36cm}{(b) Southern cap}};

\node[gate two label,align=center,anchor=center] at (251,576.5)
{retained\\northern\\cap};
\node[gate two label,align=center,anchor=center] at (1906,576.5)
{retained\\southern\\cap};

\node[gateTwoBlue,font=\fontsize{10.5}{12}\selectfont,anchor=west]
  at (365,521) {$\Sigma_N$};
\node[gateTwoBlue,font=\fontsize{10.5}{12}\selectfont,anchor=east]
  at (1792,521) {$\Sigma_S$};
\draw[gateTwoArrow,line width=0.55pt,
      -{Stealth[open,length=5.5pt,width=3.4pt]}]
  (568,832) -- (360,756);
\node[gateTwoArrow,font=\fontsize{8.5}{10}\selectfont]
  at (617,850) {north horizon};
\draw[gateTwoArrow,line width=0.55pt,
      -{Stealth[open,length=5.5pt,width=3.4pt]}]
  (1688,832) -- (1800,756);
\node[gateTwoArrow,font=\fontsize{8.5}{10}\selectfont]
  at (1658,850) {south horizon};

\node at (1078,1090) {%
  \resizebox*{10.45cm}{0.33cm}{%
    \bfseries Constant global-time slices (schematic cross sections)}};
\node[font=\fontsize{11}{13}\selectfont]
  at (355,1168.5) {\resizebox*{1.20cm}{0.24cm}{$t\to-\infty$}};
\node[font=\fontsize{11}{13}\selectfont]
  at (1078.5,1168) {\resizebox*{0.73cm}{0.26cm}{$t=0$}};
\node[font=\fontsize{11}{13}\selectfont]
  at (1801.5,1168.5) {\resizebox*{1.19cm}{0.24cm}{$t\to+\infty$}};

\foreach \gatecentre/\gateangle in
  {355/16.042818,1078.5/46.05448,1801.5/16.042818}{
  \pgfmathsetmacro{\gatecut}{164.5*cos(\gateangle)}
  \pgfmathsetmacro{\gatehalfheight}{164.5*sin(\gateangle)}
  \path[fill=gateTwoFill]
    ({\gatecentre-\gatecut},{1387.5-\gatehalfheight})
    -- ({\gatecentre-\gatecut},{1387.5+\gatehalfheight})
    arc[start angle={180-\gateangle},
        end angle={180+\gateangle},radius=164.5] -- cycle;
  \path[fill=gateTwoFill]
    ({\gatecentre+\gatecut},{1387.5-\gatehalfheight})
    -- ({\gatecentre+\gatecut},{1387.5+\gatehalfheight})
    arc[start angle=\gateangle,
        end angle={-\gateangle},radius=164.5] -- cycle;
  \draw[gateTwoHorizon,line width=0.43pt,
        dash pattern=on 1.2pt off 1.7pt]
    ({\gatecentre-164.5},1387.5) -- ({\gatecentre+164.5},1387.5);
  \draw[gateTwoOutline,line width=0.85pt]
    (\gatecentre,1387.5) circle[radius=164.5];
  \draw[gateTwoBlue,line width=1.28pt]
    ({\gatecentre-\gatecut},{1387.5-\gatehalfheight})
    -- ({\gatecentre-\gatecut},{1387.5+\gatehalfheight});
  \draw[gateTwoBlue,line width=1.28pt]
    ({\gatecentre+\gatecut},{1387.5-\gatehalfheight})
    -- ({\gatecentre+\gatecut},{1387.5+\gatehalfheight});
}
\node[font=\fontsize{9.2}{11}\selectfont]
  at (355,1585) {finite initial volume};
\node[font=\fontsize{9.2}{11}\selectfont]
  at (1078.5,1585) {maximum volume};
\node[font=\fontsize{9.2}{11}\selectfont]
  at (1801.5,1585) {finite final volume};
\end{tikzpicture}%
}
\caption{The rotationally symmetric two-way creation gate in global de Sitter coordinates. The blue curves are the images $\Sigma_N$ and $\Sigma_S$ of the common timelike junction, at $\theta=\mathfrak f(t)$ and $\theta=\pi-\mathfrak f(t)$, respectively. The shaded northern and southern caps are retained and their boundaries are identified at equal global time. The vertical coordinate in the diagram is the conformal time $\eta=\arctan[\sinh(t/L)]$; all embeddings in the text use global time. The dashed diagonals show the past and future cosmological horizons of the polar observers in the parent de Sitter regions. Each retained cap lies inside both parent horizons, which are excised in the construction. The horizontal dotted lines are common global-time slices. The lower panels give schematic cross sections of the retained caps: their angular openings vanish in the infinite past and future, while their physical volumes on these slices approach finite, positive limits. The maximum volume occurs at $t=0$. The same cap construction applies in higher dimensions and to $n$-way junctions.}
\label{fig:cosmic-creation-gate}
\end{figure}
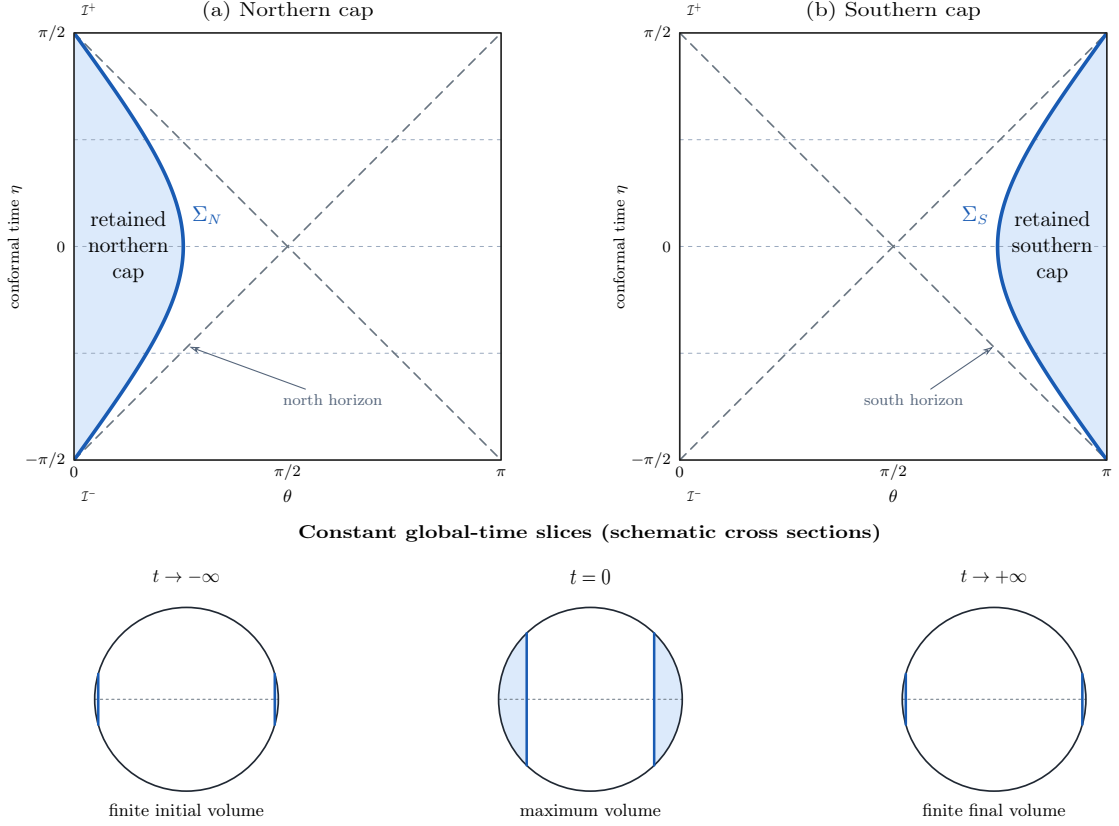

The limiting spatial volumes in the infinite past and future depend on the foliation. We use the global-time foliation defined by the geodesic congruence of comoving observers at fixed spatial coordinates on $S^2$ in each parent de Sitter region. For the rotationally invariant solution, these global times are synchronized by the junction fluid clock, providing the natural cosmological foliation for defining the volume. On static slices, associated with generally accelerated observers at fixed areal radius, the total spatial volume remains constant and equals $V_{n,3}^{\rm max}$. These conclusions extend to four and higher dimensions, as discussed in Sec.~\ref{Sec:creation-arbitrary-dimensional}. Figure~\ref{fig:cosmic-creation-gate} illustrates the construction.

\subsection{The creation gate in four dimensions}
\label{Sec:creation-four-dimensional}

We now consider $n$ identical locally dS$_4$ regions. We use the global metric and the angular coordinates $\varphi^A$, with $A=1,2$, introduced in Sec.~\ref{Sec:two_jn}. The embeddings are again given by \eqref{Eq:gate-embedding-three}, with $\phi_i=\sigma$ replaced by $\varphi_i^A=\sigma^A$. Thus all relative time and angular shifts vanish, and every junction image has $\theta_i=\theta_s=\mathfrak f(\tau)$ in the folded picture. The junction has spatial topology $S^2$ and induced metric
\begin{equation}\label{Eq:gate-induced-four}
{\rm d}s_\Sigma^2
=-\frac{1-R_0^2/L^2}
{1-(R_0^2/L^2)\operatorname{sech}^2(\tau/L)}\,{\rm d}\tau^2
+R_0^2\,{\rm d}\Omega_2^2.
\end{equation}
The temporal extrinsic curvature is unchanged, while both angular components equal the spatial component in \eqref{Eq:gate-curvatures-three}. The junction conditions therefore require
\begin{equation}\label{Eq:gate-stress-four}
\rho=\frac{n}{4\pi G_N L}\frac{\sqrt{L^2-R_0^2}}{R_0},
\qquad
P=\frac{n}{8\pi G_N L}
\left(\frac{R_0}{\sqrt{L^2-R_0^2}}
-\frac{\sqrt{L^2-R_0^2}}{R_0}\right).
\end{equation}
The additional angular direction contributes the second term in the pressure. Consequently, the equation of state gives
\begin{equation}\label{Eq:gate-eos-four}
R_0^2=L^2\frac{1+2w}{2(1+w)},
\qquad
\rho=\frac{n}{4\pi G_N L\sqrt{1+2w}},
\qquad
w>-\frac12.
\end{equation}
This lower bound is precisely the condition that the cap radius be real, positive and smaller than $L$. In particular, negative pressure is allowed by the background junction conditions. The dust solution $w=0$ has $R_0=L/\sqrt2$ and is regular, with the matter description understood as in Sec.~\ref{Sec:two_jn}.

The spatial velocity components vanish and
\begin{equation}\label{Eq:gate-fluid-four}
u^A=0,\qquad
u^\tau=\sqrt{1+(1+2w)\tanh^2(\tau/L)}.
\end{equation}
The energy density and pressure are constant. For $w\neq0$, the constant temperature is again fixed by the last expression in \eqref{Eq:gate-fluid-three}, with the four-dimensional value of $\rho$. When $w<0$, the pressure normalization $P_0$ is chosen with the same sign as $P$, so that the temperature is real and positive. The fluid expansion and acceleration vanish, and the Euler equations are solved.

The spatial volume of the $n$ retained caps is
\begin{align}\label{Eq:gate-volume-four}
V_{n,4}(\tau)
&=4\pi nL^3\cosh^3(\tau/L)
\int_0^{\mathfrak f(\tau)}\sin^2\theta\,{\rm d}\theta
\nonumber\\
&=2\pi nL^3\cosh^3(\tau/L)
\left[\mathfrak f(\tau)-\sin\mathfrak f(\tau)
\cos\mathfrak f(\tau)\right].
\end{align}
Its initial and final values are
\begin{equation}\label{Eq:gate-volume-four-limits}
V_{n,4}^{\rm in}=V_{n,4}^{\rm out}
=\frac{4\pi nL^3}{3}
\left(\frac{1+2w}{2(1+w)}\right)^{3/2}.
\end{equation}
The maximum again occurs at $\tau=0$ and is
\begin{equation}\label{Eq:gate-volume-four-maximum}
V_{n,4}^{\rm max}
=2\pi nL^3\left[
\arccos\left(\frac{1}{\sqrt{2(1+w)}}\right)
-\frac{\sqrt{1+2w}}{2(1+w)}
\right].
\end{equation}
In the limit $w\to1^-$, these quantities approach
\begin{equation}\label{Eq:gate-stiff-four}
R_0\longrightarrow\frac{\sqrt3}{2}L,\qquad
V_{n,4}^{\rm in,out}\longrightarrow\frac{\sqrt3}{2}n\pi L^3,
\qquad
V_{n,4}^{\rm max}\longrightarrow
2\pi nL^3\left(\frac{\pi}{3}-\frac{\sqrt3}{4}\right).
\end{equation}

\subsection{The creation gate in arbitrary dimensions}
\label{Sec:creation-arbitrary-dimensional}

The same construction applies to $n\geq2$ locally dS$_D$ regions for any $D\geq3$. In the global coordinates of Sec.~\ref{Sec:two_jn}, we set $t_i=\tau$, $\varphi_i^A=\sigma^A$ and $\mathfrak f_i=\mathfrak f(\tau)$, with $A=1,\cdots,D-2$ and $\mathfrak f$ given by \eqref{Eq:gate-embedding-three}. The relative embedding variables vanish in the folded picture. All the induced metrics agree, with spatial part $R_0^2{\rm d}\Omega_{D-2}^2$. There are now $D-2$ identical angular components of the extrinsic curvature, each equal to $K_i{}^\sigma{}_\sigma$ in \eqref{Eq:gate-curvatures-three}. The junction conditions give
\begin{align}\label{Eq:gate-stress-general}
\rho&=\frac{n(D-2)}{8\pi G_N L}
\frac{\sqrt{L^2-R_0^2}}{R_0},
\nonumber\\
P&=\frac{n}{8\pi G_N L}
\left(\frac{R_0}{\sqrt{L^2-R_0^2}}
-(D-3)\frac{\sqrt{L^2-R_0^2}}{R_0}\right).
\end{align}
Thus the dependence of the lower bound on dimension follows directly from the number of angular contributions to the trace of the extrinsic curvature. Imposing $P=w\rho$ determines
\begin{equation}\label{Eq:gate-radius-general}
\frac{R_0^2}{L^2}
=1-\frac{1}{(D-2)(1+w)},
\qquad
\rho=\frac{n(D-2)}
{8\pi G_N L\sqrt{(D-2)(1+w)-1}}.
\end{equation}
The condition $0<R_0<L$ is equivalent to
\begin{equation}\label{Eq:gate-w-bound}
(D-2)(1+w)>1,
\qquad\hbox{or equivalently}\qquad
w>-\frac{D-3}{D-2}.
\end{equation}
At the lower endpoint the caps collapse, their asymptotic volumes vanish and their surface energy density diverges. The endpoint is therefore excluded. For every finite $w$ strictly above it, the radius is smaller than $L$, the junction is timelike, and $\rho>0$ and $\rho+P>0$.

The remaining background variables are
\begin{equation}\label{Eq:gate-induced-general}
{\rm d}s_\Sigma^2
=-\frac{{\rm d}\tau^2}
{1+\big[(D-2)(1+w)-1\big]\tanh^2(\tau/L)}
+R_0^2{\rm d}\Omega_{D-2}^2,
\end{equation}
with $u^A=0$, $u^\tau=(-\gamma_{\tau\tau})^{-1/2}$ and constant $\rho$ and $P$. For $w\neq0$, $T$ is the constant determined by \eqref{Eq:gate-fluid-three}; $w=0$ is described by dust. The bulk Einstein equations and their constraints hold because each retained metric is de Sitter. The Gauss equation on each junction image gives the intrinsic scalar curvature $(D-2)(D-3)/R_0^2$, as also follows from \eqref{Eq:gate-induced-general}. The contracted Codazzi equations and the fluid equations hold because the fluid is homogeneous, its spatial metric is time independent, and its expansion and acceleration vanish. The global embeddings and the stress tensor \eqref{Eq:gate-stress-general} therefore satisfy all bulk, junction and fluid equations.

The global-time slices intersect the junction orthogonally to the fluid velocity $u^\mu$. The fluid proper time, with origin at $\tau=0$, is
\begin{equation}\label{Eq:gate-fluid-clock}
\tau_{\rm p}(\tau)
=\frac{L}{\sqrt{(D-2)(1+w)}}\,
\operatorname{arsinh}\left[
\sqrt{(D-2)(1+w)}\,\sinh(\tau/L)\right].
\end{equation}
It is strictly increasing in $\tau$ and runs from negative to positive infinity. 


Let $\Omega_{D-2}=2\pi^{(D-1)/2}/\Gamma((D-1)/2)$ denote the area of the unit $S^{D-2}$. The total spatial volume on this foliation is
\begin{equation}\label{Eq:gate-volume-general}
V_{n,D}(\tau)
=n\Omega_{D-2}L^{D-1}\cosh^{D-1}(\tau/L)
\int_0^{\mathfrak f(\tau)}\sin^{D-2}\theta\,{\rm d}\theta.
\end{equation}
Since $\mathfrak f(\tau)$ approaches $R_0/[L\cosh(\tau/L)]$ in both temporal limits, the initial and final volumes are
\begin{equation}\label{Eq:gate-volume-general-limits}
V_{n,D}^{\rm in}=V_{n,D}^{\rm out}
=\frac{n\Omega_{D-2}L^{D-1}}{D-1}
\left(1-\frac{1}{(D-2)(1+w)}\right)^{(D-1)/2}.
\end{equation}
These are the volumes of $n$ Euclidean $(D-1)$-balls of radius $R_0$. The spatial curvature of each parent de Sitter slice tends to zero in the same limit.

To establish the maximum for general $D$, we use \eqref{Eq:gate-radius-global} to write
\begin{equation}\label{Eq:gate-volume-angle}
V_{n,D}
=n\Omega_{D-2}R_0^{D-1}
\frac{\displaystyle\int_0^{\mathfrak f}
\sin^{D-2}\theta\,{\rm d}\theta}
{\sin^{D-1}\mathfrak f}.
\end{equation}
For $0<\mathfrak f<\pi/2$,
\begin{align}\label{Eq:gate-volume-monotonicity}
\sin^{D-1}\mathfrak f
&=(D-1)\int_0^{\mathfrak f}
\sin^{D-2}\theta\cos\theta\,{\rm d}\theta
\nonumber\\
&>(D-1)\cos\mathfrak f
\int_0^{\mathfrak f}\sin^{D-2}\theta\,{\rm d}\theta.
\end{align}
It follows that the ratio in \eqref{Eq:gate-volume-angle} is strictly increasing with $\mathfrak f$. The opening angle increases until $\tau=0$ and decreases afterwards, so that the volume has a unique maximum at $\tau=0$. Its value is
\begin{equation}\label{Eq:gate-volume-general-maximum}
V_{n,D}^{\rm max}
=n\Omega_{D-2}L^{D-1}
\int_0^{\displaystyle\arccos\left(
[(D-2)(1+w)]^{-1/2}\right)}
\sin^{D-2}\theta\,{\rm d}\theta.
\end{equation}
Equations \eqref{Eq:gate-volume-general-limits} and \eqref{Eq:gate-volume-general-maximum} give the initial, final and maximum volumes for arbitrary $n$, $w$ and $D$ in the allowed geometric range.

For fixed $n$, $D$ and $L$, both volumes increase with $w$. In the limiting case $w\to1^-$,
\begin{align}\label{Eq:gate-stiff-general}
\frac{R_0^2}{L^2}
&\longrightarrow1-\frac{1}{2(D-2)},
\nonumber\\
V_{n,D}^{\rm in,out}
&\longrightarrow
\frac{n\Omega_{D-2}L^{D-1}}{D-1}
\left(1-\frac{1}{2(D-2)}\right)^{(D-1)/2},
\nonumber\\
V_{n,D}^{\rm max}
&\longrightarrow n\Omega_{D-2}L^{D-1}
\int_0^{\displaystyle\arccos[\,1/\sqrt{2(D-2)}\,]}
\sin^{D-2}\theta\,{\rm d}\theta.
\end{align}
These are the suprema within the range $w<1$ adopted in Sec.~\ref{Sec:two_jn}. The limiting cap remains strictly within the cosmological horizons.

We examine the horizons and temporal completeness in global coordinates. The cosmological horizons of the polar observer in each parent de Sitter region satisfy
\begin{equation}\label{Eq}
L\cosh(t_i/L)\sin\theta_i=L.
\end{equation}
Since the junction has constant areal radius $R_0<L$, each retained cap lies inside both parent horizons, as shown in Fig.~\ref{fig:cosmic-creation-gate}. These horizons lie outside the fragments retained in the glued spacetime. The junction approaches the polar endpoints of the conformal diagram only at infinite global time.

The static time-translation Killing field provides a direct completeness check. In global coordinates, it is
\begin{equation}\label{Eq:gate-killing-global}
\xi_i=\cos\theta_i\,\partial_{t_i}
-\frac{\tanh(t_i/L)}{L}\sin\theta_i\,\partial_{\theta_i}.
\end{equation}
It preserves the areal radius $r=L\cosh(t_i/L)\sin\theta_i$ and is uniformly timelike throughout each retained northern cap:
\begin{equation}\label{Eq:gate-killing-norm}
-g(\xi_i,\xi_i)
=1-\frac{r^2}{L^2}
\geq1-\frac{R_0^2}{L^2}>0.
\end{equation}
At the junction, these fields induce the same tangent vector $\sqrt{1-R_0^2/L^2}\,u^\mu$, so static time translations define a common symmetry of the glued geometry. Southern fragments are treated by reflection as in \eqref{Eq:gate-unfolded-two}. The orthogonal static slices have constant volume
\begin{equation}
V_{n,D}^{\mathrm{static}}
=n\Omega_{D-2}\int_0^{R_0}
\frac{r^{D-2}\,{\rm d}r}{\sqrt{1-r^2/L^2}}
=V_{n,D}^{\max}.
\end{equation}
The static observers following the orbits of $\xi_i$ are generally accelerated, whereas the comoving observers of the global-time foliation follow timelike geodesics.

Along a future-directed causal geodesic, let $\lambda_{\rm aff}$ be an affine parameter, chosen as proper time in the timelike case, and let $E=-g(\xi,{\rm d}x/{\rm d}\lambda_{\rm aff})>0$. The tangential momentum matching at a junction preserves $E$. Causality and $0\leq\theta_i\leq\mathfrak f(t_i)$ imply
\begin{equation}\label{Eq:gate-affine-bound}
E\geq\left(1-\frac{R_0}{L}\right)
\frac{{\rm d}t_i}{{\rm d}\lambda_{\rm aff}},
\qquad
\frac{{\rm d}\lambda_{\rm aff}}{{\rm d}\tau}
\geq\frac{1-R_0/L}{E}>0.
\end{equation}
The caps are compact on every finite global-time slice, their poles are regular, and their boundaries are joined at a regular timelike junction. The piecewise geodesic equations therefore admit continuation at every finite $\tau$. The bound \eqref{Eq:gate-affine-bound} then shows that every inextendible timelike or null geodesic has infinite proper or affine length, respectively, towards both temporal ends. For a multiway junction, the continuation through a junction need not be unique, and the same bound holds for every continuation satisfying the tangential momentum matching.

The spatial volume condition and temporal completeness are therefore compatible. On the common global-time foliation synchronized by the junction fluid, the total volume has finite, strictly positive limits arbitrarily far into the past and future, while the spacetime admits causal geodesics of infinite length in both directions. In this precise sense, the spacetimes admit the notion of creation, and so we call them \textit{cosmic creation gate solutions}.

In Sec.~\ref{Sec:histories-exact-radial}, we construct exact homogeneous deformations of these gates that retain the same past volume while their future volume grows without bound.

\section{Linear perturbations of the $2$-way cosmic creation gate in three dimensions}
\label{Sec:two-way-perturbations-three}

We now study the linear perturbations of the two-way creation gate constructed in Sec.~\ref{Sec:creation-three-dimensional}. We follow the procedure of \cite{Chakraborty:2025baw}: we first solve the conditions for continuity of the induced metric, identify the worldsheet isometries directly from the relative shifts $\tau_d$ and $\sigma_d$, and then identify the ambient spacetime isometries from the transverse displacement $\theta_d$. All bulk metrics are kept locally de Sitter, and all embeddings are written in the global coordinates of Sec.~\ref{Sec:two_jn}. The perturbation parameter measures the amplitudes of the embedding and fluid fluctuations at fixed $w$, rather than an expansion in a small junction tension.

\subsection{The perturbative expansion}
\label{Sec:gate-two-expansion}

We use the northern--southern description \eqref{Eq:gate-unfolded-two}, with $\mathfrak f(\tau)$ given by \eqref{Eq:gate-embedding-three}. The expansions of the four embedding variables are
\begin{align}\label{Eq:gate-two-expansion}
&\tau_d=\epsilon\tau_{d,1}+O(\epsilon^2),\qquad
\sigma_d=\epsilon\sigma_{d,1}+O(\epsilon^2),
\nonumber\\
&\theta_s=\frac{\pi}{2}+\epsilon\theta_{s,1}+O(\epsilon^2),\qquad
\theta_d=\mathfrak f(\tau)-\frac{\pi}{2}
+\epsilon\theta_{d,1}+O(\epsilon^2).
\end{align}
Thus the images of the junction are
\begin{align}\label{Eq:gate-two-perturbed-images}
&t_1=\tau+\epsilon\tau_{d,1},\qquad
\phi_1=\sigma+\epsilon\sigma_{d,1},\qquad
\theta_1=\mathfrak f+\epsilon(\theta_{s,1}+\theta_{d,1}),
\nonumber\\
&t_2=\tau-\epsilon\tau_{d,1},\qquad
\phi_2=\sigma-\epsilon\sigma_{d,1},\qquad
\theta_2=\pi-\mathfrak f+\epsilon(\theta_{s,1}-\theta_{d,1}),
\end{align}
up to $O(\epsilon^2)$. All first-order coefficients are functions of $(\tau,\sigma)$, with period $2\pi$ in $\sigma$. In this section a dot and a prime denote derivatives with respect to the global time $\tau$ and the angle $\sigma$, respectively.

For the fluid, we write
\begin{equation}\label{Eq:gate-two-fluid-expansion}
T=T^{(0)}+\epsilon T_1+O(\epsilon^2),\qquad
u^\phi=\epsilon u_1^\phi+O(\epsilon^2),\qquad
u^\tau=u^{\tau(0)}+\epsilon u_1^\tau+O(\epsilon^2).
\end{equation}
The superscript $(0)$ denotes the background values in \eqref{Eq:gate-fluid-three}; $T_0$ remains the fixed reference temperature in the equation of state. We continue to denote the intrinsic angular velocity by $u^\phi$, as in Sec.~\ref{Sec:two_jn}, with $\sigma$ the angular coordinate on the common junction. In particular,
\begin{equation}\label{Eq:gate-two-background-values}
u^{\tau(0)}=\sqrt{1+w\tanh^2(\tau/L)},\qquad
\rho^{(0)}=\frac{1}{4\pi G_N L\sqrt w},\qquad
R_0^2=L^2\frac{w}{1+w}.
\end{equation}
The temporal velocity perturbation is determined by normalization and is not an additional independent variable.

We will also use the normal displacements already introduced in Sec.~\ref{Sec:prev}. Reflecting the second image into the folded picture, their average and relative normal displacements are
\begin{align}\label{Eq:gate-two-normal-displacements}
\eta&=L\cosh(\tau/L)\,u^{\tau(0)}\theta_{d,1},
\nonumber\\
\eta_1^{\rm rel}
&=2L\cosh(\tau/L)\,u^{\tau(0)}
\left(\theta_{s,1}-\dot{\mathfrak f}\,\tau_{d,1}\right).
\end{align}
The term involving $\tau_{d,1}$ is required because the background image moves in global coordinates. These expressions give the normal variables of Sec.~\ref{Sec:prev} in the original two-way embedding variables. The background proper clock $\tau_{\rm p}(\tau)$ is the function given in \eqref{Eq:gate-fluid-clock} with $D=3$, so that
\begin{equation}\label{Eq:gate-two-clock-derivative}
\partial_{\tau_{\rm p}}=u^{\tau(0)}\partial_\tau.
\end{equation}
The solutions below are expressed in global time through $\tau_{\rm p}(\tau)$.

\subsection{Metric continuity and the worldsheet isometries}
\label{Sec:gate-two-metric}

At first order, the three components of \eqref{Eq:hcont} give
\begin{align}\label{Eq:gate-two-metric-continuity}
&\dot\tau_{d,1}
-L^2\cosh^2(\tau/L)\dot{\mathfrak f}\,\dot\theta_{s,1}
-L\cosh(\tau/L)\sinh(\tau/L)\dot{\mathfrak f}^{\,2}\tau_{d,1}=0,
\nonumber\\
&\tau_{d,1}'-L^2\cosh^2(\tau/L)\dot{\mathfrak f}\,\theta_{s,1}'
-R_0^2\dot\sigma_{d,1}=0,
\nonumber\\
&\sigma_{d,1}'+\cot\mathfrak f\,\theta_{s,1}
+\frac{\tanh(\tau/L)}{L}\tau_{d,1}=0.
\end{align}
They involve $\theta_{s,1}$, $\tau_{d,1}$ and $\sigma_{d,1}$, while the average transverse displacement $\theta_{d,1}$ cancels. Using \eqref{Eq:gate-angle-derivative} and \eqref{Eq:gate-two-normal-displacements}, the last equation becomes
\begin{equation}\label{Eq:gate-two-angular-shift}
\sigma_{d,1}'=-\frac{\eta_1^{\rm rel}}{2L\sqrt w}.
\end{equation}
Periodicity therefore imposes
\begin{equation}\label{Eq:gate-two-relative-zero}
\int_0^{2\pi}\eta_1^{\rm rel}(\tau,\sigma)\,{\rm d}\sigma=0.
\end{equation}
In particular, there is no independent homogeneous relative normal displacement.

We first identify the rigid part of the relative shifts by setting $\theta_{d,1}=0$ and $\eta_1^{\rm rel}=0$. The images of the junction are then unchanged, and $\theta_{s,1}=\dot{\mathfrak f}\,\tau_{d,1}$. Equations \eqref{Eq:gate-two-metric-continuity} imply
\begin{equation}\label{Eq:gate-two-worldsheet-parameters}
\tau_{d,1}=\alpha_1 u^{\tau(0)},\qquad
\sigma_{d,1}=\alpha_2,\qquad
\theta_{s,1}=\alpha_1 u^{\tau(0)}\dot{\mathfrak f},
\end{equation}
where $\alpha_1$ and $\alpha_2$ are constants. We read off the corresponding worldsheet Killing vectors directly:
\begin{equation}\label{Eq:gate-two-worldsheet-killing}
l_1=u^{\tau(0)}\partial_\tau,\qquad
l_2=\partial_\sigma.
\end{equation}
Indeed, in the proper clock the background induced metric is $-{\rm d}\tau_{\rm p}^2+R_0^2{\rm d}\sigma^2$, and these are its time and angular translations. The local Lorentz boost of a flat two-dimensional metric is not periodic in $\sigma$ and does not define a global isometry of the cylinder. Thus there are two globally defined worldsheet isometries, whereas the equatorial tensile junction in \cite{Chakraborty:2025baw} had an induced dS$_2$ metric with three. This difference is already visible in the homogeneous solutions for $\tau_d$ and $\sigma_d$.

For a nonzero relative displacement, compatibility of the three equations in \eqref{Eq:gate-two-metric-continuity} gives
\begin{equation}\label{Eq:gate-two-relative-equation}
\partial_{\tau_{\rm p}}^2\eta_1^{\rm rel}
+\frac{1+w}{L^2}\partial_\sigma^2\eta_1^{\rm rel}=0.
\end{equation}
This equation, together with \eqref{Eq:gate-two-relative-zero}, determines the relative displacement degree of freedom. To reconstruct the original embedding variables, expand each first-order field in Fourier harmonics $e^{im\sigma}$, and denote its $m$-th coefficient by a final index $m$. For $m\neq0$, the solutions of \eqref{Eq:gate-two-metric-continuity} are
\begin{align}\label{Eq:gate-two-relative-reconstruction}
\sigma_{d,1;m}
&=-\frac{\eta_{1,m}^{\rm rel}}{2imL\sqrt w},
\nonumber\\
\tau_{d,1;m}
&=u^{\tau(0)}
\left[
\frac{R_0}{2m^2\sqrt{1+w}}\,
\partial_{\tau_{\rm p}}\eta_{1,m}^{\rm rel}
+\frac{L\cosh(\tau/L)\dot{\mathfrak f}}{2}\eta_{1,m}^{\rm rel}
\right],
\nonumber\\
\theta_{s,1;m}
&=u^{\tau(0)}
\left[
\frac{R_0\dot{\mathfrak f}}{2m^2\sqrt{1+w}}\,
\partial_{\tau_{\rm p}}\eta_{1,m}^{\rm rel}
+\frac{\eta_{1,m}^{\rm rel}}{2L\cosh(\tau/L)}
\right].
\end{align}
The constants in \eqref{Eq:gate-two-worldsheet-parameters} supply the remaining homogeneous solutions. Thus the relative clock and angular shifts are determined by the relative normal displacement, up to the rigid choices of worldsheet time and angle.

\subsection{Extrinsic curvature and the hybrid sound mode}
\label{Sec:gate-two-extrinsic}

Once \eqref{Eq:gate-two-metric-continuity} is satisfied, the first-order perturbation of the induced metric is determined by $\theta_{d,1}$:
\begin{align}\label{Eq:gate-two-common-metric}
\gamma_{\tau\tau}^{(1)}
&=2L^2\cosh^2(\tau/L)\dot{\mathfrak f}\,\dot\theta_{d,1},
\nonumber\\
\gamma_{\tau\sigma}^{(1)}
&=L^2\cosh^2(\tau/L)\dot{\mathfrak f}\,\theta_{d,1}',
\nonumber\\
\gamma_{\sigma\sigma}^{(1)}
&=2R_0^2\cot\mathfrak f\,\theta_{d,1}.
\end{align}
Normalization of the velocity therefore fixes
\begin{equation}\label{Eq:gate-two-velocity-normalization}
u_1^\tau
=L^2\cosh^2(\tau/L)\dot{\mathfrak f}\,
\big(u^{\tau(0)}\big)^3\dot\theta_{d,1}.
\end{equation}

We write $\rho=\rho^{(0)}+\epsilon\rho_1+O(\epsilon^2)$ and $P=P^{(0)}+\epsilon P_1+O(\epsilon^2)$, with $P^{(0)}=w\rho^{(0)}$. The extrinsic curvatures computed from \eqref{Eq:gate-two-perturbed-images} give, on substitution in \eqref{Eq:Kdisc}, the three component equations
\begin{align}\label{Eq:gate-two-extrinsic-linear}
\rho_1&=-\frac{\eta+\partial_\sigma^2\eta}{4\pi G_N R_0^2},
\nonumber\\
P_1&=\frac{1}{4\pi G_N}
\left[-\partial_{\tau_{\rm p}}^2\eta
+\frac{1+w}{L^2}\eta\right],
\nonumber\\
u_1^\phi&=\frac{1}{L\sqrt w}\,
\partial_{\tau_{\rm p}}\partial_\sigma\eta.
\end{align}
The relative displacement cancels from these equations at linear order. Imposing the same equation of state, $P_1=w\rho_1$, gives
\begin{equation}\label{Eq:gate-two-hybrid-equation}
\partial_{\tau_{\rm p}}^2\eta
-\frac{1+w}{L^2}\partial_\sigma^2\eta
-\frac{2(1+w)}{L^2}\eta=0.
\end{equation}
This is the hybrid sound degree of freedom. It combines the average transverse displacement, carried by $\theta_{d,1}$ in the northern--southern variables, with the temperature and velocity of the junction fluid. The temperature perturbation follows from the equation of state:
\begin{equation}\label{Eq:gate-two-temperature}
\frac{T_1}{T^{(0)}}
=-\frac{\sqrt w}{L}
\left(\eta+\partial_\sigma^2\eta\right).
\end{equation}
Thus $\theta_{d,1}=\eta/[L\cosh(\tau/L)u^{\tau(0)}]$, $T_1$, $u_1^\phi$ and $u_1^\tau$ are all fixed once $\eta$ is given.

The fluid equations introduce no further restrictions. Their linearized energy and momentum equations can be written as
\begin{align}\label{Eq:gate-two-euler-check}
&\partial_{\tau_{\rm p}}\rho_1+
(\rho^{(0)}+P^{(0)})
\left(\partial_\sigma u_1^\phi
+\frac{\partial_{\tau_{\rm p}}\eta}{L\sqrt w}\right)=0,
\nonumber\\
&R_0^2\partial_{\tau_{\rm p}}u_1^\phi
-\frac{\sqrt w}{L}\partial_\sigma\eta
+\partial_\sigma\left(\frac{T_1}{T^{(0)}}\right)=0.
\end{align}
Substituting \eqref{Eq:gate-two-extrinsic-linear} and \eqref{Eq:gate-two-temperature}, the first equation is an identity and the second gives the angular derivative of \eqref{Eq:gate-two-hybrid-equation}. The homogeneous radial equation is fixed directly by $P_1=w\rho_1$. This is the linearized contracted Codazzi relation. Each bulk metric is de Sitter and hence satisfies the Einstein equations and their constraints. The induced metrics and extrinsic curvatures are computed from the junction embeddings, so the Gauss constraints hold on each image. All three metric-continuity equations, all three extrinsic-curvature junction equations, the fluid equations and the gravitational constraints are therefore satisfied.

\subsection{The ambient spacetime isometries}
\label{Sec:gate-two-ambient}

We identify the ambient isometries by the same comparison with the transverse displacement as in \cite{Chakraborty:2025baw}. The $|m|=1$ solution of \eqref{Eq:gate-two-hybrid-equation}, expressed directly in $\theta_{d,1}$, is
\begin{equation}\label{Eq:gate-two-ambient-parameters}
\theta_{d,1}
=\gamma_1\sin\sigma+\gamma_2\cos\sigma
+\frac{\tanh(\tau/L)}{\cos\mathfrak f(\tau)}
\left(\gamma_3\sin\sigma+\gamma_4\cos\sigma\right),
\end{equation}
where the $\gamma_i$ are constant rigid parameters. We now read off the ambient transformations which produce these four terms.

The relevant Killing vectors of the global dS$_3$ metric can be chosen as
\begin{align}\label{Eq:gate-two-ambient-killing}
L_1&=\sin\phi\,\partial_\theta
+\cot\theta\cos\phi\,\partial_\phi,
\nonumber\\
L_2&=\cos\phi\,\partial_\theta
-\cot\theta\sin\phi\,\partial_\phi,
\nonumber\\
L_3&=L\sin\theta\sin\phi\,\partial_t
+\tanh(t/L)\cos\theta\sin\phi\,\partial_\theta
+\tanh(t/L)\csc\theta\cos\phi\,\partial_\phi,
\nonumber\\
L_4&=L\sin\theta\cos\phi\,\partial_t
+\tanh(t/L)\cos\theta\cos\phi\,\partial_\theta
-\tanh(t/L)\csc\theta\sin\phi\,\partial_\phi.
\end{align}
For the time-dependent embedding $\theta=\mathfrak f(t)$, the transverse displacement at fixed global time is
\begin{equation}\label{Eq:gate-two-isometry-graph}
\theta_{d,1}^{(L_i)}
=\left.L_i^\theta\right|_{t=\tau,\,\theta=\mathfrak f,\,\phi=\sigma}
-\dot{\mathfrak f}
\left.L_i^t\right|_{t=\tau,\,\theta=\mathfrak f,\,\phi=\sigma}.
\end{equation}
The second term accounts for the time shift generated by the ambient isometry. Using \eqref{Eq:gate-angle-derivative}, we obtain
\begin{align}\label{Eq:gate-two-isometry-shifts}
&\theta_{d,1}^{(L_1)}=\sin\sigma,\qquad
\theta_{d,1}^{(L_2)}=\cos\sigma,
\nonumber\\
&\theta_{d,1}^{(L_3)}
=\frac{\tanh(\tau/L)}{\cos\mathfrak f}\sin\sigma,\qquad
\theta_{d,1}^{(L_4)}
=\frac{\tanh(\tau/L)}{\cos\mathfrak f}\cos\sigma.
\end{align}
These are precisely the four terms in \eqref{Eq:gate-two-ambient-parameters}. Their density and temperature perturbations vanish, as follows directly from \eqref{Eq:gate-two-extrinsic-linear} and \eqref{Eq:gate-two-temperature}.

The corresponding normal amplitudes follow from the background identities
\begin{equation}
 \begin{aligned}
L\cosh(\tau/L)u^{\tau(0)}&=L\cosh\left(\frac{\sqrt{1+w}}{L}\tau_p\right),\\
 \frac{\tanh(\tau/L)}{\cos f}&=\tanh\left(\frac{\sqrt{1+w}}{L}\tau_p\right).
 \end{aligned}
\end{equation}

The same four generators identify the $|m|=1$ solutions of the relative-displacement branch. They can act independently on the two images. Acting with opposite signs on the two folded images gives, for example, the following first-order embedding coefficients per unit amplitude:
\begin{align}\label{Eq:gate-two-relative-ambient-examples}
&(\tau_{d,1},\sigma_{d,1},\theta_{s,1},\theta_{d,1})
=\left(0,\cot\mathfrak f\cos\sigma,\sin\sigma,0\right),
\nonumber\\
&(\tau_{d,1},\sigma_{d,1},\theta_{s,1},\theta_{d,1})
=\left(L\sin\mathfrak f\sin\sigma,\,
\tanh(\tau/L)\csc\mathfrak f\cos\sigma,\,
\tanh(\tau/L)\cos\mathfrak f\sin\sigma,\,0\right).
\end{align}
The first is generated by $L_1$ and the second by $L_3$; their cosine and sine partners are generated by $L_2$ and $L_4$. Direct substitution solves \eqref{Eq:gate-two-metric-continuity}, while $\eta=0$ and the junction fluid perturbations vanish. Their relative normal amplitudes are proportional to $\cosh[\sqrt{1+w}\,\tau_{\rm p}(\tau)/L]$ and $\sinh[\sqrt{1+w}\,\tau_{\rm p}(\tau)/L]$, which span the $|m|=1$ solutions of \eqref{Eq:gate-two-relative-equation}. Their explicit Killing displacements, including the clock and angular shifts, identify these solutions with de Sitter isometries acting on the individual junction images.

The ambient generators which move the gate differ from those which moved the equatorial tensile junction. In particular, the polar boost
\begin{equation}\label{Eq:gate-two-polar-boost}
L\cos\theta\,\partial_t
-\tanh(t/L)\sin\theta\,\partial_\theta
\end{equation}
produces no transverse displacement in \eqref{Eq:gate-two-isometry-graph} and restricts to $L\,l_1/\sqrt{1+w}$ on the creation gate. It is therefore a worldsheet time translation here. The two transverse boosts $L_3$ and $L_4$, which were tangent to the equatorial worldsheet, now displace the gate. Consequently the equatorial tensile background has three worldsheet generators and three transverse ambient generators, while this background has two and four, respectively. The homogeneous term proportional to $\tanh(\tau/L)$ in the transverse solution of \cite{Chakraborty:2025baw} should therefore not be carried over to the present $\theta_d$ solution. The homogeneous transverse fluctuations of the creation gate describe physical radial motion.

For the physical perturbations below, we set the rigid parameters to zero, fixing the worldsheet clock and angular origins and the ambient isometries acting on the two images.

\subsection{The physical modes}
\label{Sec:gate-two-physical-modes}

For $|m|\geq2$, the mode of the average normal displacement has
\begin{equation}\label{Eq:gate-two-hybrid-spectrum}
\omega_m^2=\frac{1+w}{L^2}(m^2-2)
=w\left(\frac{m}{R_0}\right)^2-\frac{2(1+w)}{L^2},
\end{equation}
and its general solution is
\begin{equation}\label{Eq:gate-two-hybrid-solutions}
\eta_m(\tau)
=\eta_{m,+}e^{i\omega_m\tau_{\rm p}(\tau)}
+\eta_{m,-}e^{-i\omega_m\tau_{\rm p}(\tau)}.
\end{equation}
The coefficients $\eta_{m,\pm}$ are integration constants, with the reality conditions required for a real normal displacement. These are oscillatory hybrid sound modes. The fluid coupling shifts their frequencies while preserving the short-wavelength sound speed $c_s^2=w$.

The relative normal mode has the general solution
\begin{equation}\label{Eq:gate-two-relative-solutions}
\eta_{1,m}^{\rm rel}(\tau)
=\eta_{1,m,+}^{\rm rel}
e^{|m|\sqrt{1+w}\,\tau_{\rm p}(\tau)/L}
+\eta_{1,m,-}^{\rm rel}
e^{-|m|\sqrt{1+w}\,\tau_{\rm p}(\tau)/L},
\qquad |m|\geq2,
\end{equation}
where the coefficients are again integration constants. The relative clock, angular shift and transverse displacement are reconstructed from \eqref{Eq:gate-two-relative-reconstruction}. Thus, for each generic inhomogeneous harmonic, there are two independent field degrees of freedom: the hybrid sound mode $\eta$ and the relative displacement mode $\eta_1^{\rm rel}$. The four remaining variables are determined by the junction conditions. Their explicit solutions are collected in Sec.~\ref{Sec:gate-two-reconstruction}.

In the homogeneous sector, only the average normal displacement remains after fixing the two worldsheet translations. Its solution is
\begin{equation}\label{Eq:gate-two-homogeneous-solution}
\eta_0(\tau)=\eta_{0,+}
e^{\sqrt{2(1+w)}\,\tau_{\rm p}(\tau)/L}
+\eta_{0,-}
e^{-\sqrt{2(1+w)}\,\tau_{\rm p}(\tau)/L}.
\end{equation}
It changes the common areal radius by $\delta R=\eta_0/\sqrt{1+w}$ and agrees with \eqref{Eq:fluid-radial-zero-mode} for $D=3$. The two branches grow towards opposite temporal ends. The branch with $\eta_{0,-}=0$ and $\eta_{0,+}>0$ has the exact nonlinear continuation constructed in Sec.~\ref{Sec:histories-exact-radial}. The $|m|=1$ solutions are the rigid isometries identified above. For $|m|\geq2$, the hybrid sound modes persist as oscillations and the relative modes grow towards at least one temporal end. There is therefore no nonzero physical normal mode which decays in both the infinite past and future. The angular coefficient $\theta_{d,1}$ of an oscillatory hybrid mode can nevertheless decay because of the factor $[L\cosh(\tau/L)u^{\tau(0)}]^{-1}$ in \eqref{Eq:gate-two-normal-displacements}; the proper normal amplitude $\eta$ remains oscillatory.

Finally, the principal part of \eqref{Eq:gate-two-relative-equation} is elliptic for $w>0$, whereas \eqref{Eq:gate-two-hybrid-equation} is hyperbolic. The relative mode growth rates are unbounded as $|m|$ increases, so arbitrary smooth Cauchy data do not define a well-posed evolution problem for that branch. This requires a specification of admissible perturbations and histories, which we discuss in Sec.~\ref{Sec:admissible-histories}.

\subsection{The full linearized solution}
\label{Sec:gate-two-reconstruction}

We now collect the four variables determined by the normal modes. For a harmonic $|m|\geq2$, fix the parameters of the ambient isometries and specify the constants $\eta_{m,\pm}$ and $\eta_{1,m,\pm}^{\rm rel}$ in \eqref{Eq:gate-two-hybrid-solutions} and \eqref{Eq:gate-two-relative-solutions}. The explicit solutions are
\begin{align}\label{Eq:gate-two-full-reconstruction}
\tau_{d,1;m}
&=\frac{u^{\tau(0)}}{2}
\Bigg[
\left(L\cosh(\tau/L)\dot{\mathfrak f}
+\frac{R_0}{|m|L}\right)
\eta_{1,m,+}^{\rm rel}
e^{|m|\sqrt{1+w}\,\tau_{\rm p}(\tau)/L}
\nonumber\\
&\hspace{24mm}+
\left(L\cosh(\tau/L)\dot{\mathfrak f}
-\frac{R_0}{|m|L}\right)
\eta_{1,m,-}^{\rm rel}
e^{-|m|\sqrt{1+w}\,\tau_{\rm p}(\tau)/L}
\Bigg],
\nonumber\\
\sigma_{d,1;m}
&=-\frac{1}{2imL\sqrt w}
\left[
\eta_{1,m,+}^{\rm rel}
e^{|m|\sqrt{1+w}\,\tau_{\rm p}(\tau)/L}
+\eta_{1,m,-}^{\rm rel}
e^{-|m|\sqrt{1+w}\,\tau_{\rm p}(\tau)/L}
\right],
\nonumber\\
T_{1;m}
&=\frac{T^{(0)}\sqrt w}{L}(m^2-1)
\left[
\eta_{m,+}e^{i\omega_m\tau_{\rm p}(\tau)}
+\eta_{m,-}e^{-i\omega_m\tau_{\rm p}(\tau)}
\right],
\nonumber\\
u^\phi_{1;m}
&=-\frac{m\omega_m}{L\sqrt w}
\left[
\eta_{m,+}e^{i\omega_m\tau_{\rm p}(\tau)}
-\eta_{m,-}e^{-i\omega_m\tau_{\rm p}(\tau)}
\right].
\end{align}
Here $\omega_m$ is the positive root in \eqref{Eq:gate-two-hybrid-spectrum}, and the known function $\tau_{\rm p}(\tau)$ is given by \eqref{Eq:gate-fluid-clock} with $D=3$. The time dependence is therefore explicit in global coordinates. The relative clock and angular shift depend only on the relative normal mode, while the temperature and velocity depend only on the hybrid sound mode at this order.

The two original transverse embedding coefficients follow from the normal definitions:
\begin{align}\label{Eq:gate-two-full-transverse}
\theta_{d,1;m}
&=\frac{\eta_m}{L\cosh(\tau/L)u^{\tau(0)}},
\nonumber\\
\theta_{s,1;m}
&=\dot{\mathfrak f}\,\tau_{d,1;m}
+\frac{\eta_{1,m}^{\rm rel}}{2L\cosh(\tau/L)u^{\tau(0)}}.
\end{align}
Together, \eqref{Eq:gate-two-full-reconstruction} and \eqref{Eq:gate-two-full-transverse} give all six independent first-order coefficients in the original variables of Sec.~\ref{Sec:two_jn}. Substitution in \eqref{Eq:gate-two-perturbed-images} then gives both embeddings and the identification of their points. The induced metric follows from \eqref{Eq:gate-two-common-metric}. Velocity normalization fixes the temporal fluid velocity:
\begin{equation}\label{Eq:gate-two-full-temporal-velocity}
u^\tau_{1;m}
=L^2\cosh^2(\tau/L)\dot{\mathfrak f}\,
\big(u^{\tau(0)}\big)^3
\partial_\tau\!
\left[\frac{\eta_m}{L\cosh(\tau/L)u^{\tau(0)}}\right].
\end{equation}
The density and pressure are those in \eqref{Eq:gate-two-extrinsic-linear}, with $P_1=w\rho_1$. There are no further independent integration functions. For real perturbations, the mode constants obey
\begin{equation}\label{Eq:gate-two-mode-reality}
\eta_{-m,\pm}=(\eta_{m,\mp})^*,
\qquad
\eta_{1,-m,\pm}^{\rm rel}
=(\eta_{1,m,\pm}^{\rm rel})^*.
\end{equation}

We treat the homogeneous sector separately, since \eqref{Eq:gate-two-relative-reconstruction} was derived for $m\neq0$. With $\eta_0$ given by \eqref{Eq:gate-two-homogeneous-solution}, periodicity requires $\eta_{1,0}^{\rm rel}=0$, and the complete homogeneous solution is
\begin{align}\label{Eq:gate-two-full-homogeneous}
&\tau_{d,1;0}=\alpha_1u^{\tau(0)},\qquad
\sigma_{d,1;0}=\alpha_2,\qquad
\theta_{s,1;0}=\alpha_1u^{\tau(0)}\dot{\mathfrak f},
\nonumber\\
&\theta_{d,1;0}
=\frac{\eta_0}{L\cosh(\tau/L)u^{\tau(0)}},
\qquad
T_{1;0}=-\frac{T^{(0)}\sqrt w}{L}\eta_0,
\qquad u^\phi_{1;0}=0.
\end{align}
The constants $\alpha_1$ and $\alpha_2$ are the relative worldsheet origins identified in \eqref{Eq:gate-two-worldsheet-parameters}; they may be set to zero. The two real constants $\eta_{0,\pm}$ specify the physical radial motion. Its temporal velocity is obtained by setting $m=0$ in \eqref{Eq:gate-two-full-temporal-velocity}.

Finally, the $|m|=1$ solutions are the isometries identified in Sec.~\ref{Sec:gate-two-ambient}. The average transverse displacement is given by \eqref{Eq:gate-two-ambient-parameters}, evaluated at fixed global time as in \eqref{Eq:gate-two-isometry-graph}. The relative displacements are constant linear combinations of \eqref{Eq:gate-two-relative-ambient-examples} and their angular partners, with the clock and angular shifts given by the corresponding Killing vectors. The temperature perturbation vanishes in this sector. Applying the same ambient isometry to both folded images can also induce the coordinate velocity
\[
u^\phi_{1;m}
=\frac{im}{L\sqrt w}\,
\partial_{\tau_{\rm p}}\eta_m,
\qquad |m|=1,
\]
while a purely relative isometry has $\eta=0$ and leaves the junction fluid variables unchanged. This coordinate velocity is fixed by the isometry and supplies no additional fluid mode. Velocity normalization again fixes $u_1^\tau$. Thus the radial constants, the two independent normal-mode solutions for each $|m|\geq2$, and the rigid parameters already identified give the full linear solution.

\section{Linear perturbations of the $n$-way cosmic creation gate in three dimensions}
\label{Sec:n-way-perturbations-three}

We now consider three or more bulk regions joined at the perfect-fluid creation gate. We follow the procedure of \cite{Chakraborty:2025baw}, using the embedding variables of Sec.~\ref{Sec:n_jn}. We first work out the three-way junction and then extend the construction to arbitrary $n>3$. All bulk metrics remain the global de Sitter metrics, and the expansion is in the amplitudes of the embedding and fluid perturbations at fixed $w>0$. We use the normal displacements defined in Sec.~\ref{Sec:prev}.

\subsection{The three-way junction and its perturbative expansion}
\label{Sec:gate-three-expansion}

We retain the northern fragment of each of the three copies, as in \eqref{Eq:gate-embedding-three}. Thus the background images have $t_i=\tau$, $\phi_i=\sigma$ and $\theta_i=\mathfrak f(\tau)$, with
\begin{equation}\label{Eq:gate-three-background}
\mathfrak f(\tau)=\arcsin\!\left[\frac{R_0}{L\cosh(\tau/L)}\right],
\qquad R_0^2=L^2\frac{w}{1+w},
\qquad u^{\tau(0)}=\sqrt{1+w\tanh^2(\tau/L)}.
\end{equation}
Their induced metrics agree, and their extrinsic curvatures are given by \eqref{Eq:gate-curvatures-three}. The background fluid has
\begin{equation}\label{Eq:gate-three-background-fluid}
\rho^{(0)}=\frac{3}{8\pi G_N L\sqrt w},\qquad
P^{(0)}=w\rho^{(0)},\qquad
T^{(0)}=T_0\left(\frac{w\rho^{(0)}}{P_0}\right)^{\frac{w}{1+w}},
\qquad u^{\phi(0)}=0.
\end{equation}
Thus all background junction conditions, fluid equations and bulk constraints are already satisfied by the solution of Sec.~\ref{Sec:creation-three-dimensional}.

We expand the seven embedding variables as
\begin{align}\label{Eq:gate-three-expansion}
&\theta_s=\mathfrak f+\epsilon\theta_{s,1}+O(\epsilon^2),
\nonumber\\
&\tau_{d_i}=\epsilon\tau_{d_i,1}+O(\epsilon^2),\qquad
\sigma_{d_i}=\epsilon\sigma_{d_i,1}+O(\epsilon^2),\qquad
\theta_{d_i}=\epsilon\theta_{d_i,1}+O(\epsilon^2),
\quad i=1,2.
\end{align}
The temperature and velocity are expanded as in \eqref{Eq:gate-two-fluid-expansion}, with the three-way background values above. The seven embedding variables, $T$ and $u^\phi$ are the nine independent functions. A dot and a prime in this section denote derivatives with respect to the global time $\tau$ and the angle $\sigma$, respectively. All perturbations are periodic in $\sigma$.

Inverting \eqref{Eq:xsxd} gives the identified times and angles explicitly:
\begin{align}\label{Eq:gate-three-time-angle-images}
t_1&=\tau+\epsilon(\tau_{d_1,1}-2\tau_{d_2,1}),
&\phi_1&=\sigma+\epsilon(\sigma_{d_1,1}-2\sigma_{d_2,1}),
\nonumber\\
t_2&=\tau+\epsilon(-2\tau_{d_1,1}+\tau_{d_2,1}),
&\phi_2&=\sigma+\epsilon(-2\sigma_{d_1,1}+\sigma_{d_2,1}),
\nonumber\\
t_3&=\tau+\epsilon(\tau_{d_1,1}+\tau_{d_2,1}),
&\phi_3&=\sigma+\epsilon(\sigma_{d_1,1}+\sigma_{d_2,1}).
\end{align}
The polar coordinates of the images are
\begin{align}\label{Eq:gate-three-polar-images}
\theta_1&=\mathfrak f+\epsilon(\theta_{s,1}+\theta_{d_1,1}-2\theta_{d_2,1}),
\nonumber\\
\theta_2&=\mathfrak f+\epsilon(\theta_{s,1}-2\theta_{d_1,1}+\theta_{d_2,1}),
\nonumber\\
\theta_3&=\mathfrak f+\epsilon(\theta_{s,1}+\theta_{d_1,1}+\theta_{d_2,1}).
\end{align}
Here and in subsequent first-order embedding formulas, terms of $O(\epsilon^2)$ are understood. The average time and azimuthal angle are $\tau$ and $\sigma$, as defined in \eqref{Eq:WSgauge}.

\subsection{Metric continuity and the two relative displacement modes}
\label{Sec:gate-three-metric}

We equate the induced metric of image $3$ to those of images $2$ and $1$, in this order. Each comparison gives three equations. Substituting \eqref{Eq:gate-three-time-angle-images} and \eqref{Eq:gate-three-polar-images}, and dividing out the common factor of three, gives
\begin{align}\label{Eq:gate-three-metric-continuity}
0={}&\dot\tau_{d_i,1}
-L^2\cosh^2(\tau/L)\dot{\mathfrak f}\,\dot\theta_{d_i,1}
-L\cosh(\tau/L)\sinh(\tau/L)\dot{\mathfrak f}^{\,2}\tau_{d_i,1},
\nonumber\\
0={}&\tau_{d_i,1}'
-L^2\cosh^2(\tau/L)\dot{\mathfrak f}\,\theta_{d_i,1}'
-R_0^2\dot\sigma_{d_i,1},
\nonumber\\
0={}&\sigma_{d_i,1}'+\cot\mathfrak f\,\theta_{d_i,1}
+\frac{\tanh(\tau/L)}{L}\tau_{d_i,1},
\qquad i=1,2.
\end{align}
The average transverse displacement $\theta_{s,1}$ cancels from all six equations. Each set of relative shifts therefore obeys these three equations independently at this order.

For a first-order embedding displacement, its proper normal component is obtained by contraction with \eqref{Eq:gate-global-normal}. Since the average time shift vanishes, the average normal displacement is
\begin{equation}\label{Eq:gate-three-collective-normal}
\eta=L\cosh(\tau/L)\,u^{\tau(0)}\theta_{s,1}.
\end{equation}
The normal differences $\eta_1^{\rm rel}=\eta_1-\eta_3$ and $\eta_2^{\rm rel}=\eta_2-\eta_3$ are labelled by the images, while \eqref{Eq:xsxd} defines the coordinate differences in the order $3-2$ and $3-1$, with a factor of $1/3$. Thus
\begin{align}\label{Eq:gate-three-relative-normal}
\eta_1^{\rm rel}
&=-3L\cosh(\tau/L)\,u^{\tau(0)}
\left(\theta_{d_2,1}-\dot{\mathfrak f}\,\tau_{d_2,1}\right),
\nonumber\\
\eta_2^{\rm rel}
&=-3L\cosh(\tau/L)\,u^{\tau(0)}
\left(\theta_{d_1,1}-\dot{\mathfrak f}\,\tau_{d_1,1}\right).
\end{align}
The factors of three follow from \eqref{Eq:xsxd}; the terms proportional to the relative clock shifts account for the time dependence of the background embedding.

Using \eqref{Eq:gate-angle-derivative}, the last of \eqref{Eq:gate-three-metric-continuity} becomes
\begin{equation}\label{Eq:gate-three-angular-shift}
\sigma_{d_i,1}'=\frac{\eta_{3-i}^{\rm rel}}{3L\sqrt w},
\qquad i=1,2.
\end{equation}
The angular shifts must be periodic, so
\begin{equation}\label{Eq:gate-three-relative-zero}
\int_0^{2\pi}\eta_I^{\rm rel}(\tau,\sigma)\,{\rm d}\sigma=0,
\qquad I=1,2.
\end{equation}
Thus neither relative normal displacement has a homogeneous component.

We first isolate the rigid worldsheet transformations by setting $\eta=\eta_1^{\rm rel}=\eta_2^{\rm rel}=0$. The six continuity equations then give
\begin{equation}\label{Eq:gate-three-worldsheet-parameters}
\tau_{d_i,1}=\alpha_{i,1}u^{\tau(0)},\qquad
\sigma_{d_i,1}=\alpha_{i,2},\qquad
\theta_{d_i,1}=\alpha_{i,1}u^{\tau(0)}\dot{\mathfrak f},
\qquad i=1,2,
\end{equation}
with constant $\alpha_{i,1}$ and $\alpha_{i,2}$. We read off the worldsheet generators from the clock and angular shifts: they are $u^{\tau(0)}\partial_\tau$ and $\partial_\sigma$, exactly as in \eqref{Eq:gate-two-worldsheet-killing}. There are two independent relative choices of their origins. The induced cylinder has no additional globally periodic boost isometry.

For nonzero normal displacements, differentiating the continuity equations and eliminating the clock and angular shifts gives
\begin{equation}\label{Eq:gate-three-relative-equations}
\partial_{\tau_{\rm p}}^2\eta_I^{\rm rel}
+\frac{1+w}{L^2}\partial_\sigma^2\eta_I^{\rm rel}=0,
\qquad I=1,2,
\end{equation}
where $\partial_{\tau_{\rm p}}=u^{\tau(0)}\partial_\tau$. The fluid proper time is expressed in global time by \eqref{Eq:gate-fluid-clock}.

We next determine the relative embedding variables. Expanding in Fourier harmonics $e^{im\sigma}$ and denoting each coefficient by a final index $m$, we find, for $m\neq0$,
\begin{align}\label{Eq:gate-three-relative-reconstruction}
\sigma_{d_i,1;m}
&=\frac{\eta_{3-i,m}^{\rm rel}}{3imL\sqrt w},
\nonumber\\
\tau_{d_i,1;m}
&=-\frac{u^{\tau(0)}}{3}
\left[
\frac{R_0}{m^2\sqrt{1+w}}\,
\partial_{\tau_{\rm p}}\eta_{3-i,m}^{\rm rel}
+L\cosh(\tau/L)\dot{\mathfrak f}\,\eta_{3-i,m}^{\rm rel}
\right],
\nonumber\\
\theta_{d_i,1;m}
&=-\frac{u^{\tau(0)}}{3}
\left[
\frac{R_0\dot{\mathfrak f}}{m^2\sqrt{1+w}}\,
\partial_{\tau_{\rm p}}\eta_{3-i,m}^{\rm rel}
+\frac{\eta_{3-i,m}^{\rm rel}}{L\cosh(\tau/L)}
\right],
\qquad i=1,2.
\end{align}
The angular and mixed continuity equations determine $\sigma_{d_i,1;m}$ and $\tau_{d_i,1;m}$, and \eqref{Eq:gate-three-relative-normal} then determines $\theta_{d_i,1;m}$. Substitution in the remaining temporal equation gives \eqref{Eq:gate-three-relative-equations}. Together with the homogeneous solutions \eqref{Eq:gate-three-worldsheet-parameters}, these expressions solve all six metric-continuity equations.

\subsection{The summed extrinsic curvatures and the fluid}
\label{Sec:gate-three-extrinsic}

Once metric continuity is imposed, the first-order induced metric is
\begin{align}\label{Eq:gate-three-common-metric}
\gamma_{\tau\tau}^{(1)}
&=2L^2\cosh^2(\tau/L)\dot{\mathfrak f}\,\dot\theta_{s,1},
\nonumber\\
\gamma_{\tau\sigma}^{(1)}
&=L^2\cosh^2(\tau/L)\dot{\mathfrak f}\,\theta_{s,1}',
\nonumber\\
\gamma_{\sigma\sigma}^{(1)}
&=2R_0^2\cot\mathfrak f\,\theta_{s,1}.
\end{align}
The relative clock and angular shifts cancel in the average of the three induced metric variations, and metric continuity equates each induced metric to this average. Normalization of the velocity gives
\begin{equation}\label{Eq:gate-three-velocity-normalization}
u_1^\tau=L^2\cosh^2(\tau/L)\dot{\mathfrak f}\,
\big(u^{\tau(0)}\big)^3\dot\theta_{s,1}.
\end{equation}

We next impose the second junction condition. In the folded picture, the three background embeddings have identical coordinate expressions, and their reference normals have the same orientation. The relative contributions to the image embeddings in \eqref{Eq:gate-three-time-angle-images} and \eqref{Eq:gate-three-polar-images} sum to zero. Since the first-order extrinsic-curvature variations are linear in these shifts, their relative contributions also cancel in \eqref{Eq:Kdisc-n}. The remaining contribution is three times the variation due to the average transverse displacement $\theta_{s,1}$, with \eqref{Eq:gate-three-common-metric} included in the traces. The three components give
\begin{align}\label{Eq:gate-three-extrinsic-linear}
\rho_1&=-\frac{3}{8\pi G_N R_0^2}
\left(\eta+\partial_\sigma^2\eta\right),
\nonumber\\
P_1&=\frac{3}{8\pi G_N}
\left[-\partial_{\tau_{\rm p}}^2\eta
+\frac{1+w}{L^2}\eta\right],
\nonumber\\
u_1^\phi&=\frac{1}{L\sqrt w}\,
\partial_{\tau_{\rm p}}\partial_\sigma\eta.
\end{align}
Here $\rho_1$ and $P_1$ denote the first-order density and pressure. They are determined by the fluid variables and are not additional independent fields.

The equation of state $P_1=w\rho_1$ now requires
\begin{equation}\label{Eq:gate-three-hybrid-equation}
\partial_{\tau_{\rm p}}^2\eta
-\frac{1+w}{L^2}\partial_\sigma^2\eta
-\frac{2(1+w)}{L^2}\eta=0.
\end{equation}
Thus the average normal displacement gives the hybrid sound mode. The temperature perturbation follows from $T_1/T^{(0)}=w\rho_1/[(1+w)\rho^{(0)}]$, giving
\begin{equation}\label{Eq:gate-three-temperature}
\frac{T_1}{T^{(0)}}
=-\frac{\sqrt w}{L}\left(\eta+\partial_\sigma^2\eta\right).
\end{equation}
Together with $\theta_{s,1}=\eta/[L\cosh(\tau/L)u^{\tau(0)}]$, these equations determine the average transverse displacement, temperature and velocity from $\eta$. The background density carries a factor of three, which therefore cancels from the mode equation and the velocity and fractional temperature perturbations.

The fluid equations can be checked directly. Their energy and angular momentum components reduce to
\begin{align}\label{Eq:gate-three-euler-check}
&\partial_{\tau_{\rm p}}\rho_1+
(\rho^{(0)}+P^{(0)})
\left(\partial_\sigma u_1^\phi
+\frac{\partial_{\tau_{\rm p}}\eta}{L\sqrt w}\right)=0,
\nonumber\\
&R_0^2\partial_{\tau_{\rm p}}u_1^\phi
-\frac{\sqrt w}{L}\partial_\sigma\eta
+\partial_\sigma\left(\frac{T_1}{T^{(0)}}\right)=0.
\end{align}
Substitution of \eqref{Eq:gate-three-extrinsic-linear} and \eqref{Eq:gate-three-temperature} makes the energy equation an identity; the momentum equation is the angular derivative of \eqref{Eq:gate-three-hybrid-equation}, multiplied by $L\sqrt w/(1+w)$. The homogeneous equation follows directly from $P_1=w\rho_1$ in the second junction condition. The fluid equations therefore impose no further restriction on the average transverse displacement. The bulk Einstein equations and their gravitational constraints hold because every retained region remains de Sitter. On each image, the induced metric and extrinsic curvature come from a single global embedding and therefore satisfy the Gauss and Codazzi identities. The resulting solution satisfies the six metric-continuity equations, the three extrinsic-curvature junction equations, velocity normalization and both fluid equations, as well as the bulk constraints.

\subsection{Isometries and the three physical modes}
\label{Sec:gate-three-isometries-modes}

The relative clock and angular shifts in \eqref{Eq:gate-three-worldsheet-parameters} give the worldsheet transformations. We identify the ambient isometries by comparing the transverse displacement directly with the global Killing vectors in \eqref{Eq:gate-two-ambient-killing}. For an ambient Killing vector $L_a$, the transverse displacement at fixed global time is $L_a^\theta-\dot{\mathfrak f}L_a^t$. Acting on all three images with the same amplitude therefore gives the average transverse displacement
\begin{equation}\label{Eq:gate-three-collective-isometries}
\theta_{s,1}
=\gamma_1\sin\sigma+\gamma_2\cos\sigma
+\frac{\tanh(\tau/L)}{\cos\mathfrak f}
\left(\gamma_3\sin\sigma+\gamma_4\cos\sigma\right),
\end{equation}
where the $\gamma_a$ are constant rigid parameters. These are exactly the $|m|=1$ solutions of \eqref{Eq:gate-three-hybrid-equation} after using \eqref{Eq:gate-three-collective-normal}. The average transverse displacement is carried by $\theta_{s,1}$ in the folded picture and by $\theta_{d,1}$ in the northern--southern two-way picture of Sec.~\ref{Sec:two-way-perturbations-three}. The exchange follows from reflection of the southern image.

The same Killing vectors can act with independent constant amplitudes on the three images. For example, set $\theta_{s,1}=0$ and keep only one set of relative shifts $(\tau_{d_i,1},\sigma_{d_i,1},\theta_{d_i,1})$. Two solutions are
\begin{align}\label{Eq:gate-three-relative-isometries}
(\tau_{d_i,1},\sigma_{d_i,1},\theta_{d_i,1})
&=\left(0,\cot\mathfrak f\cos\sigma,\sin\sigma\right),
\nonumber\\
(\tau_{d_i,1},\sigma_{d_i,1},\theta_{d_i,1})
&=\left(L\sin\mathfrak f\sin\sigma,\,
\tanh(\tau/L)\csc\mathfrak f\cos\sigma,\,
\tanh(\tau/L)\cos\mathfrak f\sin\sigma\right).
\end{align}
These are the restrictions of $L_1$ and $L_3$ in \eqref{Eq:gate-two-ambient-killing}; their angular partners follow from $L_2$ and $L_4$. For $i=1$, the inverse embeddings assign the respective isometry amplitudes $(1,-2,1)$ to the three images, and for $i=2$ they assign $(-2,1,1)$. Hence their mean displacement vanishes. Both sets of shifts solve \eqref{Eq:gate-three-metric-continuity} directly, while their relative normal displacements are proportional to
\begin{equation}\label{Eq:gate-three-relative-isometry-time}
\cosh\!\left(\frac{\sqrt{1+w}}{L}\tau_{\rm p}\right),
\qquad
\sinh\!\left(\frac{\sqrt{1+w}}{L}\tau_{\rm p}\right).
\end{equation}
With their sine and cosine angular dependence, these span the two relative $|m|=1$ sectors. Their transverse, clock and angular shifts thus identify these solutions directly with ambient isometries.

The background preserves the same two worldsheet generators as the two-way creation gate. In particular, the polar boost \eqref{Eq:gate-two-polar-boost} is tangent to every image and gives a worldsheet time translation. The four other transverse generators above move the circular gate. The distinction from the equatorial tensile junction therefore persists for three or more images: its worldsheet and transverse isometries cannot be carried over unchanged. When counting the physical modes, we fix the relative origins of the clock and angle and the parameters of the ambient isometries.

For each generic inhomogeneous harmonic $|m|\geq2$, the solution for the average displacement has
\begin{align}\label{Eq:gate-three-hybrid-spectrum}
\omega_m^2&=\frac{1+w}{L^2}(m^2-2)
=w\left(\frac{m}{R_0}\right)^2-\frac{2(1+w)}{L^2},
\nonumber\\
\eta_m(\tau)&=\eta_{m,+}e^{i\omega_m\tau_{\rm p}(\tau)}
+\eta_{m,-}e^{-i\omega_m\tau_{\rm p}(\tau)}.
\end{align}
The two independent relative solutions are
\begin{equation}\label{Eq:gate-three-relative-solutions}
\eta_{I,m}^{\rm rel}(\tau)
=\eta_{I,m,+}^{\rm rel}
e^{|m|\sqrt{1+w}\,\tau_{\rm p}(\tau)/L}
+\eta_{I,m,-}^{\rm rel}
e^{-|m|\sqrt{1+w}\,\tau_{\rm p}(\tau)/L},
\qquad I=1,2.
\end{equation}
The coefficients in these solutions are integration constants, with the reality conditions appropriate to real fields. Thus the nine variables contain three independent field degrees of freedom: one oscillatory hybrid sound mode and two relative displacement modes. The remaining six variables follow from \eqref{Eq:gate-three-relative-reconstruction}, \eqref{Eq:gate-three-collective-normal}, \eqref{Eq:gate-three-extrinsic-linear} and \eqref{Eq:gate-three-temperature}.

In the homogeneous sector, \eqref{Eq:gate-three-relative-zero} removes both relative normal modes. After fixing the rigid clock and angular shifts, the only remaining perturbation is
\begin{equation}\label{Eq:gate-three-radial-solution}
\eta_0(\tau)=\eta_{0,+}
e^{\sqrt{2(1+w)}\,\tau_{\rm p}(\tau)/L}
+\eta_{0,-}
e^{-\sqrt{2(1+w)}\,\tau_{\rm p}(\tau)/L}.
\end{equation}
This homogeneous average transverse displacement changes the radius of every image by $\delta R=\eta_0/\sqrt{1+w}$ and is fixed by two integration constants. It supplies no additional local field degree of freedom on the circle. The $|m|=1$ modes are the rigid ambient isometries identified above. The relative equations have elliptic principal part, with growth rates unbounded as $|m|$ increases, while the hybrid sound equation is hyperbolic. The restrictions on admissible data are discussed in Sec.~\ref{Sec:admissible-histories}. No nonzero physical normal perturbation in this linear solution decays at both temporal ends.

\subsection{The general $n$-way junction}
\label{Sec:gate-n-general}

We now consider $n>3$ northern fragments, with the same background \eqref{Eq:gate-embedding-three}. Their radius and normalized velocity remain \eqref{Eq:gate-three-background}, while
\begin{equation}\label{Eq:gate-n-background-fluid}
\rho^{(0)}=\frac{n}{8\pi G_N L\sqrt w},\qquad
P^{(0)}=w\rho^{(0)},\qquad
T^{(0)}=T_0\left(\frac{w\rho^{(0)}}{P_0}\right)^{\frac{w}{1+w}}.
\end{equation}
We use \eqref{Eq:gate-three-expansion} with $i=1,\cdots,n-1$, and expand the fluid as before. There are $3n-2$ embedding functions and two independent fluid variables.

The inverse of the cyclic definitions \eqref{Eq:xsxd} is
\begin{align}\label{Eq:gate-n-inverse-times}
t_n&=\tau+\epsilon\sum_{j=1}^{n-1}\tau_{d_j,1},
\nonumber\\
t_1&=\tau+\epsilon\left(\sum_{j=1}^{n-1}\tau_{d_j,1}
-n\tau_{d_{n-1},1}\right),
\nonumber\\
t_{i+1}&=\tau+\epsilon\left(\sum_{j=1}^{n-1}\tau_{d_j,1}
-n\tau_{d_i,1}\right),\qquad i=1,\cdots,n-2.
\end{align}
The azimuthal coordinates are given by the same expressions with $t$ replaced by $\phi$, $\tau$ by $\sigma$ and $\tau_{d_j,1}$ by $\sigma_{d_j,1}$. The polar coordinates are
\begin{align}\label{Eq:gate-n-inverse-polar}
\theta_n&=\mathfrak f+\epsilon\left(\theta_{s,1}
+\sum_{j=1}^{n-1}\theta_{d_j,1}\right),
\nonumber\\
\theta_1&=\mathfrak f+\epsilon\left(\theta_{s,1}
+\sum_{j=1}^{n-1}\theta_{d_j,1}-n\theta_{d_{n-1},1}\right),
\nonumber\\
\theta_{i+1}&=\mathfrak f+\epsilon\left(\theta_{s,1}
+\sum_{j=1}^{n-1}\theta_{d_j,1}-n\theta_{d_i,1}\right),
\qquad i=1,\cdots,n-2.
\end{align}
These embeddings satisfy the coordinate definitions \eqref{Eq:WSgauge} and are written in global coordinates.

Comparing image $n$ with each of the others gives $n-1$ copies of the three equations \eqref{Eq:gate-three-metric-continuity}. The factor of $n$ in each coordinate difference divides out. The average normal displacement is still \eqref{Eq:gate-three-collective-normal}; the relative normal displacements of Sec.~\ref{Sec:prev} are related to the coordinate variables by
\begin{align}\label{Eq:gate-n-relative-normal}
\eta_1^{\rm rel}
&=-nL\cosh(\tau/L)\,u^{\tau(0)}
\left(\theta_{d_{n-1},1}-\dot{\mathfrak f}\,\tau_{d_{n-1},1}\right),
\nonumber\\
\eta_{i+1}^{\rm rel}
&=-nL\cosh(\tau/L)\,u^{\tau(0)}
\left(\theta_{d_i,1}-\dot{\mathfrak f}\,\tau_{d_i,1}\right),
\qquad i=1,\cdots,n-2.
\end{align}
The individual normal displacements are
\begin{equation}\label{Eq:gate-n-individual-normals}
\eta_n=\eta-\frac1n\sum_{J=1}^{n-1}\eta_J^{\rm rel},
\qquad
\eta_I=\eta+\eta_I^{\rm rel}
-\frac1n\sum_{J=1}^{n-1}\eta_J^{\rm rel},
\quad I=1,\cdots,n-1.
\end{equation}

The displacements relative to their average satisfy
\begin{equation}
 \sum_{i=1}^{n}(\eta_i-\eta)=0,\qquad
 \sum_{I=1}^{n-1}\eta_I^{\mathrm{rel}}=n(\eta-\eta_n).
\end{equation}
The independent normal differences $\eta_I^{\rm rel}$ need not sum to zero.

Each angular continuity equation requires the corresponding relative displacement to have zero angular mean. Its compatibility with the temporal and mixed equations gives
\begin{equation}\label{Eq:gate-n-relative-equations}
\partial_{\tau_{\rm p}}^2\eta_I^{\rm rel}
+\frac{1+w}{L^2}\partial_\sigma^2\eta_I^{\rm rel}=0,
\qquad \int_0^{2\pi}\eta_I^{\rm rel}\,{\rm d}\sigma=0,
\qquad I=1,\cdots,n-1.
\end{equation}
For $I=2,\cdots,n-1$ and $m\neq0$, all relative coordinate variables are
\begin{align}\label{Eq:gate-n-relative-reconstruction}
\sigma_{d_{I-1},1;m}
&=\frac{\eta_{I,m}^{\rm rel}}{nimL\sqrt w},
\nonumber\\
\tau_{d_{I-1},1;m}
&=-\frac{u^{\tau(0)}}{n}
\left[
\frac{R_0}{m^2\sqrt{1+w}}\,
\partial_{\tau_{\rm p}}\eta_{I,m}^{\rm rel}
+L\cosh(\tau/L)\dot{\mathfrak f}\,\eta_{I,m}^{\rm rel}
\right],
\nonumber\\
\theta_{d_{I-1},1;m}
&=-\frac{u^{\tau(0)}}{n}
\left[
\frac{R_0\dot{\mathfrak f}}{m^2\sqrt{1+w}}\,
\partial_{\tau_{\rm p}}\eta_{I,m}^{\rm rel}
+\frac{\eta_{I,m}^{\rm rel}}{L\cosh(\tau/L)}
\right].
\end{align}
For $I=1$, the same three right hand sides give $\sigma_{d_{n-1},1;m}$, $\tau_{d_{n-1},1;m}$ and $\theta_{d_{n-1},1;m}$, respectively. The coordinate labels follow the cyclic ordering in \eqref{Eq:xsxd}, while $I$ labels the image in $\eta_I^{\rm rel}$. The homogeneous coordinate solutions are \eqref{Eq:gate-three-worldsheet-parameters}, now with $i=1,\cdots,n-1$. Thus there are $2(n-1)$ rigid relative clock and angular origins, which can be fixed.

To evaluate the second junction condition, sum the first-order extrinsic-curvature variations over all $n$ images. Their background coefficients agree. Equations \eqref{Eq:gate-n-inverse-times} and \eqref{Eq:gate-n-inverse-polar} show that the time and azimuthal variations have zero sum, while the polar variations sum to $n\theta_{s,1}$. The traces are computed with the induced metric $\gamma_{\mu\nu}$, which agrees on all images. Therefore the sum is $n$ times the variation of a single image with displacement $\theta_{s,1}$. Its three components give
\begin{align}\label{Eq:gate-n-extrinsic-linear}
\rho_1&=-\frac{n}{8\pi G_N R_0^2}
\left(\eta+\partial_\sigma^2\eta\right),
\nonumber\\
P_1&=\frac{n}{8\pi G_N}
\left[-\partial_{\tau_{\rm p}}^2\eta
+\frac{1+w}{L^2}\eta\right],
\nonumber\\
u_1^\phi&=\frac{1}{L\sqrt w}\,
\partial_{\tau_{\rm p}}\partial_\sigma\eta.
\end{align}
The hybrid sound equation is again \eqref{Eq:gate-three-hybrid-equation}, and its temperature perturbation is \eqref{Eq:gate-three-temperature}. The induced metric and $u_1^\tau$ remain \eqref{Eq:gate-three-common-metric} and \eqref{Eq:gate-three-velocity-normalization}. The fluid equations reduce to \eqref{Eq:gate-three-euler-check} with the $n$-way background density and pressure; direct substitution verifies them. The bulk Einstein equations and the Gauss--Codazzi constraints also hold for every image. The expressions above satisfy all $3(n-1)$ metric-continuity equations, the three extrinsic-curvature junction equations and the fluid equations for arbitrary $n$.

The isometries follow from the shifts used in the three-way case. The homogeneous relative clocks and angles yield the two worldsheet generators for each independent pair. Applying each of $L_1,\cdots,L_4$ with the same amplitude to all $n$ images gives \eqref{Eq:gate-three-collective-isometries}. Independent actions on the images, with amplitudes whose sum vanishes, give the $n-1$ relative $|m|=1$ sectors. The examples \eqref{Eq:gate-three-relative-isometries} apply to any one set of relative shifts, with all other relative shifts and $\theta_{s,1}$ set to zero. Inverting the coordinate differences assigns amplitude $1-n$ to the compared image and amplitude $1$ to all the others. These solutions are therefore ambient isometries acting with different constant amplitudes on the images. Increasing $n$ adds independent isometry parameters.

For $|m|\geq2$, the general solution therefore consists of the single hybrid sound mode \eqref{Eq:gate-three-hybrid-spectrum} and $n-1$ independent relative modes of the form \eqref{Eq:gate-three-relative-solutions}. Their frequencies and growth rates do not depend on $n$, whereas the background density and the density and pressure perturbations are proportional to $n$ at fixed $w$, $L$ and average normal displacement $\eta$. Once these $n$ normal fields are given, the remaining $2n$ variables are reconstructed as above, up to rigid parameters. At $m=0$ only the common radial solution \eqref{Eq:gate-three-radial-solution} remains, while $|m|=1$ consists of rigid isometry data. This verifies both the mode equations and the number of degrees of freedom stated in Sec.~\ref{Sec:prev}. The reconstruction of all embedding and fluid variables is collected in Sec.~\ref{Sec:gate-n-reconstruction}.

The $n-1$ relative modes decouple at first order because the background images are identical. At higher orders, products of relative perturbations can contribute to the equation for the average displacement. The present linear analysis therefore does not establish any correspondence with the coupled Nambu--Goto--Monge--Amp\`ere description of the equatorial, small-tension junction.

\subsection{The full linearized solution}
\label{Sec:gate-n-reconstruction}

We now give the remaining embedding and fluid variables, following Sec.~\ref{Sec:gate-two-reconstruction}. For a three-way junction, choose the solution for the average displacement \eqref{Eq:gate-three-hybrid-spectrum} and the two independent relative solutions \eqref{Eq:gate-three-relative-solutions}. Equation \eqref{Eq:gate-three-relative-reconstruction} then determines both relative clocks and both angular shifts. The two fluid variables are given by the temperature and velocity formulas in \eqref{Eq:gate-two-full-reconstruction}, using the three-way value of $T^{(0)}$ in \eqref{Eq:gate-three-background-fluid}. Their dependence on the hybrid sound mode is otherwise identical. The remaining transverse coefficients follow from \eqref{Eq:gate-three-collective-normal} and \eqref{Eq:gate-three-relative-normal}. This reconstructs all nine variables from the three independent normal fields and the rigid parameters.

For arbitrary $n>3$, choose the same solution for the average displacement and $n-1$ relative solutions of the form \eqref{Eq:gate-three-relative-solutions}, with $I=1,\cdots,n-1$. The $2(n-1)$ clock and angular shifts are determined by \eqref{Eq:gate-n-relative-reconstruction}, including its cyclic assignment for $I=1$. The temperature and velocity are again the last two relations in \eqref{Eq:gate-two-full-reconstruction}, with $T^{(0)}$ now given by \eqref{Eq:gate-n-background-fluid}. The $2n$ dependent variables are therefore
\[
\{\tau_{d_i,1},\sigma_{d_i,1}\}_{i=1}^{n-1},
\qquad T_1,\qquad u_1^\phi.
\]
The original transverse variables are recovered as
\begin{align}\label{Eq:gate-n-full-transverse}
\theta_{s,1;m}
&=\frac{\eta_m}{L\cosh(\tau/L)u^{\tau(0)}},
\nonumber\\
\theta_{d_{I-1},1;m}
&=\dot{\mathfrak f}\,\tau_{d_{I-1},1;m}
-\frac{\eta_{I,m}^{\rm rel}}{nL\cosh(\tau/L)u^{\tau(0)}},
\qquad I=2,\cdots,n-1,
\nonumber\\
\theta_{d_{n-1},1;m}
&=\dot{\mathfrak f}\,\tau_{d_{n-1},1;m}
-\frac{\eta_{1,m}^{\rm rel}}{nL\cosh(\tau/L)u^{\tau(0)}}.
\end{align}
For $n=3$, the same formula assigns $\eta_2^{\rm rel}$ to $d_1$ and $\eta_1^{\rm rel}$ to $d_2$, as required by \eqref{Eq:xsxd}. The different sign from \eqref{Eq:gate-two-full-transverse} is fixed by the folded coordinate definitions. The embeddings themselves are then \eqref{Eq:gate-n-inverse-times} and \eqref{Eq:gate-n-inverse-polar}, together with their azimuthal counterparts. The induced metric is \eqref{Eq:gate-three-common-metric}, and $u_1^\tau$ is \eqref{Eq:gate-two-full-temporal-velocity}, since the average transverse displacement is now carried by $\theta_{s,1}$.

In the homogeneous sector, all relative normal amplitudes vanish. The clocks, angular shifts and relative polar coefficients are \eqref{Eq:gate-three-worldsheet-parameters}, with $i=1,\cdots,n-1$. The hybrid sound mode is \eqref{Eq:gate-three-radial-solution}; it gives
\begin{equation}\label{Eq:gate-n-full-homogeneous}
\theta_{s,1;0}=\frac{\eta_0}{L\cosh(\tau/L)u^{\tau(0)}},
\qquad
T_{1;0}=-\frac{T^{(0)}\sqrt w}{L}\eta_0,
\qquad u^\phi_{1;0}=0.
\end{equation}
For every $n$, the branch with $\eta_{0,-}=0$ and $\eta_{0,+}>0$ is the linear limit of the exact deformed cosmic creation gate constructed in Sec.~\ref{Sec:histories-exact-radial}.

The $|m|=1$ solutions are the isometries identified above, with the fluid variables given as in Sec.~\ref{Sec:gate-two-reconstruction}. The reality conditions are \eqref{Eq:gate-two-mode-reality}, applied separately to each relative displacement. All embedding and fluid variables are therefore fixed by the normal-mode amplitudes and rigid parameters, in agreement with the number of degrees of freedom stated in Sec.~\ref{Sec:prev}. An infinite Fourier expansion requires convergence of the fields and of the derivatives entering the junction conditions on the time interval considered.

\section{Linear perturbations of the $n$-way cosmic creation gate in four and higher dimensions}
\label{Sec:higher-dimensional-perturbations}

We turn to the four-dimensional junction. We keep each bulk region locally de Sitter, thereby excluding propagating bulk gravitational perturbations, and require the junction perturbations to be smooth on the entire spatial sphere. Within this class, the linearized junction conditions leave only rigid isometries and a homogeneous average transverse displacement. There are no local junction field degrees of freedom. We first establish this result for the two-way junction, then consider $n\geq3$ and finally extend the argument to higher dimensions.

\subsection{The two-way junction in four dimensions}
\label{Sec:gate-four-two-expansion}

We use the northern--southern description of Sec.~\ref{Sec:two_jn}, with the global bulk metric
\begin{equation}\label{Eq:gate-four-global-metric}
{\rm d}s^2=-{\rm d}t^2+L^2\cosh^2(t/L)
\left({\rm d}\theta^2+\sin^2\theta\,
\Omega_{AB}{\rm d}\varphi^A{\rm d}\varphi^B\right),
\qquad A,B=1,2.
\end{equation}
Here $\Omega_{AB}$ is the metric on the unit $S^2$. We denote its covariant derivative by $\mathcal D_A$ and its Laplacian by $\mathcal D^2=\Omega^{AB}\mathcal D_A\mathcal D_B$. Angular indices on the first-order shifts will be raised and lowered with $\Omega_{AB}$. The background is the creation gate of Sec.~\ref{Sec:creation-four-dimensional}:
\begin{align}\label{Eq:gate-four-two-background}
&\mathfrak f(\tau)=\arcsin\!\left[\frac{R_0}{L\cosh(\tau/L)}\right],
\qquad
R_0^2=L^2\frac{1+2w}{2(1+w)},\qquad w>-\frac12,
\nonumber\\
&u^{\tau(0)}=\sqrt{1+(1+2w)\tanh^2(\tau/L)},\qquad
\rho^{(0)}=\frac{1}{2\pi G_NL\sqrt{1+2w}},
\qquad P^{(0)}=w\rho^{(0)}.
\end{align}
For $w\neq0$, the constant $T^{(0)}$ is fixed by the equation of state as in \eqref{Eq:gate-fluid-three}. The case $w=0$ is described by the dust density in place of the temperature. All the geometric equations below are regular at $w=0$.

The first-order images of the junction are
\begin{align}\label{Eq:gate-four-two-images}
&t_{1,2}=\tau\pm\epsilon\tau_{d,1},\qquad
\varphi_{1,2}^A=\sigma^A\pm\epsilon\sigma_{d,1}^A,
\nonumber\\
&\theta_1=\mathfrak f+\epsilon(\theta_{s,1}+\theta_{d,1}),\qquad
\theta_2=\pi-\mathfrak f+\epsilon(\theta_{s,1}-\theta_{d,1}),
\end{align}
up to $O(\epsilon^2)$. Thus $\theta_s=\pi/2+\epsilon\theta_{s,1}$ and
$\theta_d=\mathfrak f-\pi/2+\epsilon\theta_{d,1}$. We also expand
\[
T=T^{(0)}+\epsilon T_1,\qquad
u^A=\epsilon u_1^A,\qquad
u^\tau=u^{\tau(0)}+\epsilon u_1^\tau.
\]
There are five independent embedding coefficients and three fluid coefficients, giving the eight variables counted in Sec.~\ref{Sec:two_jn}. The normalization of $u^\mu$ fixes $u_1^\tau$. A dot denotes $\partial_\tau$, while derivatives with respect to the proper clock will be written explicitly. In particular,
\[
\partial_{\tau_{\rm p}}=u^{\tau(0)}\partial_\tau,
\]
with $\tau_{\rm p}(\tau)$ given by \eqref{Eq:gate-fluid-clock} at $D=4$. All bulk coordinates remain global coordinates.

Reflecting the second image into the folded picture, the normal displacements are
\begin{align}\label{Eq:gate-four-two-normals}
\eta&=L\cosh(\tau/L)u^{\tau(0)}\theta_{d,1},
\nonumber\\
\eta_1^{\rm rel}
&=2L\cosh(\tau/L)u^{\tau(0)}
\left(\theta_{s,1}-\dot{\mathfrak f}\,\tau_{d,1}\right).
\end{align}
These are the normal variables of Sec.~\ref{Sec:prev}, with the same orientation convention as in Section~\ref{Sec:two-way-perturbations-three}.

\subsection{Metric continuity and the relative shifts}
\label{Sec:gate-four-two-continuity}

The six components of the first junction condition give
\begin{align}\label{Eq:gate-four-two-continuity}
0={}&\dot\tau_{d,1}
-L^2\cosh^2(\tau/L)\dot{\mathfrak f}\,\dot\theta_{s,1}
-L\cosh(\tau/L)\sinh(\tau/L)\dot{\mathfrak f}^{\,2}\tau_{d,1},
\nonumber\\
0={}&\mathcal D_A\tau_{d,1}
-L^2\cosh^2(\tau/L)\dot{\mathfrak f}\,\mathcal D_A\theta_{s,1}
-R_0^2\Omega_{AB}\dot\sigma_{d,1}^B,
\nonumber\\
0={}&\mathcal D_A\sigma_{d,1\,B}
+\mathcal D_B\sigma_{d,1\,A}
+2\Omega_{AB}
\left(\cot\mathfrak f\,\theta_{s,1}
+\frac{\tanh(\tau/L)}{L}\tau_{d,1}\right),
\end{align}
where $\sigma_{d,1\,A}=\Omega_{AB}\sigma_{d,1}^B$. The average transverse variable $\theta_{d,1}$ cancels. Taking the trace and the traceless part of the last equation, respectively, gives
\begin{align}\label{Eq:gate-four-relative-angular}
&\mathcal D_A\sigma_{d,1}^A
=-\frac{\eta_1^{\rm rel}}{L\sqrt{1+2w}},
\nonumber\\
&\mathcal D_A\sigma_{d,1\,B}
+\mathcal D_B\sigma_{d,1\,A}
-\Omega_{AB}\mathcal D_C\sigma_{d,1}^C=0.
\end{align}
Thus, at every time, the relative angular shift is a smooth conformal Killing field on the round sphere. Taking covariant derivatives of its defining equation and using the curvature of the unit sphere gives $(\mathcal D^2+2)\mathcal D_A\sigma_{d,1}^A=0$. Its divergence therefore contains only $\ell=1$ harmonics. Subtracting the corresponding gradient leaves a divergence-free conformal Killing field, which is a rotation. Hence the full smooth space consists of the three rotations and the gradients of the three $\ell=1$ harmonics. The trace condition implies
\begin{equation}\label{Eq:gate-four-relative-angular-spectrum}
(\mathcal D^2+2)\eta_1^{\rm rel}=0.
\end{equation}
In particular, there is no homogeneous relative normal displacement and no relative harmonic with $\ell\geq2$. This restriction follows from metric continuity itself.

We first set $\eta_1^{\rm rel}=\eta=0$ to identify the rigid worldsheet transformations. Choose a real basis $Y_{1a}$ of the $\ell=1$ harmonics, with $a=1,2,3$, and define the three rotation fields
\[
\mathcal R_a^A=\varepsilon^{AB}\mathcal D_BY_{1a},
\]
where $\varepsilon^{AB}$ is the contravariant area form of the unit sphere. The continuity equations then give
\begin{equation}\label{Eq:gate-four-worldsheet-parameters}
\tau_{d,1}=\alpha_1u^{\tau(0)},\qquad
\sigma_{d,1}^A=\sum_{a=1}^3\alpha_{a+1}\mathcal R_a^A,\qquad
\theta_{s,1}=\alpha_1u^{\tau(0)}\dot{\mathfrak f}.
\end{equation}
The $\alpha_i$ are constants. Reading off the generators from the clock and angular shifts gives
\begin{equation}\label{Eq:gate-four-worldsheet-generators}
l_1=u^{\tau(0)}\partial_\tau,\qquad
l_{a+1}=\mathcal R_a^A\partial_{\sigma^A},\qquad a=1,2,3.
\end{equation}
They are precisely the time translation and rotations of the induced metric
$-{\rm d}\tau_{\rm p}^2+R_0^2{\rm d}\Omega_2^2$. This identifies the worldsheet isometries directly in the embedding variables.

For nonzero $\eta_1^{\rm rel}$, all six equations in \eqref{Eq:gate-four-two-continuity} are solved by
\begin{align}\label{Eq:gate-four-relative-reconstruction}
\sigma_{d,1}^A
&=\frac{\mathcal D^A\eta_1^{\rm rel}}{2L\sqrt{1+2w}}
+\sum_{a=1}^3\alpha_{a+1}\mathcal R_a^A,
\nonumber\\
\tau_{d,1}
&=\frac{u^{\tau(0)}}2
\left[
L\cosh(\tau/L)\dot{\mathfrak f}\,\eta_1^{\rm rel}
+\frac{R_0^2}{L\sqrt{1+2w}}\,
\partial_{\tau_{\rm p}}\eta_1^{\rm rel}
\right]+\alpha_1u^{\tau(0)},
\nonumber\\
\theta_{s,1}
&=\dot{\mathfrak f}\,\tau_{d,1}
+\frac{\eta_1^{\rm rel}}{2L\cosh(\tau/L)u^{\tau(0)}},
\end{align}
provided
\begin{equation}\label{Eq:gate-four-relative-time-equation}
\partial_{\tau_{\rm p}}^2\eta_1^{\rm rel}
-\frac{2(1+w)}{L^2}\eta_1^{\rm rel}=0.
\end{equation}
The spatial equation fixes the gradient part of the angular shift. The curl of the mixed equation makes its rotational part time independent. The mixed equation then fixes the nonconstant part of the clock shift, and the temporal equation gives \eqref{Eq:gate-four-relative-time-equation}. Its homogeneous clock solution is the first term in \eqref{Eq:gate-four-worldsheet-parameters}. The relative normal displacement is therefore fixed by six integration constants:
\begin{equation}\label{Eq:gate-four-relative-solution}
\eta_1^{\rm rel}
=\sum_{a=1}^3
\left[
\eta_{1,a,+}^{\rm rel}
e^{\sqrt{2(1+w)}\,\tau_{\rm p}/L}
+\eta_{1,a,-}^{\rm rel}
e^{-\sqrt{2(1+w)}\,\tau_{\rm p}/L}
\right]Y_{1a}.
\end{equation}
We identify these constants with ambient isometries below. These expressions are regular at every finite global time for $w>-1/2$, including $\tau=0$, where $\dot{\mathfrak f}$ vanishes.

\subsection{The second junction condition and the average transverse displacement}
\label{Sec:gate-four-two-extrinsic}

After continuity is imposed, the variation of the induced metric is
\begin{align}\label{Eq:gate-four-common-metric}
\gamma_{\tau\tau}^{(1)}
&=2L^2\cosh^2(\tau/L)\dot{\mathfrak f}\,\dot\theta_{d,1},
\nonumber\\
\gamma_{\tau A}^{(1)}
&=L^2\cosh^2(\tau/L)\dot{\mathfrak f}\,\mathcal D_A\theta_{d,1},
\nonumber\\
\gamma_{AB}^{(1)}
&=2R_0^2\cot\mathfrak f\,\theta_{d,1}\Omega_{AB}.
\end{align}
The relative variations cancel from the sum in the second junction condition. Its temporal component, spatial trace and mixed components give, respectively,
\begin{align}\label{Eq:gate-four-extrinsic-linear}
\rho_1&=-\frac{1}{4\pi G_NR_0^2}
(\mathcal D^2+2)\eta,
\nonumber\\
P_1&=\frac{1}{4\pi G_N}
\left[
-\partial_{\tau_{\rm p}}^2\eta
+\frac{\mathcal D^2\eta}{2R_0^2}
+\left(\frac{2(1+w)}{L^2}+\frac1{R_0^2}\right)\eta
\right],
\nonumber\\
u_1^A&=\frac{1}{L\sqrt{1+2w}}\,
\mathcal D^A\partial_{\tau_{\rm p}}\eta.
\end{align}
The mixed junction equations determine both velocity components, leaving no independent divergence-free fluid velocity.

The two traceless spatial equations impose an additional restriction:
\begin{equation}\label{Eq:gate-four-collective-hessian}
\left(\mathcal D_A\mathcal D_B
-\frac12\Omega_{AB}\mathcal D^2\right)\eta=0.
\end{equation}
The traceless spatial part of the second junction condition excludes the generic inhomogeneous sector because a perfect fluid has no spatial anisotropic stress at first order about the resting background. To see its kernel explicitly, expand in scalar harmonics $Y_{\ell m}$ on the unit sphere. Integration by parts gives
\begin{align}\label{Eq:gate-four-hessian-norm}
&\int_{S^2}
\left|\mathcal D_A\mathcal D_BY_{\ell m}
-\frac12\Omega_{AB}\mathcal D^2Y_{\ell m}\right|^2
{\rm d}\Omega_2
\nonumber\\
&\hspace{15mm}
=\frac12(\ell-1)\ell(\ell+1)(\ell+2)
\int_{S^2}|Y_{\ell m}|^2\,{\rm d}\Omega_2.
\end{align}
Every $\ell\geq2$ coefficient therefore vanishes. The average normal displacement contains only $\ell=0$ and $\ell=1$ harmonics.

The equation of state $P_1=w\rho_1$, together with the background radius, gives
\begin{equation}\label{Eq:gate-four-collective-equation}
\partial_{\tau_{\rm p}}^2\eta
-\frac{1+w}{L^2}\mathcal D^2\eta
-\frac{4(1+w)}{L^2}\eta=0.
\end{equation}
The full junction system requires both this evolution equation and the angular condition \eqref{Eq:gate-four-collective-hessian}. For $w\neq0$, the temperature is determined by
\begin{equation}\label{Eq:gate-four-temperature}
\frac{T_1}{T^{(0)}}
=-\frac{w}{L\sqrt{1+2w}}(\mathcal D^2+2)\eta.
\end{equation}
The $\ell=1$ modes of the average displacement have $\rho_1=P_1=T_1=0$. Their proper-time dependence obeys the same equation as \eqref{Eq:gate-four-relative-time-equation}.

\subsection{Identification of the ambient isometries}
\label{Sec:gate-four-ambient-isometries}

We identify the remaining $\ell=1$ solutions by comparing their embedding shifts with the Killing vectors of the global metric, as in Sections~\ref{Sec:two-way-perturbations-three} and \ref{Sec:n-way-perturbations-three}. In angular coordinates $(\varphi^1,\varphi^2)$, choose
\[
Y_{11}=\sin\varphi^1\cos\varphi^2,\qquad
Y_{12}=\sin\varphi^1\sin\varphi^2,\qquad
Y_{13}=\cos\varphi^1.
\]
The six ambient generators which displace the rotationally symmetric hypersurface are
\begin{align}\label{Eq:gate-four-ambient-generators}
L_a^{\rm rot}
&=Y_{1a}\partial_\theta
+\cot\theta\,\mathcal D^AY_{1a}\partial_{\varphi^A},
\nonumber\\
L_a^{\rm boost}
&=L\sin\theta\,Y_{1a}\partial_t
+\tanh(t/L)\cos\theta\,Y_{1a}\partial_\theta
+\tanh(t/L)\csc\theta\,
\mathcal D^AY_{1a}\partial_{\varphi^A},
\qquad a=1,2,3.
\end{align}
The harmonics and their derivatives are evaluated at the bulk angles $\varphi^A$, which equal $\sigma^A$ on the junction images. The transverse displacements at fixed global time are
\begin{equation}\label{Eq:gate-four-ambient-graph}
\theta_{d,1}^{(L_a^{\rm rot})}=Y_{1a},
\qquad
\theta_{d,1}^{(L_a^{\rm boost})}
=\frac{\tanh(\tau/L)}{\cos\mathfrak f}\,Y_{1a}.
\end{equation}
These follow directly from $L^\theta-\dot{\mathfrak f}L^t$. Thus the complete average $\ell=1$ displacement is
\begin{equation}\label{Eq:gate-four-collective-isometries}
\theta_{d,1}^{\rm iso}
=\sum_{a=1}^3
\left(\gamma_a+
\gamma_{a+3}\frac{\tanh(\tau/L)}{\cos\mathfrak f}\right)Y_{1a},
\end{equation}
with constant $\gamma_1,\cdots,\gamma_6$. Using the known proper clock,
\[
L\cosh(\tau/L)u^{\tau(0)}
=L\cosh\!\left(\frac{\sqrt{2(1+w)}}L\tau_{\rm p}\right),
\]
and multiplying this identity by $\tanh(\tau/L)/\cos\mathfrak f$ gives
$L\sinh[\sqrt{2(1+w)}\,\tau_{\rm p}/L]$. These normal amplitudes span the solutions for the average $\ell=1$ displacement of \eqref{Eq:gate-four-collective-equation}.

Independent actions on the two folded images identify the relative sector. Acting with opposite amplitudes gives the following two sets of relative shifts for each $a$, with $\theta_{d,1}=0$:
\begin{align}\label{Eq:gate-four-relative-isometries}
(\tau_{d,1},\sigma_{d,1}^A,\theta_{s,1})
&=\left(0,\cot\mathfrak f\,\mathcal D^AY_{1a},Y_{1a}\right),
\nonumber\\
(\tau_{d,1},\sigma_{d,1}^A,\theta_{s,1})
&=\left(L\sin\mathfrak f\,Y_{1a},
\,\tanh(\tau/L)\csc\mathfrak f\,\mathcal D^AY_{1a},
\,\tanh(\tau/L)\cos\mathfrak f\,Y_{1a}\right).
\end{align}
They are the restrictions of $L_a^{\rm rot}$ and $L_a^{\rm boost}$. Their relative normal amplitudes are proportional to the corresponding hyperbolic cosine and sine above, and they span \eqref{Eq:gate-four-relative-solution}. Thus every relative $\ell=1$ solution is generated by a de Sitter isometry acting on the junction images. Together with \eqref{Eq:gate-four-worldsheet-parameters}, it accounts for every solution of the first junction condition.

The remaining ambient generators are the three rotations tangent to the sphere and the polar boost \eqref{Eq:gate-two-polar-boost}. They restrict to the worldsheet generators \eqref{Eq:gate-four-worldsheet-generators}; the polar boost is $L\,l_1/\sqrt{2(1+w)}$ on the background. Thus there are four worldsheet generators and six transverse ambient generators. This classification follows from the embedding shifts themselves. In particular, the growing proper normal amplitudes in the $\ell=1$ sector describe rigid isometries.

\subsection{The radial zero mode and completion of the two-way solution}
\label{Sec:gate-four-radial-completion}

After fixing the rigid isometries, the only remaining normal displacement is homogeneous on $S^2$. Equation \eqref{Eq:gate-four-collective-equation} gives
\begin{equation}\label{Eq:gate-four-radial-solution}
\eta_0(\tau)
=\eta_{0,+}e^{2\sqrt{1+w}\,\tau_{\rm p}(\tau)/L}
+\eta_{0,-}e^{-2\sqrt{1+w}\,\tau_{\rm p}(\tau)/L}.
\end{equation}
The two real constants $\eta_{0,\pm}$ fix the radial motion. In the original variables the solution is
\begin{align}\label{Eq:gate-four-radial-reconstruction}
&\theta_{d,1}=\frac{\eta_0}{L\cosh(\tau/L)u^{\tau(0)}},
\qquad
\theta_{s,1}=\tau_{d,1}=\sigma_{d,1}^A=0,
\qquad u_1^A=0,
\nonumber\\
&\rho_1=-\frac{\eta_0}{2\pi G_NR_0^2},\qquad
P_1=w\rho_1,\qquad
\frac{T_1}{T^{(0)}}
=-\frac{2w}{L\sqrt{1+2w}}\eta_0
\quad(w\neq0).
\end{align}
The metric is \eqref{Eq:gate-four-common-metric}, and velocity normalization fixes
\begin{equation}\label{Eq:gate-four-temporal-velocity}
u_1^\tau=L^2\cosh^2(\tau/L)\dot{\mathfrak f}\,
\big(u^{\tau(0)}\big)^3
\partial_\tau
\left[\frac{\eta}{L\cosh(\tau/L)u^{\tau(0)}}\right].
\end{equation}
This expression also applies when rigid modes of the average displacement are retained. The common areal radius changes by
\begin{equation}\label{Eq:gate-four-radial-radius}
\delta R=\frac{\eta_0}{\sqrt{2(1+w)}},\qquad
\partial_{\tau_{\rm p}}^2\delta R
-\frac{4(1+w)}{L^2}\delta R=0.
\end{equation}
The zero mode describes homogeneous radial motion, with one branch growing towards each temporal end, and supplies no freely specifiable field on the sphere.

For general allowed perturbations of the average displacement, the fluid energy and momentum equations are
\begin{align}\label{Eq:gate-four-euler-check}
&\partial_{\tau_{\rm p}}\rho_1
+(\rho^{(0)}+P^{(0)})
\left(\mathcal D_Au_1^A
+\frac{2}{L\sqrt{1+2w}}\partial_{\tau_{\rm p}}\eta\right)=0,
\nonumber\\
&R_0^2\Omega_{AB}\partial_{\tau_{\rm p}}u_1^B
-\frac{\sqrt{1+2w}}L\mathcal D_A\eta
+\mathcal D_A\left(\frac{T_1}{T^{(0)}}\right)=0.
\end{align}
The energy equation is an identity after \eqref{Eq:gate-four-extrinsic-linear} is substituted. Taking the divergence of \eqref{Eq:gate-four-collective-hessian} gives
$\mathcal D_A(\mathcal D^2+2)\eta=0$. The temperature is therefore spatially homogeneous. The remaining momentum equation follows from \eqref{Eq:gate-four-collective-equation} on the allowed $\ell=0,1$ harmonics. For dust, its last term is absent, and the remaining terms give the geodesic fluid equation. Thus the fluid equations add no restriction beyond the junction conditions.

There are six metric-continuity equations and six extrinsic-curvature junction equations. We have solved all their components, including both traceless spatial equations. The bulk Einstein equations and gravitational constraints hold in each de Sitter region. The Gauss and Codazzi identities hold on each image because its induced metric and extrinsic curvature are computed from the same global embedding. The full linear solution is therefore the radial mode \eqref{Eq:gate-four-radial-solution}, together with the rigid worldsheet and ambient transformations already identified. There are therefore no local junction field degrees of freedom in the two-way four-dimensional construction.

\subsection{The $n$-way junction in four dimensions}
\label{Sec:gate-four-n-way}

We now retain $n\geq3$ northern fragments and use the junction coordinates and relative embedding variables defined in Sec.~\ref{Sec:n_jn}, replacing each azimuthal shift by the angular vector $\sigma_{d_i}^A$. The background remains \eqref{Eq:gate-four-two-background}, except that
\begin{equation}\label{Eq:gate-four-n-background}
\rho^{(0)}=\frac{n}{4\pi G_NL\sqrt{1+2w}},
\qquad P^{(0)}=w\rho^{(0)}.
\end{equation}
There are $4n$ independent variables and $6n$ junction equations. The global embeddings are the inverses given in \eqref{Eq:gate-n-inverse-times} and \eqref{Eq:gate-n-inverse-polar}, with the same inverse applied to each angular component. In the folded picture, the average transverse displacement is $\theta_{s,1}$, so
\[
\eta=L\cosh(\tau/L)u^{\tau(0)}\theta_{s,1}.
\]
The relative normals are related to the cyclic coordinate differences by \eqref{Eq:gate-n-relative-normal}, with the four-dimensional $u^{\tau(0)}$.

Each comparison with image $n$ gives six equations of the form \eqref{Eq:gate-four-two-continuity} for one relative clock, angular shift and polar displacement. Consequently every relative angular shift is a smooth conformal Killing field. The trace condition and temporal compatibility require
\begin{equation}\label{Eq:gate-four-n-relative}
(\mathcal D^2+2)\eta_I^{\rm rel}=0,\qquad
\partial_{\tau_{\rm p}}^2\eta_I^{\rm rel}
-\frac{2(1+w)}{L^2}\eta_I^{\rm rel}=0,
\qquad I=1,\cdots,n-1.
\end{equation}
Comparison with \eqref{Eq:gate-four-ambient-generators} shows that all these relative solutions are generated by de Sitter isometries acting on the junction images. The angular Killing parts and homogeneous clock shifts give the relative worldsheet isometries.

The reconstruction can be stated explicitly. For $I=2,\cdots,n-1$,
\begin{align}\label{Eq:gate-four-n-relative-reconstruction}
\sigma_{d_{I-1},1}^A
&=-\frac{\mathcal D^A\eta_I^{\rm rel}}{nL\sqrt{1+2w}}
+\sum_{a=1}^3\alpha_{I-1,a+1}\mathcal R_a^A,
\nonumber\\
\tau_{d_{I-1},1}
&=-\frac{u^{\tau(0)}}n
\left[
L\cosh(\tau/L)\dot{\mathfrak f}\,\eta_I^{\rm rel}
+\frac{R_0^2}{L\sqrt{1+2w}}\,
\partial_{\tau_{\rm p}}\eta_I^{\rm rel}
\right]+\alpha_{I-1,1}u^{\tau(0)},
\nonumber\\
\theta_{d_{I-1},1}
&=\dot{\mathfrak f}\,\tau_{d_{I-1},1}
-\frac{\eta_I^{\rm rel}}{nL\cosh(\tau/L)u^{\tau(0)}}.
\end{align}
The $\alpha_{i,b}$ are constant relative isometry parameters. For $I=1$, the same formulas use the cyclic coordinate label $d_{n-1}$ and constants $\alpha_{n-1,b}$. The sign and factor of $n$ follow from the coordinate definitions \eqref{Eq:xsxd}. These expressions solve all $6(n-1)$ first junction equations in terms of constant isometry parameters.

In the second junction condition, the relative variations again have zero sum. The temporal and spatial trace equations are \eqref{Eq:gate-four-extrinsic-linear} with their factor $1/(4\pi G_N)$ replaced by $n/(8\pi G_N)$; the velocity formula is unchanged. The two traceless spatial equations remain \eqref{Eq:gate-four-collective-hessian}, and the equation of state gives \eqref{Eq:gate-four-collective-equation}. The six $\ell=1$ constants in $\theta_{s,1}$ multiply the de Sitter Killing displacements applied with the same amplitude to all $n$ images. After fixing the rigid parameters, the remaining average transverse displacement is homogeneous and given by \eqref{Eq:gate-four-radial-solution}. The embedding and fluid variables are
\begin{align}\label{Eq:gate-four-n-radial}
&\theta_{s,1}=\frac{\eta_0}{L\cosh(\tau/L)u^{\tau(0)}},
\qquad
\tau_{d_i,1}=\sigma_{d_i,1}^A=\theta_{d_i,1}=0,
\qquad u_1^A=0,
\nonumber\\
&\rho_1=-\frac{n}{4\pi G_NR_0^2}\eta_0,\qquad
P_1=w\rho_1,\qquad
\frac{T_1}{T^{(0)}}
=-\frac{2w}{L\sqrt{1+2w}}\eta_0
\quad(w\neq0).
\end{align}
The induced metric is \eqref{Eq:gate-four-common-metric} with $\theta_{d,1}$ replaced by $\theta_{s,1}$, and normalization gives \eqref{Eq:gate-four-temporal-velocity}. The fluid equations are \eqref{Eq:gate-four-euler-check} with the $n$-way background density and pressure. Direct substitution verifies the fluid equations, and all bulk and image constraints also hold. Increasing $n$ therefore adds independent parameters specifying ambient isometries of the images but introduces no local physical junction mode in four dimensions.

\subsection{Generalization to higher dimensions}
\label{Sec:gate-higher-generalization}

The argument extends to every $D\geq4$. The background radius and proper clock are \eqref{Eq:gate-radius-general} and \eqref{Eq:gate-fluid-clock}, with $w>-(D-3)/(D-2)$. On the round $S^{D-2}$, the traceless spatial part of the second junction condition is
\[
\left(\mathcal D_A\mathcal D_B
-\frac{1}{D-2}\Omega_{AB}\mathcal D^2\right)\eta=0.
\]

For a scalar harmonic $Y_\ell$ satisfying $\mathcal D^2Y_\ell=-\ell(\ell+D-3)Y_\ell$, integration by parts as in \eqref{Eq:gate-four-hessian-norm} gives
\begin{equation}
 \begin{aligned}
 &\int_{S^{D-2}}\left|\mathcal D_A\mathcal D_B Y_\ell-\frac{\Omega_{AB}}{D-2}\mathcal D^2Y_\ell\right|^2d\Omega_{D-2}\\
 &\quad=\frac{D-3}{D-2}\ell(\ell+D-3)(\ell-1)(\ell+D-2)
 \int_{S^{D-2}}|Y_\ell|^2d\Omega_{D-2}.
 \end{aligned}
\end{equation}

Only the $\ell=0,1$ harmonics satisfy the traceless spatial junction condition. The traceless spatial part of metric continuity requires every relative angular shift to be a conformal Killing field. Its divergence contains only $\ell=1$, and the trace fixes the corresponding relative normal displacement. The remaining equations give
\begin{align}\label{Eq:gate-higher-normal-equations}
&\partial_{\tau_{\rm p}}^2\eta
-\frac{1+w}{L^2}\mathcal D^2\eta
-\frac{2(D-2)(1+w)}{L^2}\eta=0,
\nonumber\\
&(\mathcal D^2+D-2)\eta_I^{\rm rel}=0,\qquad
\partial_{\tau_{\rm p}}^2\eta_I^{\rm rel}
-\frac{(D-2)(1+w)}{L^2}\eta_I^{\rm rel}=0.
\end{align}
The relative equation applies to every independent pair. There is no relative homogeneous normal mode.

To identify the $\ell=1$ solutions, we again compare the transverse displacements with the ambient Killing vectors. There are $D-1$ real $\ell=1$ harmonics. Replacing $Y_{1a}$ in \eqref{Eq:gate-four-ambient-generators} by these harmonics gives $D-1$ transverse rotations and $D-1$ transverse boosts, with displacements $Y_{1a}$ and $\tanh(\tau/L)Y_{1a}/\cos\mathfrak f$ at fixed global time. Their normal amplitudes span the solutions for the average and relative $\ell=1$ displacements of \eqref{Eq:gate-higher-normal-equations}. The homogeneous relative clock shifts and the sphere rotations give the worldsheet isometries. Thus the entire inhomogeneous sector consists of rigid transformations.

The remaining metric and fluid variables are determined. In particular, the mixed junction equations give
\begin{equation}\label{Eq:gate-higher-fluid-velocity}
u_1^A=
\frac{\mathcal D^A\partial_{\tau_{\rm p}}\eta}
{L\sqrt{(D-2)(1+w)-1}},
\end{equation}
which leaves no independent vector mode. After fixing the rigid isometries, only the homogeneous average transverse displacement remains. In terms of the common areal radius,
\begin{align}\label{Eq:gate-higher-radial-completion}
&\delta R=\frac{\eta_0}{\sqrt{(D-2)(1+w)}},
\qquad
\partial_{\tau_{\rm p}}^2\delta R
-\frac{2(D-2)(1+w)}{L^2}\delta R=0,
\nonumber\\
&\delta R=C_+e^{\Gamma_D\tau_{\rm p}}
+C_-e^{-\Gamma_D\tau_{\rm p}},
\qquad
\Gamma_D=\frac{\sqrt{2(D-2)(1+w)}}L,
\nonumber\\
&\frac{\rho_1}{\rho^{(0)}}=-(D-2)(1+w)\frac{\delta R}{R_0},
\qquad P_1=w\rho_1,\qquad
\frac{T_1}{T^{(0)}}=-(D-2)w\frac{\delta R}{R_0}\quad(w\neq0),
\qquad u_1^A=0.
\end{align}
These formulas also follow directly from the homogeneous fluid conservation equation and the radial junction equation. The density and velocity expressions also apply to dust. The two constants $C_\pm$ describe global radial motion; there is no local field degree of freedom on $S^{D-2}$. This proves the higher-dimensional statement in Sec.~\ref{Sec:prev} for both two-way and general $n$-way junctions within the locally de Sitter class.

For $C_-=0$ and $C_+>0$, the exact continuation is the deformed cosmic creation gate of Sec.~\ref{Sec:histories-exact-radial}, with finite past volume and unbounded future volume.

\section{The general structure of creation: conditioned admissible histories and the Cauchy conundrum}
\label{Sec:admissible-histories}

The spatial volume of the rotationally invariant cosmic creation gate approaches finite, positive limits in both the infinite past and future. Our study of its perturbations raises a broader question: which cosmological histories admit a controlled, finite-volume past while remaining free to expand in the future? The modes that grow towards the past require fine-tuned initial conditions. We use this requirement to characterize admissible histories and disturbances, and to define a causal response without imposing a bound on the future spatial volume.

We continue to work within the restricted class in which all bulk regions remain locally de Sitter, thereby excluding propagating bulk gravitational modes in every dimension. The junction embeddings, the identifications of their points and the perfect-fluid fields must satisfy all junction conditions and gravitational constraints. We hold the rigid isometry parameters fixed when comparing physical perturbations. We examine creation histories within this restricted class.

\subsection{The condition on the past}
\label{Sec:histories-past-condition}

 We first state the conditions under which regular deformations of the rotationally invariant cosmic creation gates admit the interpretation of creation. We measure the spatial volume on the natural cosmological global-time foliation synchronized by the junction fluid clock, as discussed in  Sec.~\ref{Sec:creation-gates}, and first require
\begin{equation}\label{Eq:histories-past-volume}
0<V_-:=\lim_{\tau_{\rm p}\to-\infty}V(\tau_{\rm p})<\infty.
\end{equation}
The induced geometry and fluid variables must remain regular, and the embeddings must remain single-valued. The limiting volume $V_-$ is determined by the history. We impose no corresponding bound on the future volume, which may grow indefinitely.

Under a second criterion, the spatial volume can continue to oscillate towards the infinite past, while remaining bounded and bounded away from zero:
\begin{equation}\label{Eq:histories-bounded-past-volume}
0<\liminf_{\tau_{\rm p}\to-\infty}V(\tau_{\rm p})
\leq\limsup_{\tau_{\rm p}\to-\infty}V(\tau_{\rm p})<\infty.
\end{equation}
The same requirements of regularity and single-valuedness apply to this second criterion. A history can then remain well defined arbitrarily far into the past even when its volume has no single past limit. The first criterion implies the second, and the exact cosmic creation gates and their homogeneous deformations satisfy both.

For the perturbative discussion, we assume that the solutions admit expansions analytic in the mode amplitudes. The expansions must converge on the global time band considered, and their convergence towards the infinite past must justify evaluating the volume limit in \eqref{Eq:histories-past-volume} order by order when the first criterion is imposed. For the second criterion, convergence must give uniform control of the past volume and its finite, strictly positive bounds. We also require convergence of any infinite angular harmonic sum and of the derivatives entering the junction conditions. A future-growing perturbation may eventually require continuation beyond the expansion about the rotationally invariant cosmic creation gate. In the following subsection, we study an exact homogeneous deformation in which the future volume grows indefinitely. For the physical inhomogeneous modes, convergence and preservation of single-valuedness to all orders remain assumptions to be established.

\subsection{Exact $n$-way deformed cosmic creation gates in $D\geq3$ from homogeneous backreaction}
\label{Sec:histories-exact-radial}

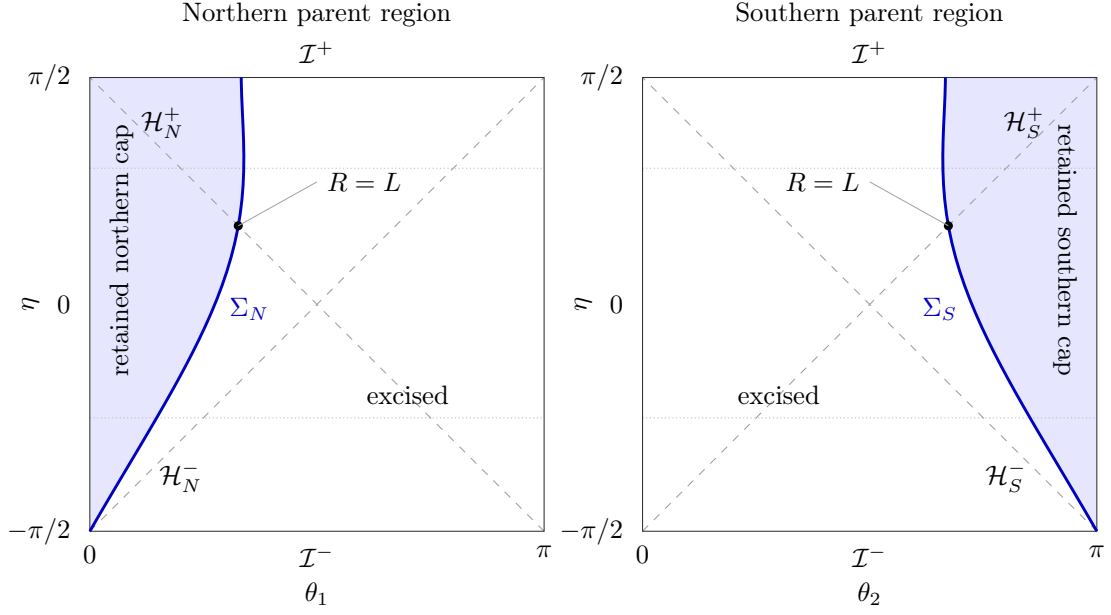
\begin{figure}[t]
\centering
\begin{tikzpicture}[x=6cm,y=6cm,font=\small]
\def\deformedGatePoints{
(0.00000,0.00000) (0.01610,0.02780)
(0.03231,0.05556) (0.04861,0.08324)
(0.06499,0.11093) (0.08149,0.13870)
(0.09795,0.16642) (0.11447,0.19430)
(0.13095,0.22228) (0.14743,0.25054)
(0.16358,0.27861) (0.17956,0.30688)
(0.19523,0.33527) (0.21051,0.36373)
(0.22535,0.39234) (0.23993,0.42165)
(0.25400,0.45141) (0.26731,0.48137)
(0.27965,0.51127) (0.29118,0.54177)
(0.30166,0.57262) (0.31102,0.60398)
(0.31898,0.63517) (0.32563,0.66672)
(0.33093,0.69865) (0.33485,0.73087)
(0.33741,0.76287) (0.33876,0.79524)
(0.33905,0.82755) (0.33848,0.85976)
(0.33731,0.89201) (0.33583,0.92406)
(0.33439,0.95611) (0.33344,0.98811)
(0.33333,1.00000)
}
\foreach \side/\offset in {0/0,1/7.3}{
\begin{scope}[xshift=\offset cm]
  \ifnum\side=0
    \path[fill=blue!10]
      plot coordinates {\deformedGatePoints}
      -- (0,1) -- cycle;
  \else
    \begin{scope}[xshift=6cm,xscale=-1]
      \path[fill=blue!10]
        plot coordinates {\deformedGatePoints}
        -- (0,1) -- cycle;
    \end{scope}
  \fi

  \foreach \height in {0.25,0.80}
    \draw[gray!55,densely dotted]
      (0,\height) -- (1,\height);
  \draw[gray!75,dashed] (0,0) -- (1,1);
  \draw[gray!75,dashed] (1,0) -- (0,1);
  \draw[black!80] (0,0) rectangle (1,1);

  \ifnum\side=0
    \draw[blue!75!black,line width=1.1pt]
      plot coordinates {\deformedGatePoints};
    \fill (0.326820,0.673180) circle (1.7pt);
    \draw[gray!70]
      (0.326820,0.673180) -- (0.50,0.77)
      node[right,black] {$R=L$};
    \node[blue!75!black,right]
      at (0.285,0.49) {$\Sigma_N$};
    \node[rotate=90]
      at (0.070,0.62) {retained northern cap};
    \node at (0.70,0.30) {excised};
    \node at (0.16,0.90) {$\mathcal H_N^+$};
    \node at (0.20,0.12) {$\mathcal H_N^-$};
    \node at (0.50,1.14) {Northern parent region};
    \node[below] at (0.50,-0.09) {$\theta_1$};
  \else
    \begin{scope}[xshift=6cm,xscale=-1]
      \draw[blue!75!black,line width=1.1pt]
        plot coordinates {\deformedGatePoints};
    \end{scope}
    \fill (0.673180,0.673180) circle (1.7pt);
    \draw[gray!70]
      (0.673180,0.673180) -- (0.50,0.77)
      node[left,black] {$R=L$};
    \node[blue!75!black,left]
      at (0.715,0.49) {$\Sigma_S$};
    \node[rotate=-90]
      at (0.930,0.62) {retained southern cap};
    \node at (0.30,0.30) {excised};
    \node at (0.84,0.90) {$\mathcal H_S^+$};
    \node at (0.80,0.12) {$\mathcal H_S^-$};
    \node at (0.50,1.14) {Southern parent region};
    \node[below] at (0.50,-0.09) {$\theta_2$};
  \fi

  \node[above] at (0.50,1.01) {$\mathcal I^+$};
  \node[below] at (0.50,-0.01) {$\mathcal I^-$};
  \node[below] at (0,-0.01) {$0$};
  \node[below] at (1,-0.01) {$\pi$};
  \node[left] at (-0.02,1) {$\pi/2$};
  \node[left] at (-0.02,0.5) {$0$};
  \node[left] at (-0.02,0) {$-\pi/2$};
  \node[rotate=90] at (-0.13,0.5) {$\eta$};
\end{scope}
}
\end{tikzpicture}
\caption{The two-way deformed cosmic creation gate in global
de Sitter coordinates. The shaded northern and southern
fragments belong to separate parent spacetimes; points on
the blue junction images $\Sigma_N$ and $\Sigma_S$ are
identified at equal global time $\tau$. The vertical
coordinate is $\eta=\arctan[\sinh(\tau/L)]$, and radial null
rays have slopes $\pm1$. Dashed lines mark the cosmological
horizons of the corresponding parent polar observers.
Each junction image crosses its future horizon at $R=L$
(black dots), remaining timelike throughout. In the infinite
past, the angular openings vanish while the total spatial
volume approaches $2\pi R_0^2$. In the infinite future,
the openings approach nonzero values and the spatial volume
grows without bound. The trajectories are obtained from
the exact radial equation for $D=3$, $w=1/2$ and
$R_0=L/\sqrt3$, with $\theta_s\to\pi/3$ towards the future;
the second image is obtained by reflection.
Dotted horizontal lines indicate synchronized
global-time slices.}
\label{fig:deformed-cosmic-creation-gate}
\end{figure}

We now study an exact rotationally invariant deformation of the cosmic creation gates of Sec.~\ref{Sec:creation-gates}, for any $n\geq2$ and $D\geq3$. The solution approaches the corresponding cosmic creation gate in the infinite past, retaining its finite, strictly positive limiting spatial volume, while its future volume grows without bound. We refer to this solution as the \textit{deformed cosmic creation gate}. For the two-way junction, this solution is depicted in Fig.~\ref{fig:deformed-cosmic-creation-gate}. We derive its evolution directly from the global embeddings and junction conditions, and then identify it as the nonlinear backreaction of the homogeneous hybrid sound mode that decays towards the infinite past and grows towards the future. The junction fluid expands for every allowed $w$. Its temperature decreases for $w>0$ and increases for the allowed negative values of $w$, as we establish below.

Let us retain $n\geq2$ northern fragments, as in Sec.~\ref{Sec:creation-gates}, and describe every junction image by the same time-dependent function $\theta_i=\theta_s(\tau)$. In the folded picture, all relative embedding variables vanish, and the images of the junction are
\begin{equation}\label{Eq:histories-global-images}
t_i=\tau,\qquad \varphi_i^A=\sigma^A,\qquad
\theta_i=\theta_s(\tau),\qquad u^A=0,
\qquad i=1,\cdots,n.
\end{equation}
For $D=3$, $\varphi_i^A=\sigma^A$ means $\phi_i=\sigma$. The northern--southern two-way description follows by the reflection in \eqref{Eq:gate-unfolded-two}. A dot in this subsection denotes a derivative with respect to global time $\tau$.

The metric induced on each image is
\begin{equation}\label{Eq:histories-global-induced}
{\rm d}s_\Sigma^2
=-\left[1-L^2\cosh^2(\tau/L)\dot\theta_s^{\,2}\right]{\rm d}\tau^2
+L^2\cosh^2(\tau/L)\sin^2\theta_s\,{\rm d}\Omega_{D-2}^2.
\end{equation}
These metrics agree exactly, so the first junction condition is satisfied at every stage of the evolution. The fluid follows the curves of fixed angular coordinates on the junction. Its velocity normalization determines $u^\tau$ and relates its proper clock to global time:
\begin{equation}\label{Eq:histories-evolving-clock}
u^\tau=\left[1-L^2\cosh^2(\tau/L)\dot\theta_s^{\,2}\right]^{-1/2},
\qquad
\frac{{\rm d}\tau_{\rm p}}{{\rm d}\tau}=\frac1{u^\tau}.
\end{equation}

Using the orientation of the normals specified in Sec.~\ref{Sec:setup}, we obtain the nonzero mixed components of the extrinsic curvature:
\begin{align}\label{Eq:histories-global-curvatures}
K_i{}^\tau{}_\tau
&=L\cosh(\tau/L)(u^\tau)^3
\left[\ddot\theta_s+\frac2L\tanh(\tau/L)\dot\theta_s
-L\cosh(\tau/L)\sinh(\tau/L)\dot\theta_s^{\,3}\right],
\nonumber\\
K_i{}^A{}_B
&=u^\tau\left[\sinh(\tau/L)\dot\theta_s
+\frac{\cot\theta_s}{L\cosh(\tau/L)}\right]\delta^A_B.
\end{align}
The temporal and angular components of the second junction condition then give the fluid density and pressure:
\begin{equation}\label{Eq:histories-global-stress}
\rho=\frac{n}{8\pi G_N}K_i{}^A{}_A,
\qquad
P=-\frac{n}{8\pi G_N}
\left(K_i{}^\tau{}_\tau+\frac{D-3}{D-2}K_i{}^A{}_A\right),
\end{equation}
The repeated angular index is traced, and any image $i$ may be used because the embeddings are identical. The mixed components vanish. Imposing $P=w\rho$ gives the exact second-order equation for the transverse embedding $\theta_s(\tau)$:
\begin{equation}\label{Eq:histories-global-eos}
K_i{}^\tau{}_\tau
+\frac{(D-2)(1+w)-1}{D-2}K_i{}^A{}_A=0,
\end{equation}
Together with the curvatures in \eqref{Eq:histories-global-curvatures}, this determines the evolution of the junction in global coordinates.

The energy component of the Euler equations is
\[
\dot\rho+(D-2)(1+w)\rho
\left[\frac1L\tanh(\tau/L)+\dot\theta_s\cot\theta_s\right]=0.
\]
We integrate this equation and require that the fluid approach the constant density and temperature of the cosmic creation gate of Sec.~\ref{Sec:creation-gates} in the infinite past. This fixes the integration constant and gives
\begin{align}\label{Eq:histories-global-fluid}
\rho(\tau)&=\rho^{(0)}
\left[\frac{R_0}{L\cosh(\tau/L)\sin\theta_s}\right]^{(D-2)(1+w)},
\nonumber\\
T(\tau)&=T^{(0)}
\left[\frac{R_0}{L\cosh(\tau/L)\sin\theta_s}\right]^{(D-2)w}.
\end{align}
Here $\rho^{(0)}$ and $T^{(0)}$ are the corresponding values on that cosmic creation gate, and the temperature formula applies for $w\neq0$. For dust, the density supplies the fluid variable in place of $T$. The spatial Euler equation is satisfied because the fluid variables are homogeneous and the intrinsic fluid acceleration vanishes. The embedding equation \eqref{Eq:histories-global-eos}, the fluid solution \eqref{Eq:histories-global-fluid} and velocity normalization therefore determine all junction fields. The bulk Einstein equations and their constraints hold in each de Sitter fragment. The junction embeddings also ensure that the Gauss and Codazzi identities hold on every image.

To integrate the embedding equation once, we use the areal radius of the junction spheres, which is determined by $\theta_s(\tau)$:
\begin{equation}\label{Eq:histories-radius-from-embedding}
R(\tau)=L\cosh(\tau/L)\sin\theta_s(\tau),\qquad
\frac{{\rm d}R}{{\rm d}\tau_{\rm p}}
=u^\tau\left[\sinh(\tau/L)\sin\theta_s
+L\cosh(\tau/L)\cos\theta_s\dot\theta_s\right].
\end{equation}
Multiplying the angular curvature in \eqref{Eq:histories-global-curvatures} by $R$ and using the density equation in \eqref{Eq:histories-global-stress} gives
\begin{equation}\label{Eq:histories-angular-radius}
\frac{8\pi G_N R\rho}{n(D-2)}
=u^\tau\left[\cos\theta_s
+L\cosh(\tau/L)\sinh(\tau/L)\sin\theta_s\dot\theta_s\right].
\end{equation}
Squaring this expression and subtracting the square of ${\rm d}R/{\rm d}\tau_{\rm p}$ eliminates the terms linear in $\dot\theta_s$. The remaining terms combine into
\[
\begin{aligned}
&\left(\frac{8\pi G_N R\rho}{n(D-2)}\right)^2
-\left(\frac{{\rm d}R}{{\rm d}\tau_{\rm p}}\right)^2\\
&\quad=(u^\tau)^2
\left[1-L^2\cosh^2(\tau/L)\dot\theta_s^{\,2}\right]
\left[1-\cosh^2(\tau/L)\sin^2\theta_s\right]
=1-\frac{R^2}{L^2},
\end{aligned}
\]
where the last equality follows from velocity normalization. The angular junction condition therefore implies
\begin{equation}\label{Eq:histories-radial-exact}
\left(\frac{8\pi G_N R\rho}{n(D-2)}\right)^2
=1+\left(\frac{{\rm d}R}{{\rm d}\tau_{\rm p}}\right)^2-\frac{R^2}{L^2}.
\end{equation}
Fluid conservation gives $\rho=\rho^{(0)}(R_0/R)^{(D-2)(1+w)}$, while \eqref{Eq:gate-radius-general} fixes $\rho^{(0)}$. Consequently,
\begin{equation}\label{Eq:histories-density-radius}
\left(\frac{8\pi G_N R\rho}{n(D-2)}\right)^2
=\left(1-\frac{R_0^2}{L^2}\right)
\left(\frac{R_0}{R}\right)^{2[(D-2)(1+w)-1]}
=\frac{\left(R_0/R\right)^{2[(D-2)(1+w)-1]}}
{(D-2)(1+w)}.
\end{equation}
Substitution in \eqref{Eq:histories-radial-exact} gives the first integral of the global embedding equation:
\begin{equation}\label{Eq:histories-radial-first-integral}
\left(\frac{{\rm d}R}{{\rm d}\tau_{\rm p}}\right)^2
=\frac{R^2}{L^2}-1
+\frac{1}{(D-2)(1+w)}
\left(\frac{R_0}{R}\right)^{2[(D-2)(1+w)-1]}.
\end{equation}
We retain the positive-density orientation specified by the unsquared equation \eqref{Eq:histories-angular-radius}. On the expanding branch, differentiation of \eqref{Eq:histories-radial-first-integral}, together with the value of $R_0$, yields
\begin{equation}\label{Eq:histories-radial-acceleration}
\frac{{\rm d}^2R}{{\rm d}\tau_{\rm p}^2}
=\frac{1}{L^2}\left[
R-R_0\left(\frac{R_0}{R}\right)^{2(D-2)(1+w)-1}\right].
\end{equation}
This equation and fluid conservation give the remaining spatial junction equation. At $R=R_0$, their continuous limit recovers the cosmic creation gate of Sec.~\ref{Sec:creation-gates}.

For $w>-(D-3)/(D-2)$, the right-hand side of \eqref{Eq:histories-radial-first-integral} has a double zero at $R_0$ and increases strictly for $R>R_0$. Near $R_0$ it takes the form
\begin{equation}\label{Eq:histories-exact-past-branch}
\left(\frac{{\rm d}R}{{\rm d}\tau_{\rm p}}\right)^2
=\Gamma_D^2(R-R_0)^2+O\!\left((R-R_0)^3\right),
\qquad
\Gamma_D=\frac{\sqrt{2(D-2)(1+w)}}L.
\end{equation}
The positive square root determines an exact expanding solution, given implicitly by
\begin{equation}\label{Eq:histories-expanding-quadrature}
\tau_{\rm p}-\tau_{{\rm p}*}
=\int_{R(\tau_{{\rm p}*})}^{R(\tau_{\rm p})}
\frac{{\rm d}\widetilde R}{\displaystyle
\sqrt{\frac{\widetilde R^2}{L^2}-1
+\frac{1}{(D-2)(1+w)}
\left(\frac{R_0}{\widetilde R}\right)^{2[(D-2)(1+w)-1]}}}.
\end{equation}
Here $\tau_{{\rm p}*}$ is any finite time on the solution. The integral diverges logarithmically at both $R=R_0$ and $R=\infty$. The junction therefore approaches the cosmic creation gate in the infinite past and expands to arbitrarily large radius in the infinite future. Its density remains positive throughout the evolution and tends to zero towards the future.

The global embedding is shown in Fig.~\ref{fig:deformed-cosmic-creation-gate}. Here $\theta_s(\tau)$ is the folded embedding variable of this subsection. The northern--southern description has $\theta_1=\theta_s(\tau)$ and $\theta_2=\pi-\theta_s(\tau)$, with their boundary points identified at equal global time. In terms of the conformal time $\eta=\arctan[\sinh(\tau/L)]$, \eqref{Eq:histories-radius-from-embedding} and \eqref{Eq:histories-evolving-clock} give
\begin{equation}\label{Eq:histories-angular-reconstruction}
\sin\theta_s(\tau)
=\frac{R\bigl(\tau_{\rm p}(\tau)\bigr)}{L\cosh(\tau/L)},
\qquad
\left|\frac{{\rm d}\theta_s}{{\rm d}\eta}\right|
=L\cosh(\tau/L)|\dot\theta_s|<1.
\end{equation}
The reconstruction uses the evolving fluid clock in \eqref{Eq:histories-evolving-clock}. We follow the continuous angular branch with $0<\theta_s<\pi$. For the solution drawn in the figure, $\theta_s<\pi/2$ throughout, so the principal inverse sine applies. Global time increases strictly along the timelike junction, and its embedding remains single-valued.

In the infinite past, $R\to R_0$ and hence
\begin{equation}\label{Eq:histories-past-angular-opening}
\theta_s(\tau)\sim\frac{2R_0}{L}e^{\tau/L}
\sim\frac{R_0}{L}\left(\eta+\frac{\pi}{2}\right),
\qquad \tau\to-\infty.
\end{equation}
The junction therefore approaches the same polar endpoint of $\mathcal I^-$ as the cosmic creation gate of Sec.~\ref{Sec:creation-gates}, with conformal slope $R_0/L<1$. Its angular opening vanishes while its areal radius approaches the finite value $R_0$.

Towards the future, \eqref{Eq:histories-global-eos} gives $K_i{}^\tau{}_\tau\propto\rho\to0$. The bulk acceleration of the fluid consequently decays, and the global embedding has the asymptotic behaviour
\[
R\sim {\rm const.}\,e^{\tau_{\rm p}/L},\qquad
\tau=\tau_{\rm p}+O(1),\qquad
0<\lim_{\tau\to+\infty}\theta_s(\tau)<\pi.
\]
Let $\theta_\infty=\lim_{\tau\to+\infty}\theta_s(\tau)$. Expanding the global embedding equation \eqref{Eq:histories-global-eos} towards the infinite future gives
\begin{equation}\label{Eq:histories-future-angular-opening}
\theta_s(\tau)
=\theta_\infty+2\cot\theta_\infty\,e^{-2\tau/L}
+o\!\left(e^{-2\tau/L}\right).
\end{equation}
The junction thus reaches $\mathcal I^+$ at a nonzero angular opening, with vanishing conformal slope.

The growth of the junction radius also determines the expansion of the fluid itself. From \eqref{Eq:histories-global-induced} and \eqref{Eq:histories-evolving-clock}, the induced volume element satisfies $\sqrt{-\gamma}\propto R^{D-2}/u^\tau$. Since $u^A=0$, the fluid expansion is
\begin{equation}\label{Eq:histories-fluid-expansion}
\nabla_\mu u^\mu
=\frac{1}{\sqrt{-\gamma}}\partial_\tau
\left(\sqrt{-\gamma}\,u^\tau\right)
=\frac{D-2}{R}\frac{{\rm d}R}{{\rm d}\tau_{\rm p}}>0.
\end{equation}
The fluid therefore expands for every allowed equation of state, including negative $w$. Its expansion tends to zero in the infinite past and to $(D-2)/L$ in the infinite future.

For $w\neq0$, the temperature in \eqref{Eq:histories-global-fluid} can be written directly as a function of the evolving fluid proper time:
\begin{align}\label{Eq:histories-temperature-evolution}
T(\tau_{\rm p})
&=T^{(0)}\left[\frac{R_0}{R(\tau_{\rm p})}\right]^{(D-2)w},
\nonumber\\
\frac{{\rm d}\ln T}{{\rm d}\tau_{\rm p}}
&=-w\,\nabla_\mu u^\mu,
\qquad
\frac{{\rm d}\ln\rho}{{\rm d}\tau_{\rm p}}
=-(1+w)\,\nabla_\mu u^\mu.
\end{align}
The temperature response to expansion depends on the sign of $w$. For $w>0$, the fluid cools monotonically from $T^{(0)}$ in the infinite past to zero in the infinite future. This is necessarily the behaviour in three dimensions, where the cosmic creation gate requires $w>0$. For the allowed range $-(D-3)/(D-2)<w<0$ in four and higher dimensions, the expanding fluid instead heats monotonically, with its temperature growing without bound in the infinite future. In particular,
\begin{align}\label{Eq:histories-temperature-asymptotics}
&\lim_{\tau_{\rm p}\to-\infty}T(\tau_{\rm p})=T^{(0)}>0,
\nonumber\\
&T(\tau_{\rm p})\sim {\rm const.}\,
e^{-(D-2)w\tau_{\rm p}/L},
\qquad \tau_{\rm p}\to+\infty,
\nonumber\\
&\rho(\tau_{\rm p})\sim {\rm const.}\,
e^{-(D-2)(1+w)\tau_{\rm p}/L}\longrightarrow0,
\qquad \tau_{\rm p}\to+\infty.
\end{align}
Every allowed $w$ satisfies $w>-1$, so the density decreases in both cases. For negative $w$, the equation of state relates this decreasing density to an increasing temperature. The temperature is finite at every finite proper time, and its divergence occurs only in the infinite future.

For negative \(w\), this thermodynamic parametrization has negative local heat capacity and negative sound-speed squared, signalling thermodynamic and gradient instabilities in an ordinary local fluid interpretation. The classical junction solutions and their temperature evolution remain valid.

The total spatial volume of the glued spacetime is measured on the same global-time slices as in Sec.~\ref{Sec:creation-gates}. Direct integration over the retained fragments gives
\begin{equation}\label{Eq:histories-exact-global-volume}
V(\tau)=n\Omega_{D-2}L^{D-1}\cosh^{D-1}(\tau/L)
\int_0^{\theta_s(\tau)}\sin^{D-2}\theta\,{\rm d}\theta.
\end{equation}
For the two-way three-dimensional solution in Fig.~\ref{fig:deformed-cosmic-creation-gate}, this reduces to
\begin{equation}\label{Eq:histories-two-way-global-volume}
V_{2,3}(\tau)
=4\pi L^2\cosh^2(\tau/L)\bigl[1-\cos\theta_s(\tau)\bigr].
\end{equation}
The past and future limits are therefore
\begin{equation}\label{Eq:histories-radial-volumes}
V\longrightarrow\frac{n\Omega_{D-2}}{D-1}R_0^{D-1}
\quad\hbox{in the infinite past},\qquad
V\longrightarrow\infty
\quad\hbox{in the infinite future}.
\end{equation}
The deformed cosmic creation gate retains the finite past volume of the undeformed gate and has an unbounded expanding future.

Figure~\ref{fig:deformed-cosmic-creation-gate} shows how this evolution appears in the parent conformal diagrams. For the northern polar observer, the past and future cosmological horizons are
\begin{equation}\label{Eq:histories-parent-horizons}
\mathcal H_N^-:\quad \theta_1=\frac{\pi}{2}+\eta,
\qquad
\mathcal H_N^+:\quad \theta_1=\frac{\pi}{2}-\eta.
\end{equation}
The timelike bound in \eqref{Eq:histories-angular-reconstruction} makes both $\eta-\theta_s$ and $\eta+\theta_s$ strictly increasing. Since $\eta-\theta_s$ increases from $-\pi/2$, the past horizon lies outside the retained cap at every finite time. Meanwhile, $\eta+\theta_s$ increases from $-\pi/2$ to $\pi/2+\theta_\infty$, so the junction crosses the future horizon exactly once. At the crossing, $R=L$ and \eqref{Eq:histories-radial-first-integral} gives $\mathrm{d}R/\mathrm{d}\tau_{\rm p}>0$. The crossing occurs at finite proper time and the junction remains timelike. Reflection gives the corresponding horizons and crossing for the southern fragment.

The plotted solution has $D=3$, $w=1/2$ and $\theta_\infty=\pi/3$. Its angular opening first increases, reaches $\theta_s\simeq1.0652$ at $\tau/L\simeq1.24$, and then decreases towards $\pi/3$, approaching it from above as predicted by \eqref{Eq:histories-future-angular-opening}. The expanding branch of \eqref{Eq:histories-radial-first-integral} keeps the areal radius increasing through this angular turnaround, and the fluid continues to expand according to \eqref{Eq:histories-fluid-expansion}. Equation \eqref{Eq:histories-two-way-global-volume} gives the two volume limits: the vanishing past opening gives $V_-=2\pi R_0^2$, while the nonzero future opening gives unbounded growth of the spatial volume. For arbitrary $n$ and $D$, \eqref{Eq:histories-exact-global-volume} gives the same finite past limit and unbounded future volume.

The angular evolution described above applies to every rotationally invariant $n$-way deformed cosmic creation gate. In the folded picture, all junction images have $t_i=\tau$ and $\theta_i=\theta_s(\tau)$, with $i=1,\cdots,n$. At fixed $D$, $w$ and $L$, and for the same integration data on the expanding branch, $\theta_s(\tau)$ is independent of $n$. The required fluid density and pressure are proportional to $n$, and this overall factor cancels from the global embedding equation \eqref{Eq:histories-global-eos} upon imposing $P=w\rho$. Each image therefore has the same past angular asymptotics, crosses its parent future horizon exactly once, and approaches the same nonzero $\theta_\infty$. For $D=3$, $w=1/2$ and $\theta_\infty=\pi/3$, the angular maximum described above occurs for every $n$, with the same embedding data as the two-way example in Fig.~\ref{fig:deformed-cosmic-creation-gate}. The total spatial volume is $n$ times that of a single retained fragment, as in \eqref{Eq:histories-exact-global-volume}.

To identify the homogeneous hybrid sound mode underlying this exact history, we write the radius as a function of the evolving fluid proper time and expand in $\epsilon$ as
\[
R=R_0+\epsilon R_1+\epsilon^2R_2+O(\epsilon^3).
\]
At first order, \eqref{Eq:histories-radial-acceleration} gives
\begin{equation}\label{Eq:histories-homogeneous-linear}
(\partial_{\tau_{\rm p}}^2-\Gamma_D^2)R_1=0,
\qquad
R_1=C_+e^{\Gamma_D\tau_{\rm p}}+C_-e^{-\Gamma_D\tau_{\rm p}}.
\end{equation}
In the original global embedding variables, the same perturbation is
\begin{equation}\label{Eq:histories-homogeneous-normal}
\theta_s=\mathfrak f+\epsilon\theta_{s,1}+O(\epsilon^2),
\qquad
R_1=L\cosh(\tau/L)\cos\mathfrak f\,\theta_{s,1}
=\frac{\eta_0}{\sqrt{(D-2)(1+w)}}.
\end{equation}
Here $\eta_0$ is the homogeneous average proper transverse displacement defined in the preceding perturbation analysis. Since the undeformed radius $R_0$ is constant, the evolving fluid clock and the cosmic creation gate clock give the same radius perturbation at first order. For $D=3$, the equation for $\eta_0$ is precisely the $m=0$ hybrid sound equation, with $\omega_0^2=-2(1+w)/L^2$. In four and higher dimensions it is the homogeneous mode in \eqref{Eq:gate-higher-radial-completion}.

The fluid perturbations follow by expanding \eqref{Eq:histories-global-fluid}:
\begin{equation}\label{Eq:histories-homogeneous-fluid}
\frac{\rho_1}{\rho^{(0)}}=-(D-2)(1+w)\frac{R_1}{R_0},
\qquad P_1=w\rho_1,\qquad
\frac{T_1}{T^{(0)}}=-(D-2)w\frac{R_1}{R_0}\quad(w\neq0).
\end{equation}
The angular velocity vanishes, the temporal velocity is fixed by normalization, and all relative embedding variables remain zero. Thus both the embedding and fluid perturbations reproduce the homogeneous hybrid sound mode. Approach to the cosmic creation gate in the infinite past requires $C_-=0$, while $C_+>0$ selects the expanding branch.

At second order, the same exact embedding equation gives
\begin{equation}\label{Eq:histories-homogeneous-second-order}
(\partial_{\tau_{\rm p}}^2-\Gamma_D^2)R_2
=-\frac{(D-2)(1+w)\,[2(D-2)(1+w)-1]}{L^2R_0}\,R_1^2.
\end{equation}
The square of the first-order mode therefore sources the next correction to the radius. Defining $\epsilon C_+$ as the coefficient of the leading past exponential, we obtain
\begin{equation}\label{Eq:histories-homogeneous-nonlinear-branch}
R=R_0+\epsilon C_+e^{\Gamma_D\tau_{\rm p}}
-\epsilon^2\frac{2(D-2)(1+w)-1}{6R_0}
C_+^2e^{2\Gamma_D\tau_{\rm p}}+O(\epsilon^3).
\end{equation}
This expansion describes the departure from the cosmic creation gate while the displacement is small compared with $R_0$. The exact solution \eqref{Eq:histories-expanding-quadrature} continues the evolution beyond that regime.

The deformed cosmic creation gate therefore gives an exact nonlinear continuation of the homogeneous hybrid sound mode that decays towards the past and grows towards the future. Its backreaction determines the coupled evolution of the embedding, density and temperature, while every bulk region remains locally de Sitter. The junction fluid expands throughout this evolution, cooling for positive $w$ and heating for allowed negative $w$. The past volume remains finite and strictly positive, and the future volume grows without bound. This gives \textit{an explicit nonlinear creation history in which the condition on the past selects the admissible branch and the growing mode develops into indefinite cosmological expansion.}

\subsection{Inhomogeneous backreaction and the volume in the infinite past}
\label{Sec:histories-linear-selection}
\label{Sec:histories-cauchy}

We now return to the inhomogeneous perturbations of the rotationally invariant cosmic creation gate of Sec.~\ref{Sec:creation-gates}. These physical modes exist only for $D=3$ within the class of locally de Sitter bulk regions considered here. We examine how their backreaction affects the first criterion of a finite, strictly positive limiting spatial volume in the infinite past. At second order, this gives an additional restriction on the hybrid sound amplitudes, supplementing the exclusion of modes that grow towards the past. We also determine an amplitude-dependent correction to the limiting volume. The homogeneous deformations of the preceding subsection preserve the past volume of the cosmic creation gate, whereas the inhomogeneous sound waves generally change it.

We use the global coordinates and embedding variables of Sec.~\ref{Sec:n-way-perturbations-three}. In this subsection, $\tau_{\rm p}(\tau)$ is the proper clock of the undeformed cosmic creation gate, given in \eqref{Eq:gate-fluid-clock}. It labels the global-time slices on which all perturbation coefficients and volumes are evaluated.

The condition on the infinite past already restricts the linear modes. The homogeneous hybrid sound mode must have $C_-=0$, as in the preceding subsection. For each relative displacement, the coefficient $\eta_{I,m,-}^{\rm rel}$ of the branch growing towards the past must also vanish. Equivalently, the displacement and its velocity on any finite initial slice obey
\begin{equation}\label{Eq:histories-radial-data}
\partial_{\tau_{\rm p}}\delta R=\Gamma_D\delta R,
\qquad
\partial_{\tau_{\rm p}}\eta_{I,m}^{\rm rel}
=\kappa_m\eta_{I,m}^{\rm rel},
\end{equation}
Here $\Gamma_D$ is given in \eqref{Eq:histories-exact-past-branch}, with $D=3$, and
\begin{equation}\label{Eq:histories-relative-selection}
\kappa_m=\frac{|m|\sqrt{1+w}}L,\qquad
|m|\geq2,\qquad I=1,\cdots,n-1.
\end{equation}
These conditions retain the branches that decay towards the past and allow their amplitudes to grow towards the future.

The hybrid sound modes with $|m|\geq2$ remain bounded and continue to oscillate towards the infinite past. Since the volume depends nonlinearly on the embedding, these bounded waves can produce a second-order volume with no definite past limit. To determine whether the limit exists, we must include the change in the junction embedding forced by their backreaction.

Let us retain $n\geq2$ identical northern fragments and perturb the undeformed cosmic creation gate by taking $\theta_i=\theta_s(\tau,\sigma)$ on every junction image. With all relative embedding variables set to zero, the images are
\begin{equation}\label{Eq:histories-inhomogeneous-images}
t_i=\tau,\qquad \phi_i=\sigma,\qquad
\theta_i=\theta_s(\tau,\sigma),\qquad i=1,\cdots,n.
\end{equation}
Their induced metrics agree exactly, so the first junction condition is satisfied at every order. We take $0<w<1$ and finitely many first-order sound harmonics with $|m|\geq2$; the result extends continuously to $w=1$. The expansion of the average transverse embedding variable is
\[
\theta_s=\mathfrak f(\tau)+\epsilon\theta_{s,1}(\tau,\sigma)
+\epsilon^2\theta_{s,2}(\tau,\sigma)+O(\epsilon^3).
\]
The temperature and angular velocity $u^\phi$ are expanded to the same order. Their coefficients are determined by the second junction condition, while $u^\tau$ follows from normalization, as in the linear calculation of Sec.~\ref{Sec:n-way-perturbations-three}.

On each global-time slice, the areal radius is determined by the transverse embedding:
\begin{align}\label{Eq:histories-quadratic-radius}
R&=L\cosh(\tau/L)\sin\theta_s
=R_0+\epsilon R_1+\epsilon^2R_2+O(\epsilon^3),
\nonumber\\
R_1&=L\cosh(\tau/L)\cos\mathfrak f\,\theta_{s,1}
=\frac{\eta}{\sqrt{1+w}},
\nonumber\\
R_2&=L\cosh(\tau/L)\cos\mathfrak f\,\theta_{s,2}
-\frac{R_0}{2}\theta_{s,1}^2.
\end{align}
The coefficient $R_2$ multiplies $\epsilon^2$ without an additional factor of $1/2$. In particular, it includes both the contribution of $\theta_{s,2}$ and the quadratic contribution of $\theta_{s,1}$. The time dependence on the prescribed global-time slices follows by substituting the known function $\tau_{\rm p}(\tau)$.

Integrating the spatial volume of the retained fragments gives
\begin{equation}\label{eq:quad_exact_global_volume}
V_{n,3}(\tau)=n\int_0^{2\pi}L^2\cosh^2(\tau/L)
\left[1-\sqrt{1-\frac{R(\tau,\sigma)^2}
{L^2\cosh^2(\tau/L)}}\right]{\rm d}\sigma.
\end{equation}
Let $\langle X\rangle=(2\pi)^{-1}\int_0^{2\pi}X\,{\rm d}\sigma$ denote the angular average. We write $R_{2,0}(\tau_{\rm p})=\langle R_2\rangle$ for the average second-order radius and $C(\tau_{\rm p})=\langle R_1^2\rangle$ for the mean square of the first-order displacement. Since the first-order sound harmonics have zero angular average, the first correction to the total volume is quadratic. Towards the infinite past,
\begin{equation}\label{Eq:histories-quadratic-volume}
V_{n,3}=n\pi R_0^2+\epsilon^2V_2+O(\epsilon^3)+o(1),
\qquad
\frac{V_2}{2\pi n}=R_0R_{2,0}+\frac C2.
\end{equation}
The term $C/2$ is the direct quadratic contribution of the first-order waves, while $R_0R_{2,0}$ is the contribution of their induced average displacement. A finite past-volume limit requires the persistent oscillations in their sum to vanish. The average $R_{2,0}$ is fixed by the junction conditions, so the possibility of this cancellation is a dynamical question.

At second order, the temporal and mixed components of the second junction condition determine the density and angular velocity in terms of the embedding perturbations. Substituting these expressions in the angular component and imposing $P=w\rho$ gives the equation for $\theta_{s,2}$. Its angular average determines the embedding contribution to the volume. Using the first-order sound equation and integrating angular derivatives by parts, we obtain
\begin{align}\label{eq:quad_averaged_equation}
&(\partial_{\tau_{\rm p}}^2-\Gamma_3^2)
\left(R_{2,0}+\frac{\sqrt w}{2}\partial_{\tau_{\rm p}}C\right)
\nonumber\\
&\quad=-\frac{1+w}{L^2R_0}C
-\frac{(1+w)R_0}{L^2}\partial_{\tau_{\rm p}}^2C
+\frac{(1+w)^2}{L^2R_0}\langle(\partial_\sigma R_1)^2\rangle
\nonumber\\
&\qquad\quad-\frac{1-w}{R_0}
\langle(\partial_{\tau_{\rm p}}\partial_\sigma R_1)^2\rangle+\cdots.
\end{align}
All terms on the right are determined by the first-order waves. Here $\cdots$ denotes terms that vanish towards the infinite past at the indicated perturbative order. The term involving $\partial_{\tau_{\rm p}}C$ on the left arises from evaluating the radius on fixed global-time slices. The last term on the right includes the quadratic contribution of the first-order fluid velocity in \eqref{Eq:gate-n-extrinsic-linear}. Both contributions enter the volume calculation.

The averaged equation determines $R_{2,0}$ up to the homogeneous solutions $e^{\pm\Gamma_3\tau_{\rm p}}$. The branch growing towards the past is excluded, and the branch decaying there cannot cancel an oscillation of fixed amplitude. Thus any persistent oscillation in the second-order volume is determined by the first-order sound amplitudes.

To evaluate this oscillation, we use the Fourier solution in \eqref{Eq:gate-three-hybrid-spectrum}. For each positive $m$, the coefficients $\eta_{m,+}$ and $\eta_{m,-}$ multiply the two frequency branches of the sound mode. Their squared magnitudes give a constant contribution to $C$, while their interference gives an oscillation at frequency $2\omega_m$:
\begin{align}\label{eq:quad_mode_average}
C&=\sum_{m\geq2}\big(C_{m,0}+C_{m,\mathrm{osc}}\big),
\nonumber\\
C_{m,0}&=\frac{2}{1+w}
\left(|\eta_{m,+}|^2+|\eta_{m,-}|^2\right),
\nonumber\\
C_{m,\mathrm{osc}}
&=\frac{4}{1+w}\operatorname{Re}
\left[\eta_{m,+}(\eta_{m,-})^*
e^{2i\omega_m\tau_{\rm p}}\right],
\qquad \omega_m^2=\frac{1+w}{L^2}(m^2-2).
\end{align}
Products with different values of $|m|$ integrate to zero. The gradient terms in the averaged junction equation are also expressed in terms of the constant and oscillating contributions:
\[
\begin{aligned}
\langle(\partial_\sigma R_1)^2\rangle
&=\sum_{m\geq2}m^2\big(C_{m,0}+C_{m,\mathrm{osc}}\big),\\
\langle(\partial_{\tau_{\rm p}}\partial_\sigma R_1)^2\rangle
&=\sum_{m\geq2}m^2
\left[\omega_m^2\big(C_{m,0}+C_{m,\mathrm{osc}}\big)
+\frac12\partial_{\tau_{\rm p}}^2C_{m,\mathrm{osc}}\right].
\end{aligned}
\]
Each harmonic therefore induces a constant average displacement and an average displacement oscillating at $2\omega_m$. Solving \eqref{eq:quad_averaged_equation} at these frequencies and combining the result with the direct quadratic contribution in \eqref{Eq:histories-quadratic-volume} gives
\begin{align}\label{eq:quad_global_volume_oscillation}
\frac{V_{2,\mathrm{osc}}}{2\pi n}
=\sum_{m\geq2}\Bigg[&
\frac{-(1-w)m^4+(3-7w)m^2+8w-2}{4m^2-6}
C_{m,\mathrm{osc}}
\nonumber\\
&-\frac{R_0\sqrt w}{2}\partial_{\tau_{\rm p}}C_{m,\mathrm{osc}}
\Bigg].
\end{align}
This expression includes the full oscillating contribution to the volume at the order calculated. For a nonzero $C_{m,\mathrm{osc}}$, the two terms for a given $m$ are a sinusoid and its time derivative, which differ in phase by a quarter period. Their sum has a nonzero amplitude because the derivative coefficient is nonzero for $w>0$. The induced average displacement therefore leaves a persistent volume oscillation. Since the frequencies $2\omega_m$ are distinct, contributions from different harmonics cannot cancel it. The past-volume condition \eqref{Eq:histories-past-volume} consequently requires
\begin{equation}\label{eq:quad_amplitude_condition}
\boxed{\eta_{m,+}(\eta_{m,-})^*=0,\qquad m\geq2.}
\end{equation}

The physical meaning of this condition is simple. For each angular harmonic, only one direction of propagation can be present towards the infinite past at this order; the direction may be chosen independently for different harmonics. A travelling wave moves around the junction with a constant angular mean square. Waves travelling in opposite directions at the same harmonic produce the interference responsible for $C_{m,\mathrm{osc}}$. For example, the standing wave $\eta=\eta_*\cos(m\sigma)\cos(\omega_m\tau_{\rm p})$, with real nonzero $\eta_*$, has $\eta_{m,+}=\eta_{m,-}=\eta_*/4$ and fails \eqref{eq:quad_amplitude_condition}. The travelling wave $\eta=\eta_*\cos(m\sigma-\omega_m\tau_{\rm p})$ has $\eta_{m,+}=0$ and $\eta_{m,-}=\eta_*/2$ and satisfies it. Simply put, for the first criterion, each $m$-harmonic should be chiral for $\vert m \vert \geq 2$ with the direction of propagation allowed to differ between harmonics.

Once the persistent oscillation is removed, the constant part of the averaged junction equation fixes the limiting spatial volume to second order:
\begin{align}\label{eq:quad_allowed_past_volume}
V_-={}&n\pi R_0^2+
\frac{2\pi n\epsilon^2}{1+w}
\sum_{m\geq2}(m^2-1)\big[(1-w)m^2-2\big]
\left(|\eta_{m,+}|^2+|\eta_{m,-}|^2\right)
+O(\epsilon^3).
\end{align}
The correction is finite for the finite superposition of harmonics considered here, and sufficiently small amplitudes preserve the positivity of the past volume at this order. The allowed sound waves therefore generally change the limiting past volume. Their first-order amplitudes persist towards the infinite past, allowing a constant quadratic correction to survive there. The homogeneous deformation of Sec.~\ref{Sec:histories-exact-radial}, by comparison, decays towards the past and leaves the limiting volume unchanged.

The admitted homogeneous and relative displacement modes also decay towards the past. Their quadratic products with bounded sound waves vanish in that limit and cannot cancel the persistent contribution in \eqref{eq:quad_global_volume_oscillation}. The sound-amplitude restriction therefore remains necessary for a definite past-volume limit when these decaying modes are included at first order.

The perturbative expansion about the cosmic creation gate cannot remain valid for all time when future-growing modes are present. It describes the evolution while the perturbations remain small, and their subsequent growth requires a nonlinear continuation. If infinitely many angular harmonics are included, the sums defining the fields and the derivatives entering the junction conditions must converge throughout the interval considered. The relative modes require sufficiently rapid decay of their Fourier coefficients because their growth rates \(\kappa_m\) increase without bound with \(|m|\). Analyticity in the mode amplitudes alone does not ensure this convergence.

The first past-volume criterion restricts the initial displacements and velocities, as well as the amplitudes of sound waves present in the infinite past. In addition to excluding modes that grow towards the past, the second-order junction equations require \eqref{eq:quad_amplitude_condition} and give the correction \eqref{eq:quad_allowed_past_volume} to the initial volume. These are necessary second-order conditions on the inhomogeneous sound perturbations of the undeformed cosmic creation gate considered here. Whether the amplitudes satisfying them admit a nonlinear continuation that remains regular and single-valued, with a finite, positive past-volume limit, must be established from the full junction equations.

At higher orders, we may continue to impose the first criterion by requiring the persistent volume oscillations to cancel. Even when every first-order sound harmonic is chiral, products of different harmonics can produce further oscillating terms. Consider four positive angular harmonics satisfying
\[
m_1+m_2=m_3+m_4.
\]
When the two harmonics in each pair $(m_1,m_3)$ and $(m_2,m_4)$ propagate in opposite directions, their quartic product can have a nonzero frequency while surviving angular averaging. For example, take $m_1=2$ and $m_2=5$ travelling in one direction, and $m_3=3$ and $m_4=4$ travelling in the opposite direction. Since $2+5=3+4$, their product contains an angularly constant contribution oscillating at frequency $\omega_2+\omega_5+\omega_3+\omega_4$.

Such contributions are allowed in the fourth-order junction equations. The full calculation must determine whether they survive in the volume after all induced embedding and fluid corrections are included. At each order, we can require any persistent volume oscillations to cancel and derive further conditions on the mode amplitudes. Whether these successive conditions are compatible with nontrivial inhomogeneous solutions of all the junction conditions must be investigated. Alternatively, we may use the second criterion and allow persistent volume oscillations, provided the full history remains regular and its past volume stays between finite, strictly positive bounds.

\subsection{The Cauchy conundrum, causal response and conditioned admissible histories}
\label{Sec:histories-retarded-response}
\label{Sec:histories-synthesis}

The requirement of creation is a condition on the entire past development of a cosmological history. Its data on a finite-time slice must satisfy both the local gravitational and junction constraints and the restrictions needed for a regular, finite-volume past. These restrictions correlate the initial displacements, velocities and sound amplitudes. Arbitrarily chosen data generally excite modes that grow towards the past or produce a spatial volume incompatible with the chosen criterion. The \textit{Cauchy conundrum} is to reconcile this selection of cosmological histories with a causal response to physical disturbances introduced later.

The homogeneous equation has an ordinary local initial-value problem, as discussed above. The condition on the infinite past selects the branch $C_-=0$ and hence the relation between displacement and velocity in \eqref{Eq:histories-radial-data}. The relative modes obey analogous restrictions, and \eqref{eq:quad_amplitude_condition} gives an additional second-order condition on sound waves present towards the past when a definite past-volume limit is required. We therefore formulate the response problem by fixing an earlier admissible history and introducing a source on the junction at a later time. The induced changes in the embedding and fluid fields vanish before the source begins to act. Their subsequent evolution is determined by the sourced junction equations, with no bound imposed on the future spatial volume.

The second criterion also selects conditioned admissible histories. It allows sound waves travelling in both directions at a given harmonic, provided their backreaction remains regular and the past volume obeys \eqref{Eq:histories-bounded-past-volume}. The earlier history may then contain persistent volume oscillations. The response to a later junction source must leave this entire earlier history unchanged. The source must satisfy the same conservation and junction conditions described below, and the response remains subject to the stated convergence and regularity assumptions.

All bulk fragments continue to be locally de Sitter. We represent an infinitesimal junction disturbance by an additional surface stress $\delta S_{\mu\nu}^{\rm src}$, supported on the junction. While the source acts, the stress entering the second junction condition is
\begin{equation}\label{Eq:histories-junction-source}
S_{\mu\nu}^{\rm total}
=S_{\mu\nu}^{\rm fluid}+\delta S_{\mu\nu}^{\rm src},
\qquad
S_{\mu\nu}^{\rm fluid}
=(\rho+P)u_\mu u_\nu+P\gamma_{\mu\nu},
\qquad P=w\rho.
\end{equation}
The source describes additional junction matter or its interaction with the fluid. Its energy and momentum exchange with the fluid must obey the conservation identity implied by the Gauss--Codazzi constraint:
\begin{equation}\label{Eq:histories-source-conservation}
\nabla_\mu S_{\rm total}^{\mu\nu}=0,
\qquad
\nabla_\mu S_{\rm fluid}^{\mu\nu}
=-\nabla_\mu\delta S_{\rm src}^{\mu\nu}.
\end{equation}
The source therefore enters both the metric junction condition and the forced fluid equations. When it vanishes, these reduce to the perfect-fluid junction equations studied above.

We call a junction disturbance \textit{admissible} if its source vanishes before a finite junction time and is smooth when it begins to act. Its exchange of energy and momentum with the fluid must satisfy \eqref{Eq:histories-source-conservation}, and the resulting embedding and fluid fields must obey all remaining junction conditions. The source must also have sufficient regularity for the required derivatives, harmonic sums and retarded time integrals to exist. These conditions restrict the allowed source components and angular profiles. The forcing term in each mode equation follows from the surface stress $\delta S_{\mu\nu}^{\rm src}$ through the junction conditions.

For an explicit response, take the undeformed cosmic creation gate of Sec.~\ref{Sec:creation-gates} as the earlier history. We use its proper clock $\tau_{\rm p}(\tau)$ in \eqref{Eq:gate-fluid-clock}; the embeddings remain expressed in the global coordinates. Let a homogeneous source begin at $\tau_{{\rm p}a}$, with surface energy density $\delta\rho_{\rm src}$ and isotropic pressure $\delta P_{\rm src}$ measured in the fluid frame of that gate. The temporal and angular junction conditions determine the total density and pressure. Subtracting $w$ times the total density from the total pressure gives $\delta P_{\rm src}-w\delta\rho_{\rm src}$, because the fluid component still obeys $P=w\rho$. Linearizing these conditions using \eqref{Eq:histories-global-stress} gives the equation for the induced change in the radius shared by all junction images:
\begin{align}\label{Eq:histories-forced-homogeneous}
&(\partial_{\tau_{\rm p}}^2-\Gamma_D^2)\delta R_{\rm ret}
=\mathcal F_0(\tau_{\rm p}),
\nonumber\\
&\mathcal F_0
=-\frac{8\pi G_N}{n\sqrt{(D-2)(1+w)}}
\left(\delta P_{\rm src}-w\delta\rho_{\rm src}\right).
\end{align}
Here $\Gamma_D$ is defined in \eqref{Eq:histories-exact-past-branch}. The remaining embedding and fluid changes follow from the temporal junction condition, the equation of state and velocity normalization.

We define the retarded response by requiring that the sourced and unsourced histories agree before $\tau_{{\rm p}a}$. In particular, $\delta R_{\rm ret}$ and $\partial_{\tau_{\rm p}}\delta R_{\rm ret}$ both vanish at source onset. The solution is
\begin{equation}\label{Eq:histories-retarded-radial}
\delta R_{\rm ret}(\tau_{\rm p})
=\int_{-\infty}^{\tau_{\rm p}}
\frac{\sinh[\Gamma_D(\tau_{\rm p}-\tau_{\rm p}')]}{\Gamma_D}
\,\mathcal F_0(\tau_{\rm p}')\,{\rm d}\tau_{\rm p}',
\qquad
\mathcal F_0=0\quad\hbox{for }\tau_{\rm p}<\tau_{{\rm p}a}.
\end{equation}
The earlier history fixes the amplitude of the unforced mode that decays towards the past, leaving no freedom to add it to the response. The source and the unchanged earlier data therefore determine a unique homogeneous retarded solution.

If the source ceases to act at $\tau_{{\rm p}b}$, the subsequent displacement is
\begin{align}\label{Eq:histories-retarded-future}
\delta R_{\rm ret}(\tau_{\rm p})
={}&\frac{e^{\Gamma_D\tau_{\rm p}}}{2\Gamma_D}
\int_{\tau_{{\rm p}a}}^{\tau_{{\rm p}b}}
e^{-\Gamma_D\tau_{\rm p}'}
\mathcal F_0(\tau_{\rm p}')\,{\rm d}\tau_{\rm p}'
\nonumber\\
&-\frac{e^{-\Gamma_D\tau_{\rm p}}}{2\Gamma_D}
\int_{\tau_{{\rm p}a}}^{\tau_{{\rm p}b}}
e^{\Gamma_D\tau_{\rm p}'}
\mathcal F_0(\tau_{\rm p}')\,{\rm d}\tau_{\rm p}',
\qquad \tau_{\rm p}>\tau_{{\rm p}b}.
\end{align}
The two coefficients in this expression are fixed by the source pulse. A generic admissible source excites the branch growing towards the future. Requiring its coefficient to vanish would impose an additional integral condition on the source. We impose no such restriction and allow the future volume to grow. The exponential decomposition in \eqref{Eq:histories-retarded-future} applies after the pulse; the complete forced solution remains zero before the source begins. Thus its subsequent growth leaves the prescribed past unchanged.

In three dimensions, the junction source can also depend on the angular coordinate $\sigma$. Let $\delta\eta_{\rm ret}$ be the induced change in the average proper transverse displacement. The density and pressure variations in \eqref{Eq:gate-n-extrinsic-linear} give
\begin{equation}\label{Eq:histories-forced-sound}
\left[\partial_{\tau_{\rm p}}^2
-\frac{1+w}{L^2}\partial_\sigma^2-\Gamma_3^2\right]
\delta\eta_{\rm ret}
=-\frac{8\pi G_N}{n}
\left(\delta P_{\rm src}-w\delta\rho_{\rm src}\right).
\end{equation}
The mixed junction condition determines the angular fluid velocity, including the source momentum. For each physical sound harmonic $|m|\geq2$, the left-hand side of \eqref{Eq:histories-forced-sound} is $(\partial_{\tau_{\rm p}}^2+\omega_m^2)\delta\eta_{m,\rm ret}$. Integrating its source against the sine kernel below gives the retarded sound response.

The relative displacement equations \eqref{Eq:gate-n-relative-equations} follow from metric continuity. Adding $\delta S_{\mu\nu}^{\rm src}$ to the second junction condition does not introduce a forcing term into these linear metric-continuity equations about the undeformed cosmic creation gate. Their induced response is therefore zero when the earlier relative displacements and velocities are held fixed. When the folded embeddings are identical in every parent region, the relative embedding variables can consistently remain zero at every order. When the earlier history contains relative displacements, the nonlinear continuity conditions can instead produce higher-order forcing fixed by the lower-order embeddings. The retarded response to this higher-order forcing is computed with $G_m^{\rm rel}$ below. The two kernels are
\begin{align}\label{Eq:histories-mode-kernels}
G_m^{\rm sound}(\tau_{\rm p},\tau_{\rm p}')
&=\Theta(\tau_{\rm p}-\tau_{\rm p}')
\frac{\sin[\omega_m(\tau_{\rm p}-\tau_{\rm p}')]}{\omega_m},
\nonumber\\
G_m^{\rm rel}(\tau_{\rm p},\tau_{\rm p}')
&=\Theta(\tau_{\rm p}-\tau_{\rm p}')
\frac{\sinh[\kappa_m(\tau_{\rm p}-\tau_{\rm p}')]}{\kappa_m}.
\end{align}
Here $\Theta$ is the step function, $\omega_m$ is given in \eqref{Eq:gate-two-hybrid-spectrum}, and $\kappa_m$ in \eqref{Eq:histories-relative-selection}. Equation \eqref{Eq:histories-forced-sound} supplies the sound source. Any higher-order relative forcing must be derived from metric continuity and integrated with $G_m^{\rm rel}$, keeping the earlier amplitudes of the source-free relative modes fixed.

The angular profiles must satisfy the convergence requirements discussed in the preceding subsection. For finitely many angular harmonics, the retarded response follows directly from these kernels. When a relative displacement kernel is required at higher orders, its exponential growth with $|m|$ imposes an additional restriction on the Fourier coefficients as we have seen explicitly in the case of the sound modes at second order.\footnote{For positive time separation, the relative displacement kernel does not have a Fourier sum defining the usual Green function for an angular delta-function source. It applies to the restricted angular profiles for which the sum and its required derivatives converge.} The sound-amplitude restriction follows from the first criterion, while convergence of the relative displacement response is required under either criterion. Sources extending to the infinite past would also require convergence of the retarded time integrals and compatibility with the prescribed past. The sources considered here begin at a finite junction time, so their retarded response leaves the earlier history unchanged.

Under the first criterion, the sound-amplitude condition \eqref{eq:quad_amplitude_condition} concerns waves already present towards the infinite past. A later junction source can generate sound waves in either direction without changing those past amplitudes or the limiting past volume. The junction embedding and fluid follow the earlier history until the source acts; their subsequent response is determined by the sourced junction equations.

Under the convergence and regularity assumptions stated above, the induced changes remain zero before the source begins to act at every perturbative order. At each order, the forced embedding equations and the continuity conditions contain terms determined by the junction source and lower-order fields. If the source and induced changes vanish before $\tau_{{\rm p}a}$, the changes in these forcing terms also vanish there. Solving with retarded kernels and the same earlier data preserves this property at the next order. The remaining junction and fluid equations must be satisfied at each step through the source conservation and compatibility conditions. Convergence then gives the retarded response on the domain where the resulting geometry is regular and its embeddings remain single-valued.

We call these solutions \textit{conditioned admissible histories}. Their initial data satisfy the gravitational and junction constraints together with the restrictions imposed by a regular, finite-volume past on the natural cosmological foliation. Admissible junction sources determine their subsequent response, while their future spatial volume may grow without bound. The exact homogeneous deformation establishes this possibility nonlinearly, and the second-order inhomogeneous calculation gives an additional restriction on the waves present towards the past when a definite past-volume limit is required.

Observers, their clocks and the matter responsible for a junction disturbance belong to the same gravitational system. The fluid clock measures when the source acts and the subsequent evolution of the junction. The junction conditions relate this evolution to the energy and momentum exchanged between the source and the fluid. The finite-volume past selects the histories under consideration; admissible junction sources and the gravitational equations determine how those histories can subsequently evolve. A controlled past and a causal response allowing future growth are compatible properties of the same Lorentzian history.

\section{Conclusions}\label{Sec:conclusions}

In this paper, we have shown that multiple locally de Sitter universes can be created with a finite total initial spatial volume within classical Lorentzian gravity, provided we allow gravitational junctions. We have constructed rotationally invariant solutions joining any finite number $n\geq2$ of locally dS$_D$ regions, with $D\geq3$, at a common timelike junction supporting a perfect fluid with equation of state $P=w\rho$. For $w>-(D-3)/(D-2)$, these \textit{cosmic creation gates} are geodesically complete in both temporal directions, while their total spatial volume approaches finite, strictly positive values in the infinite past and future on the natural global-time foliation synchronized by the junction fluid clock. We have determined the initial volume and the maximal volume attained at an intermediate time in terms of the equation of state, the de Sitter radius, the spacetime dimension and the number of regions joined.

Furthermore, we have constructed exact \textit{deformed cosmic creation gates} which approach the cosmic creation gates in the infinite past, retaining their finite, strictly positive limiting spatial volume, while their future volume grows without bound. These solutions also describe creation and provide the nonlinear continuation of the homogeneous hybrid sound mode whose amplitude decays towards the past and grows towards the future. The junction fluid expands throughout this evolution. For $w>0$, its temperature decreases from a finite past value to zero in the infinite future; for the allowed negative values of $w$ in four and higher dimensions, it instead grows without bound. The energy density decreases in both cases. These geometries provide an explicit nonlinear realization of creation with a controlled finite-volume past and an indefinitely expanding future.

We have also determined the linearized perturbations of the undeformed cosmic creation gate while keeping all bulk regions locally de Sitter. In three dimensions, the generic inhomogeneous sector contains one hybrid sound mode and \(n-1\) relative transverse modes, while in four and higher dimensions only the homogeneous hybrid sound mode survives, apart from isometries. A controlled, finite-volume past requires fine-tuned initial conditions that exclude modes growing towards the infinite past. Our second-order calculation in the inhomogeneous sound sector reveals a further restriction when a definite past-volume limit is required: waves traveling in opposite directions at the same angular harmonic produce persistent volume oscillations that prevent a definite past limit. The resulting necessary second-order condition requires each harmonic present in the infinite past to be chiral, with the direction of propagation allowed to differ between harmonics. The allowed waves generally give an amplitude-dependent correction to the limiting volume, preserving its positivity for sufficiently small amplitudes at the order calculated. The homogeneous deformations leave the past volume unchanged. However, we can also impose a second criterion: persistent volume oscillations are allowed as long as the volume remains within finite, strictly positive bounds and the geometry and fluid remain regular.

These results clarify the Cauchy conundrum: data satisfying all local gravitational constraints and junction conditions need not describe a history with a regular, finite-volume past. Creation restricts their entire past development, and further conditions on the allowed data may arise at higher orders. We call the histories satisfying these requirements \textit{conditioned admissible histories}. To formulate their causal response to later disturbances, we fix an earlier admissible history and introduce smooth junction sources compatible with total energy-momentum conservation and all junction conditions. The induced response vanishes before the sources act, preserving the prescribed past, and may subsequently grow without any imposed bound on the future spatial volume. Under the stated convergence and regularity assumptions, this retarded construction extends order by order over the global time band where the perturbative expansion remains valid; subsequent evolution may require a nonperturbative continuation. 

When the bulk and junction curvature radii remain large compared with the Planck length, a semiclassical description can remain valid arbitrarily far into the past, even at what we interpret as the beginning of time, provided a suitable quantum state exists whose renormalized bulk and junction backreaction remains controlled. We discuss this quantum-state requirement more below towards the end of this section.

In the emergent-universe scenario of Ellis and Maartens \cite{Ellis:2002we,Banerjee:2007sg}, a closed universe approaches an Einstein static state of finite spatial volume in the infinite past and subsequently undergoes inflation. With suitable curvature scales, such a history can remain within the domain of classical gravity throughout its evolution. Our construction realizes a finite-volume past through a perfect-fluid gravitational junction joining locally de Sitter regions. The junction fluid clock synchronizes the natural cosmological foliation, while the junction conditions determine the average and relative transverse dynamics of the regions joined. Several universes can therefore be created together, with their admissible cosmological histories restricted by the requirement of a controlled, finite-volume past.

An immediate outstanding problem is to construct the full nonlinear inhomogeneous creation histories and establish convergence of their analytic expansions and global single-valuedness of their embeddings. The question is which combinations of the hybrid sound and relative transverse modes, consistent with the chosen past-volume criterion, admit a continuation with a finite, strictly positive past volume and a regular, possibly expanding future. For the first criterion, the second-order amplitude restrictions obtained here and any further restrictions at higher orders must be compatible with a nonlinear continuation.

A further question is how these solutions are described in the dS/CFT correspondence. This would extend the analysis of interface CFTs in AdS \cite{Bachas:2020yxv,Chakraborty:2025dmc,Banerjee:2025zuw,Chakraborty:2026wip,Arav:2026bbn} and the maps between Euclidean CFTs discussed in \cite{Chakraborty:2025baw}. The condition of a finite initial volume and the associated fine-tuning of initial data would then require an interpretation in this language, which could also provide a more non-perturbative description of the creation mechanism.

For cosmology, creation must be accompanied by a quantum scalar field whose fluctuations can seed structure formation in one or more of the universes created at the junction. This field could play the role of the inflaton in generating primordial fluctuations, without having to drive the inflationary expansion itself, which is already provided by the de Sitter geometry. In a forthcoming work \cite{Chakrabarti:2026spacelike}, we study quantum fields at a spacelike version of the two-way junction considered here. We find that a junction preparing a quantum state with finite backreaction can also produce a physical clock in the matter sector, initiating cosmological evolution in the future universe while the incoming de Sitter region remains empty. An extension to the timelike junctions studied here could allow multiway junctions in which one of the locally dS$_3$ universes remains empty while the others are populated by quantum fields and physical clocks. The relative clocks already present in the gravitational junction \cite{Chakraborty:2025baw} should then be related to those carried by the bulk quantum matter fields. This would require solving the quantum matching conditions together with the semiclassical gravitational backreaction.

Fundamentally, the fine-tuned nature of the classical solutions describing creation suggests that the initial quantum states of the fields seeding structure formation must also be carefully selected. This is explicitly realized in \cite{Chakrabarti:2026spacelike}, where the bare transparent junction leads to uncontrolled backreaction on the background geometry. A controlled construction is obtained by coupling the bulk quantum field to ancillary quantum degrees of freedom on the junction. Choosing the initial state of these ancillary degrees of freedom and conditioning on their final state defines a conditional transition operator for the bulk field. Extending this construction to multiway junctions would provide a quantum description of the creation of \(n\) universes together with the preparation of fluctuations that can seed structure formation in at least one of them.

In the timelike creation solutions, the spatial volume retained by the moving junction can increase even while the parent global de Sitter geometry is contracting in the infinite past. Thus, the expansion of the spatial volume need not coincide with the local Hubble expansion measured by observers. Once quantum fields and physical clocks are included, the cosmological evolution measured by these clocks and the resulting redshift--distance relation remain to be determined. In particular, the observed cosmological evolution need not be determined by the scale factor of the parent de Sitter spacetime alone. We leave these questions for future investigation.

Whether this distinction could shed light on the Hubble tension between
the expansion rates inferred from the cosmic microwave background and
local observations \cite{Planck:2018cosmo,Riess:2022shoes} remains an open
question. Addressing it would require understanding how the junction
dynamics and physical clocks affect the observed redshift--distance
relation.

\begin{acknowledgments}
We are deeply grateful to Radouane Gannouji and Jorge Nore\~na for valuable discussions. We acknowledge the use of ChatGPT (OpenAI) for assistance with calculations and with revising the manuscript. The authors take full responsibility for the scientific results and the final text. AM and MM acknowledge support from FONDECYT regular grant no. 1240955 and ``Doctorado Nacional'' grant no. 21250596 of La Agencia Nacional de Investigaci\'{o}n y Desarrollo (ANID), Chile, respectively. Avik Chakraborty appreciates the warm hospitality extended by AM and Instituto de F\'{\i}sica, Pontificia Universidad Cat\'{o}lica de Valpara\'{\i}so, Chile, where all the work was carried out.
\end{acknowledgments}

\bibliographystyle{JHEP}
\bibliography{Refs}
\end{document}